# Hocus-Socus: An Error Catastrophe for Complex Hebbian Learning Implies Neocortical Proofreading

## Cox, Kingsley J.A.[1,2] and Adams, Paul R.[1,2]

[1]Dept Neurobiology, Stony Brook University, NY
[2]Kalypso Institute, Stony Brook, NY

Correspondence to: P.R. Adams[1]  Correspondence and requests for materials should be addressed to P.R.A. (email: padams@notes.sunysb.edu)

**The neocortex is widely believed to be the seat of intelligence and "mind". However, it's unclear what "mind" is, or how the special features of neocortex enable it, though likely "connectionist" principles are involved \*[A]. The key to intelligence[1] is learning relationships between large numbers of signals (such as pixel values), rather than memorizing explicit patterns. Causes (such as objects) can then be inferred from a learned internal model. These relationships fall into 2 classes: simple pairwise or second-order correlations (socs), and complex, and vastly more numerous, higher-order correlations (hocs[B]), such as the product of 3 or more pixels averaged over a set of images. Thus if 3 pixels correlate, they may give an "edge". Neurons with "Hebbian" synapses (changing strength in response to input-output spike-coincidences) are sensitive to such correlations, and it's likely that learned internal models use such neurons. Because output firing depends on input firing via the relevant connection strengths, Hebbian learning provides, in a feedback manner, sensitivity to input correlations. Hocs are vital, since they express "interesting" structure[2] (e.g. edges), but their detection requires nonlinear rules operating at synapses of individual neurons. Here we report that in single model neurons learning from hocs fails, and defaults to socs, if nonlinear Hebbian rules are not sufficiently connection-specific. Such failure would inevitably occur if a neuron's input synapses were too crowded, and would undermine biological connectionism. Since the cortex must be hoc-sensitive to achieve the type of learning enabling mind, we propose it uses known, detailed but poorly understood circuitry and physiology to "proofread" Hebbian connections. Analogous DNA proofreading allows evolution of complex genomes (i.e. "life").**

This view, combining insights from synapse biophysics, molecular evolution, neocortical anatomy and neural learning theory, seems as unpromising as the notion that life is the outcome of amplified molecular accidents, to which it is closely linked[C]. Recent data suggest that Hebbian adjustments are highly[3], but not completely[4,5] specific, because of excellent (~99%) confinement of calcium[6,7] and its effects[8] by spines[D]. Since biological processes are usually error-tolerant the observed specificity might suffice for learning hocs, but this has never been tested, and there is a highly relevant case where extraordinary accuracy is essential, DNA replication. Darwinian evolution, a type of chemical complex learning from the world[9], is only possible because error rates for base copying are comparable to reciprocal genome lengths[E,9,10]. Accurate replication (per-base error rates $<10^{-9}$) is achieved by multiple mechanisms[11] which evolved in a series of information-enriching transitions[12]: relatively (~99%) selective base-pairing; selectivity conferred by replicases; proofreading; mismatch repair. The largest contributor is proofreading, reducing the error rate from $< 10^{-3}$ to $<< 10^{-6}$, since 2 independent pairing events must concur.

---

*superscripted letters refer to Supplementary Notes; the Supplement also contains additional material.

We proposed[F,13,14,15] that key neocortical circuitry accomplishes a conceptually-identical proofreading operation on 2 independent measures of spike-pairing, allowing large improvements in Hebbian accuracy, and otherwise usually impossible feats of learning. We now show that the required circuitry closely matches recent data[16] and describe computational results providing the crucial missing link: complex learning by a model neuron typically collapses to simple learning, if Hebbian specificity falls below a threshold comparable to that expected (and observed) from biophysics. This test requires a model where learning (i.e. convergence to stable weights) depends on input correlations generated in a defined manner, and on a nonlinear Hebb rule. Independent Component Analysis (ICA)[17,18] meets these requirements. This model learns implicit target weights using the hocs in an ensemble of input patterns. We used the simplest possible "crosstalk" model[G,19], corresponding to the usual genetic assumption of base- and position-independent copying-accuracy, though we obtained similar results when "hotspots"[F] were introduced. As in Darwinian evolution[9,10,E], the threshold depends on the learning task[H], but typically falls within a biophysically-plausible range, so if the neocortex is to solve diverse problems, it must have wetware overcoming this limitation.

Complex learning power reflects the number of inputs whose hocs can be exploited by Hebbian rules, and is therefore best done in individual neurons, rather than dendritic segments[I]. Our model is based on single-unit ICA[J,17,20], a minimal hoc-based abstraction of object identification. Input vectors $\mathbf{x}$ (with pixel-like elements $x$,) generated by linearly combining n independently-fluctuating unknown object–like "sources" using an approximately orthogonal[K] square "mixing" matrix $\mathbf{M_o}$ are applied to the adjustable weights $w$ of a neuron whose output $y$ (the inferred "object") is the weighted input sum $\mathbf{x^T w}$ (Fig.1a). $x$ and $y$ are mean rates rather than detailed descriptions of firing times which may be necessary to predict real neuron output, since this "connectionist" model doesn't respond to temporal sequencing. Timing would make it even more difficult for real synapses to achieve high specificity[L]. The $i$th weight adjustment is made using the nonlinear Hebbian rule $\Delta w_i = +/- k\ x_i\ f(y)$. k is a small learning constant, and f any sufficiently smooth nonlinearity; we usually used the statistically robust[17] tanh which for typical superGaussian sources requires a negative sign ("antiHebb") in the rule[17,20]. In real neurons this multiplication could be implemented by spike coincidence detection[M]. Linear Hebb rules are only sensitive to pairwise correlations[19,21]; nonlinearity provides additional sensitivity to hocs[O]. Hebbian rules produce weights that grow or shrink without limit, and require stabilization: we divided the weight vector by its new length after each adjustment[20]. Similar "normalization" could be achieved by a variety of mechanisms and is "multiplicative", confining the weight vector to a unit sphere[22] (Fig. 1b).

The 1-unit rule also requires that inputs be preprocessed, or "whitened", to remove socs; we found that partial whitening typically sufficed[P]. A random $\mathbf{M}$ was used to generate an initial batch (typically $10^3$) of mix vectors, for which a small-sample covariance matrix $\mathbf{C_s}$ was calculated[P]. $\mathbf{M_o}$ was formed using $\mathbf{M_o} = \mathbf{C_s}^{-1/2}\mathbf{M}$, so $\mathbf{C_L}$, the large-sample ($10^5$) covariance matrix of the imperfectly decorrelated, "off-white", mix vectors, is close to a scaled identity matrix $\mathbf{I}$ (Fig. 1c), to an extent that depends on the small-sample size. In practice perfect decorrelation cannot be achieved using reasonable samples, or with biological crosstalk and finite k[Q19]. If the vector $\mathbf{w}$ converges to a row of $\mathbf{M_o}^{-1}$, the output tracks a source; to simplify model learning and its interpretation, in most cases all sources but one were Gaussian so only one equilibrium, extracting the nonGaussian source, is stable[R,17,20,23] (Figs. 1b, 2a,b).

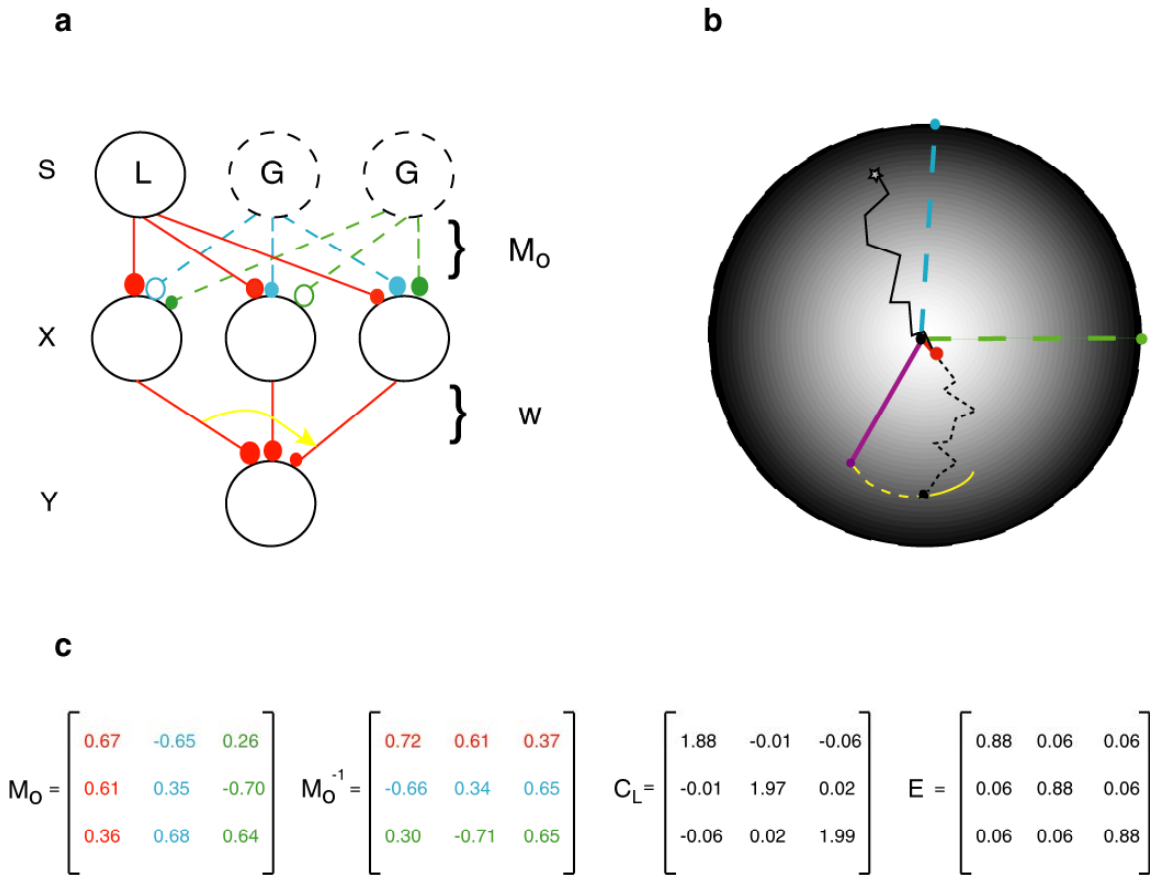

**Figure 1  The ICA-with-crosstalk model: structure, behaviour, parameters**

**a** shows a model neuron (output y) receiving input from 3 mix signals x via adjustable connection weights w. The mix signals are formed by combining 3 independently and symmetrically fluctuating sources s via a set of fixed mixing coefficients (different size colored dots; open dots are negative), the elements of the almost orthogonal matrix $\mathbf{M_o}$. In practice ICA is done in 2 stages: initial linear PCA ("whitening") followed by nonlinear learning; these are combined in this figure by replacing $\mathbf{M}$ by $\mathbf{M_o}$. The first column of $\mathbf{M_o}$ corresponds to the red coefficients, arising from the nonGaussian (typically Laplacian) source shown as a solid circle ("L"). The other sources are Gaussian (dotted circles; "G").

**b** diagrams schematically the 3 weights, zigzagging (solid: subthreshold crosstalk; dotted: suprathreshold) under the influence of successive patterns, and confined by normalisation to the unit sphere. For low crosstalk weights zigzag to the IC (red dot), above threshold to the approximate PC (solid yellow line). The 3D weight surface has been rotated so the direction of the red column of $\mathbf{M_o}$ points straight at the reader, so the direction of the first row of $\mathbf{M_o}^{-1}$ (to which it is almost parallel) points almost to the reader (short solid red line starting at the black dot origin and terminating on the sphere as a red dot). The other directions of almost orthogonal rows of $\mathbf{M_o}^{-1}$ are also shown as blue and green dotted lines. The red dot is the target weight vector that allows the neuron to track the nonGaussian source (the "IC"). The purple line shows the least PC, and the yellow line the loci of the terminations of the least eigenvectors of $\mathbf{EC}$ on the sphere (i.e. the stable weights obtained using purely Gaussian sources at various errors). Just suprathreshold error triggers a movement (dotted zigzags) from the approximate IC to the square; further increase in error moves the learned average weights along the solid yellow line; the dotted yellow line is not stable when sources are nonGaussian.

**c** shows the mixing matrix $\mathbf{M_o}$ and its inverse, the unmixing matrix $\mathbf{M_o}^{-1}$; the red row is the only stable IC and corresponds roughly to the red coefficients in the first column of $\mathbf{M_o}$. Since $\mathbf{M_o}$ is only approximately orthogonal, the covariance matrix $\mathbf{C_L}$ of even a large (100,000) batch of $\mathbf{x}$ has offdiagonal elements very small but nonzero, and slightly unequal diagonal elements. The error matrix $\mathbf{E}$ (which has equal diagonal elements Q and equal offdiagonal elements (1-Q)/(n-1)) is shown with entries corresponding to the threshold in Fig. 2. For further detail see Supplementary Legends.

We introduced crosstalk by modifying the rule to $\Delta\mathbf{w} = +/-k\mathbf{E} \mathbf{x} f(y)$, where $\mathbf{E}$ is a symmetric "error matrix" assigning a fraction (1-Q) of an adjustment to the other weights, dividing it up according to the offdiagonal elements of $\mathbf{E}$[S,T,19]. Zero crosstalk, assumed in standard models, implies Q ("quality") = 1. Usually we set offdiagonal elements of $\mathbf{E}$, and also diagonal elements, to be equal (Fig. 1c), corresponding to the standard connectionist assumption that all connections of a given type are equivalent, and to spatiotemporal averaging of varying synaptic configurations[T,19].

In most tests the nonGaussian source had a Laplacian distribution[U]. With zero error the rule converged to the weights corresponding to this source, the "IC" (Figs. 1, 2). A low level of error ("crosstalk", expressed as a per-connection quantity that is independent of n[T,19]) produces only slight degradation of learning, but, crucially, above a narrow threshold range, weights snap from the IC to a new average direction[V]. If the nonlinear rule fails to learn from hocs above a threshold, this new direction could correspond to mere soc learning. We tested this using the same $\mathbf{M_o}$ but with all sources Gaussian, so the mix vectors exhibit only socs[W]. Now the error-free nonlinear rule learned the least eigenvector of $\mathbf{C_L}$, as expected for an antiHebbian rule[P]. As crosstalk increased, the learned vector gradually moved away from this direction, and above the nonGaussian threshold the weights learned for either mixed Laplacian-Gaussian sources or pure Gaussian sources were identical (Figs. 1b, 2a,b). Furthermore, the learned vector for Gaussian sources tracked the expected theoretical curve[19] (corresponding to the least eigenvector of $\mathbf{EC}$) for a linear rule (Fig. 2a), although the rule is nonlinear. Crucially, minor crosstalk makes the nonlinear rule behave linearly, ignoring hocs. This was true for different $\mathbf{M}$s (though the threshold varied; for 4 cases studied in detail the average threshold was 0.04 +/- 0.03 (SD)), and for different source distributions or degrees of whitening (being more error sensitive for lower kurtosis sources or less whitening[X]).

We are not proposing the brain does ICA, though it may do something similar[Y,24,25,26]. Instead our results suggest a principle: nonlinear Hebbian rules become insensitive to hocs above a threshold crosstalk level[Z]. A normalized nonlinear correctly-signed rule automatically learns ICs if inputs are generated by square linear mixing. If inputs are generated differently, for example by rectangular mixing, nondeterministically or nonlinearly, a single neuron may not learn any stable weight vector[27], but if it does, enough crosstalk will cause failure[AA].

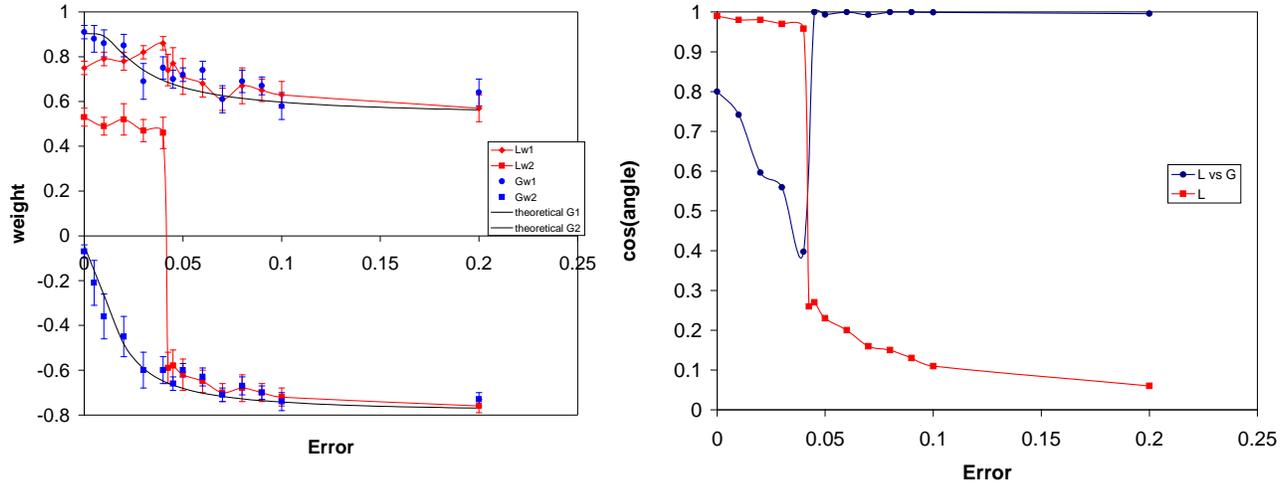

**Figure 2  Crosstalk causes hoc learning to collapse to soc learning in 1-unit ICA.**

Calculations were done using the conditions in Fig. 1, with 3 inputs and weights, using f(y)= tanh(y), explicit normalisation and antiHebb learning. The left hand plot shows two of the steady-state averaged weights using 1 Laplacian source (red) or all Gaussian sources (blue; error bars show SD). At the threshold per connection error b = 0.0425 the Laplacian weights snap to match the all-Gaussian weights. The per connection error b is related to Q (Fig. 1) by b = (1-Q)/nQ, and expresses the expected dependence of crosstalk on biophysical parameters[19]. The black lines show theoretical weights calculated from the least eigenvectors of $EC_L$[19]; $C_L$ was estimated from a sample of 100,000 input vectors. The right hand plots show the cosine of the angle between the Laplacian and Gaussian weight vectors as a function of error (blue line); at the critical error the Laplacian weight vector jumps to the Gaussian weight vector. The red line show how the cosine of the angle between the weight vector and the first row of $M_o^{-1}$ (Laplacian source) changes with error (again with sharp change at b = 0.0425). See Supplementary Legends for details. k = 0.002; similar results were obtained with k = 0.0002

Why does crosstalk prevent learning from hocs[BB]? A weight vector parallel to a row of $M_o$ is a stable equilibrium of an accurate averaged nonlinear rule[17,20] but the rule may have other equilibria. For a linear rule and any nonwhite input distribution the eigenvectors of $C$ (Principal Components; PCs) are equilibria, and the greatest (or for antiHebb, least) is, typically, stable[19,21,P]; this should also be true for nonlinear rules with Gaussian inputs. We confirmed that the leading or (for antiHebb rules, least) eigenvector of $C$ is stable for Gaussian inputs, even for a cubic nonlinearity with no linear term. However for nonlinear rules, sufficiently nonGaussian inputs destabilize this PC, and stabilize the IC[17,20]. Suprathreshold error apparently nulls the nonlinearity, destabilizing the IC and restabilizing the approximate PC (least or largest eigenvector of $EC$), because error moves the equilibrium weights slightly away from the IC, eventually invalidating the stability proof[17,20]. Further study should reveal what factors other than source kurtosis and orthogonality[X] of $M_o$ (e.g. $M$, n and bit resolution) set the threshold and its sharpness, and why. However, either these factors reflect the task specifics (and cannot be circumvented) or unavoidable neural limitations.

These results suggest that individual neurons typically cannot learn connection weights that reflect input hocs unless the necessarily nonlinear Hebbian rule is highly (~95%) accurate[T]. Indeed, they underestimate the problem since they ignore additional effects of crosstalk on biological prewhitening[T,19]. Even if only ~1% of the calcium entering a spine escapes to the shaft[7] but there are 10 or more synapses within range of that calcium[28,29], such accuracy may be unobtainable[19CC]. Crowded synapses are inevitable if neurons learn from many inputs[W]. Since neocortical neurons manifestly do learn from hocs, even though their numerous inputs may obey more less favourable statistics than for ICA, the neocortex must presumably use a non-synaptic (neuronal) strategy for increasing Hebbian accuracy[13,14,15]. The root problem is coincidence-detection failure: because of intracellular messenger diffusion from nearby synapses experiencing coincidences, a connection may register a "false spike-pair", analogous to incorrect base-pairing. The obvious way to overcome this uses an additional, independent measure of near-coincident firing of input and output neurons contributing to the synapse; double mistakes should be rare. We have suggested[13,14,15] the neocortex might contain (in layer 6) dedicated "Hebbian neurons" detecting coincidences across connections, using branches of the relevant axons, and supplying these independently detected coincidences in near real-time (<100 msec[DD]), to the relevant (probably thalamocortical) connections, so if the "second" (neuronal) coincidence confirms a "first" (synaptic) coincidence, the relevant weight is allowed to change. Fig. 3a diagrams the necessary wetware. Selective confirmation delivery to the relevant connection could be achieved by applying it pre- and postsynaptically (e.g. to a relay and its layer 4 target), requiring that both sides of the connection receive it (Fig. 3a). This "proofreading" strategy would seem to need a dedicated proofreading neuron for every anatomical feedforward connection (recurrent connections may not need proofreading if they learn socs) even if comprised only of silent synapses. However, since coincidences across connections are probably rare (antiHebbian learning and NMDAR maturation tend to reduce them), a proofreading neuron could monitor many connections in a distributed manner (Fig. 3b): while it could wrongly enable strength changes at connections not experiencing genuine coincidences, this would vanish in the sparse coincidence limit (much as interference in associative memories vanishes for sparse patterns). The diagram in Fig. 3b (see also Supplement Fig. 1), which goes beyond our previous sketch[15], matches known but mysterious "universal" thalamocorticothalamic circuitry and physiology[30], and could form the backbone of the cortical "column". There is remarkably close agreement between these requirements for distributed proofreading and recent counterintuitive data on the pattern of CT feedback[16,EE].

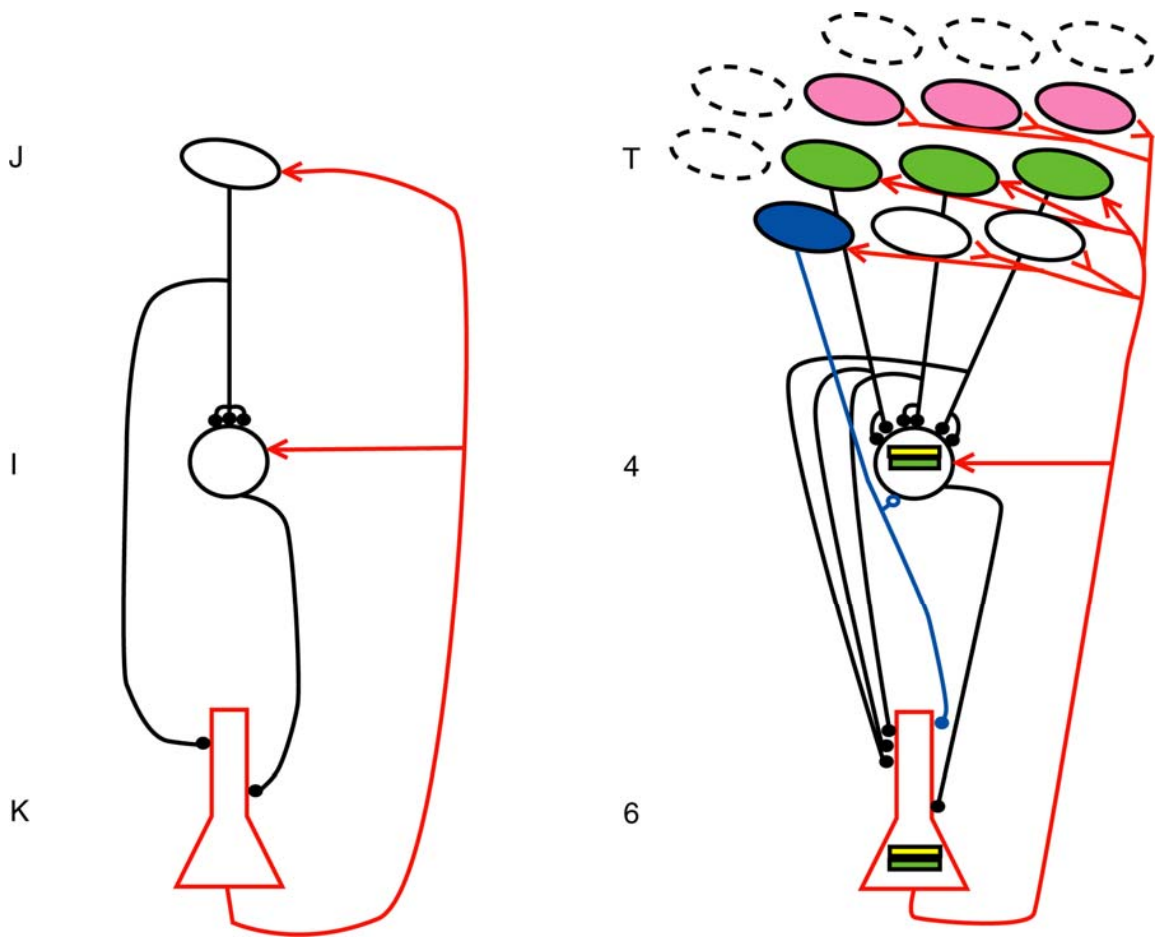

**Figure 3    Proposed thalamocortical circuitry for dedicated or distributed proofreading.**

Fig. **3a**  (Left) Dedicated Proofreading. An input J connects to a neuron I and also, via a weaker connection, to a dedicated proofreading partner K. The K cell also receives weak input from the I cell. K feeds back to both J and I via modulatory connections (red arrows). K fires when the J and I cells fire near-coincident spikes and shifts the target J-I connection to "plasticity approved" mode by conjoint modulation of the input and output sides.

**3b**.  (Right) Distributed "PushPull" Proofreading. The diagram makes the identification J = thalamic relay (T), I = layer 4, K = layer 6 CT cell and shows one possible version of distributed proofreading, for concreteness drawn for an orientation-tuned layer 4 simple cell (i.e. responding to a horizontal edge) ; only off relays, which generate the green off-lobe of the layer 4 RF, are shown; the green, "overlapping' and "matching"[16],  relays contribute to the off-lobe; the on-lobe is shown yellow, and the corresponding "overlapping" but "nonmatching" [ZZ] off-relays are shown pink – these do not connect to the layer 4 or 6 cells shown). The layer 6 cell firing modulates its partner layer 4 cell directly to briefly enable thalamocortical plasticity postsynaptically. It also modulates the set of thalamic relays (green and blue ovals) that innervate (by silent or nonsilent synapses) its partner layer 4 cell, via a TRN inhibitory cell (omitted), which shifts relays to burst mode, briefly enabling thalamocortical plasticity presynaptically (red arrows). Both pre- and post-enabling are required for the strength change triggered by T-4 spike-coincidence to be expressed; such dual-enabling occurs if the 6–cell rapidly confirms the spike-coincidence "seen" by the relevant thalamocortical synapses. Enablement should be executed before the typical arrival of the next coincidence (~100 msec -10 sec). The dotted ovals correspond to "unavailable" relays that cannot reach the dendrites of the illustrated layer 4 cell. This is a "functional" diagram; see Supplement for an anatomical diagram showing the intervening TRN cell, which innervates all the nondotted relays.

Currently unconnected "incipient" relays[14] , including "nonmatching" relays (pink) and nonoverlapping, open undotted relays , that could form synapses on the 4-cell receive direct depolarizing modulation (reversed red arrows) which maintains them in tonic, plasticity-disabled, mode (unless they receive enabling signals from other 6-cells monitoring the connections they do form, on other 4-cells). Some "nonoverlapping" connected relays (e.g. blue oval), make only silent synapses (open blue dot) and therefore do not contribute to the receptive field. These silent connections must be monitored and should receive enabling input. The scheme closely fits recent results[16,EE]. See Supplement for details.

Our results also suggest a generalization of Eigen's "error threshold"[9,10] (setting the maximum size of genomes) to other forms of learning: learned information depends on the reciprocal of the learning error rate. This seems true for socs[FF,19]. For hocs, the learned information at zero error, the product of the vector dimension (n) and the $w$ bit resolution, evaporates at the threshold. Thus the effect of error on soc and hoc learning is quantitatively the same but qualitatively different, being gradual (and tolerable) for the former and abrupt (and catastrophic) for the latter[FF].

This explanation of the neocortical basis of sophisticated learning by individual neurons, the key to intelligence and "mind", is simple, and parallels that accepted as the key to "life"[GG]. In intelligent brains neurons must learn from hocs; perfect Hebbian synapses could accomplish this, but in practice crosstalk usually makes this impossible. A "proofreading" mechanism, conceptually identical to that allowing the evolution of complex genomes ("life"), would allow such learning and matches known, but enigmatic, thalamocortical anatomy and physiology. Nevertheless, even with proofreading, cortical neurons could probably only handle around 1000 inputs (as typically observed), since otherwise synapses become so crowded that crosstalk would increase to the point where hoc learning fails[CC]. This would vastly restrict the learning power of neurons and brains. Evolution may provide useful analogies for understanding learning and intelligence. Perhaps further major evolutionary transitions after DNA/protein[12] provide useful clues about mechanisms enabling human levels of mind[HH,II].

**Acknowledgements**

We thank the following for comments on a draft of this paper: H. Barlow, P. Dayan, J. Diamond, S.Y Ge, C. Gilbert, G. Hinton, C. Jahr, D. Johnston, A. Maffei, R. Malenka, M. Maravall, H. Markram, D. McKinnon, S. Nelson, R. Nicoll, E. Oja, D. Purves, R. O'Reilly, M. Rattray, M. Sherman, S. Siegelbaum, G. Stuart, C. Thornton, A. Treves, L. Turin, R. Yuste, Q. Zhou. We also thank the following for other input: J.M. Alonso, M. Bethge, M. Deschenes, T. Elliott, A. Hyvarinen, W. Kim, M. Larkum, K. Likharev, K. Martin, A. Radulescu, T. Sejnowski, M. Sherman, A. Sillito, G. Stuart, H. Swadlow, R. Yuste, Q. Zhou

# Supplementary Information

The Supplementary Information is arranged in five sections as follows:

1 Supplementary Figure
2 Supplementary Figure Legends
3 Nonmathematical Guide to PCA, ICA and figures 1,2
4 Supplementary Notes
5 References to the Supplementary Information

## <u>Section 1</u>

**Supplementary Figure**

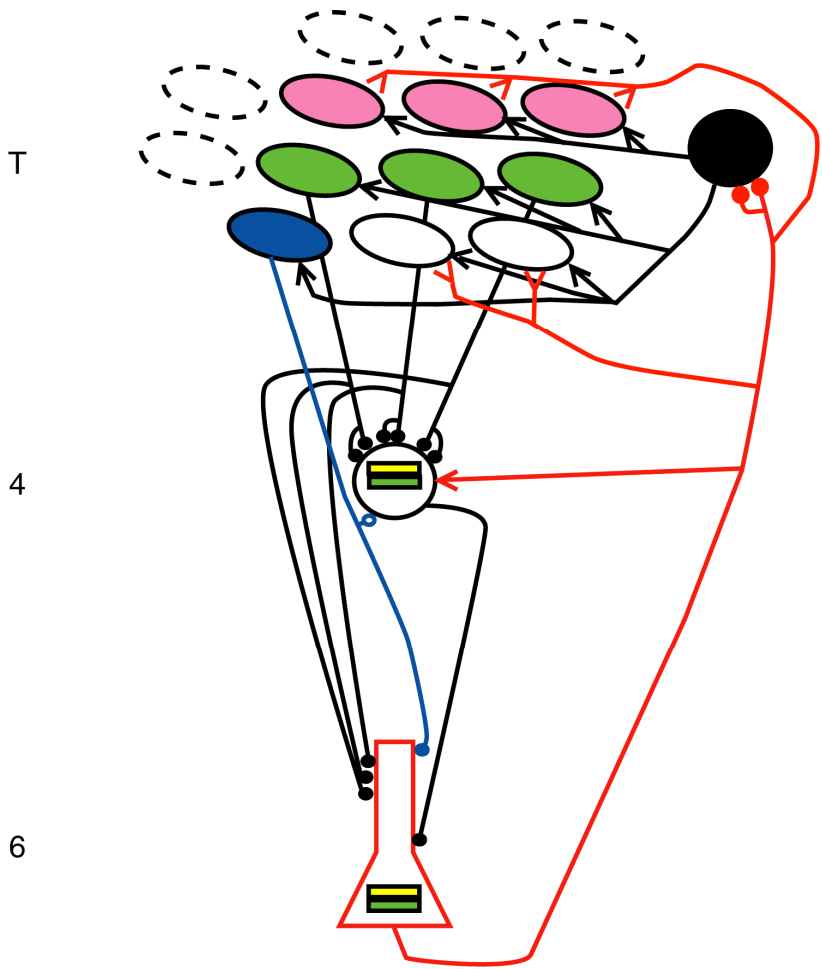

**Supplementary Figure Legend**

This figure illustrates the likely anatomical basis for the functional push-pull "presynaptic unitary distributed proofreading" ("prud") circuit shown in Fig. 3b. It should be read in conjunction with Supplementary Note EE. The main difference from Fig. 3b is the explicit depiction of the TRN inhibitory neuron (solid black) which mediates the disynaptic CT- induced hyperpolarisation which shifts relays to burst mode (and would therefore half-enable the plasticity of the connections they make in layer 4; we use the term "half-enable" because presynaptic enabling via relays is not enough to confirm a coincidence). Symbols are the same as in that figure, except the functional red (plasticity-enabling) arrows have been replaced by black (plasticity-enabling) "anatomical" arrows.

In both Fig. 3b and this Supplementary Fig. the direction of the feedback arrows indicates whether the feedback could either enable plasticity (i.e. confirm the occurrence of a synaptic coincidence and allow strengthening to proceed)) or disenable plasticity (maintain the connection in an implastic, default state). Forward arrows mean "enabling" and reverse arrows mean "disenabling". In the case of CT feedback enabling is done by hyperpolarisation (va TRN), which shifts relays to burst mode. CT disenabling is done by direct depolarisation, which shifts relays to tonic mode. However, in Fig. 3b the enabling arrows are drawn as functional red "direct" connections but in reality, as shown here, there is an intervening inhibitory neuron, so the enabling connection is anatomically indirect, and is represented by a excitatory synapses onto TRN followed by enabling (forward) black arrows. The feedback to layer 4 is shown as purely enabling (direct red arrow), even though this is an excitatory synapse (onto dendritic shafts via "drumsticks"), because here enabling is not done by hyperpolarisation into burst mode, but via mGluRs. There could be additional postsynaptic disenabling of other layer 4 cells, but this is not shown in the figures. It is theoretically possible that the feedback to layer 4 could follow a similar pushpull scheme to that shown for feedback to thalamus: enabling could be done globally, e.g. via inhibition, with disenabling done locally, via drumsticks. In this case the pattern of feedback to layer 4 shown in both Figs. 3b and here would be functional, not anatomical.

Thus green and blue ovals are relays that currently connect anatomically to the indicated layer 4 cell, and empty and pink nondotted ovals are currently unconnected ("incipient") relays which could become connected (to the indicated layer 4 cell) merely by spine addition to the current, relatively stable, axodendritic geometry. We refer to the total pool of incipient and connected relays (i.e. the current set of relays whose layer 4 axons pass within a spine length of the dendrites of a given layer 4 target neuron) as the "available" pool. Dotted ovals represent "unconnectable", unavailable, relays whose layer 4 axons arborize so far from the dendrites of the indicated layer 4 cell they cannot form synapses on it by creation of a spine (of course this population shifts early in development as the newly arrives layer 4 cells and thalamic axons adjust their arborisations; during this period there must be appropriate rewiring of TRN connections). Unavailable relays do not get CT feedback from the relevant layer 6 CT cell, and these would account for the relays that do not get feedback in Wang et al.[1]

 For concreteness, we suppose that the layer 4 cell is an orientation-tuned spiny stellate "simple" cell in the striate cortex of a cat, which responds selectively to an edge that aligns with its RF. The RF is composed of flanking extended lobes, shown diagrammatically inside the layer 4 simple cell and its layer 6 simple partner (off-lobe green, on-lobe yellow). All the relays shown are off –center relays; the off-center relays that contribute to the RF of the layer 4 cell are shown in green. These are, in the terminology of Wang et al.[1], "overlapping" (they fall within the RF) and "matching" (their polarity conforms with the RF) relays. The pink off relays have their centers at the same visual location as the on relays which generate the on-lobe of the cortical RF (they are "overlapping"), but they do not contribute to the RF (they are nonmatching), and almost certainly do not even silently connect to the 4 or 6 cell (such disconnection would be favoured by persistently anticorrelated firing). However, the pink relays do contribute to the RFs of other layer 4 cells (not shown) and make silent (or very weak) connections on still others. The blue relay, although nonoverlapping and therefore not contributing to the RF, nevertheless makes a silent connection (at least one silent synapse, shown as an open blue dot) on the layer 4 cell, and a semidriving connection on the layer 6 cell (we hypothesize that layer 6 cells require the conjunction of thalamic and layer 4 input to fire; these connections are therefore "semidriving"). Thus the figure shows the 5 possible classes of off relay: overlapping and matching (green; drivingly connected), overlapping and nonmatching (pink; connectable but unconnected), nonoverlapping yet silently connected (blue), nonoverlapping, connectable but unconnected (open solid ovals) and unavailable (dotted). A critical issue is whether the proofreading hypothesis can successfully account for the various patterns of TC feedback reported by Wang et al.[1,EE]. We know of no other hypothesis that can do so.

 It is known that CT axons contact both TRN neurons (which contact relays) and relays. The detailed pattern of connections has not yet been fully worked out, although all available information is consistent with that depicted here[2,3,4]. The populations of dotted (unavailable) and undotted (available) relays exchange early in development as axon and dendrites grow or retreat, but within a topographic framework which is probably set by initial processes in the subplate and perireticular nucleus[4]. Once these axodendritic patterns have stabilized, further adjustments in connectivity are made within the available pool by (probably random) spine addition/ deletion at close axodendritic approaches. Spine addition leads to new, but silent, synapses. Synapses that stay silent for long periods are deleted. The lifetime of silent synapses (which reflects the inverse of the sum of the rates of unsilencing plus removal)  seems to vary with maturation: in the adult they last longer, so spine turnover is reduced[6]. In our proofreading model these synapses are unsilenced by batches of spike coincidences, but only if their plasticity is enabled by CT feedback (i.e. if the synaptic coincidences are appropriately confirmed). After an initial period of axodendritic growth, further adjustments in the TRN connections are not required, although ongoing adjustments of CT-relay connections, by "drumstick" formation/deletion, within the available pool, are. We hypothesize that in the mature animal drumstick turnover is the main mechanism that allows CT feedback connectivity to be updated to keep up with ongoing adjustments in TC connectivity, so geometric rewiring of axons or dendrites is not required. It is known that the spatial distribution of CT axon arbors in TRN and in dorsal thalamus proper are rather similar, in agreement with the scheme shown[3]. TRN axons within dorsal thalamus have narrow distributions and contact many of the relays whose dendrites are available to them, but it is not known whether they contact all available dendrites (as shown here)[2]. CT axons in dorsal thalamus have wide distributions [3,4], because they have to "hunt" for the currently incipient relays, which tend to lie either orthogonal to the direction of the tuning of the layer 6 cell (after all, these are available but disconnected, and are systematically uncorrelated with that 6 cell) or parallel to it (these connect to nonmatching off relays, shown in pink). If a CT axon excites a TRN cell which then inhibits the relay targeted by that CT axon, the arrangement is referred to as "feedforward inhibition" or "closed-loop inhibition"[2]. If a CT axon excites a TRN cell which inhibits a relay that does not get direct excitation from a CT axon, the arrangement is referred to as "lateral inhibition" or "open-loop" inhibition. In the scheme shown inhibition is a mixture of open and closed types, in agreement with data [2,7.] It is also possible that some relays receive only direct CT input, but this would imply highly selective, and malleable, targeting of inhibitory input, and is not consistent with the typical high divergence of inhibitory neurons.

"Pushpull" refers to the idea that a CT axon "pushes" (directly depolarizes, tonic-enables and plasticity-disables) some relays and "pulls" (indirectly hyperpolarizes, burst-enables and plasticity-enables) others, such that the balance between the net contribution from various CT inputs determines the ongoing relay mode (burst or tonic) and thus influences the plasticity of its central terminals.

"prud" refers to the particular, extreme, version of distributed proofreading shown here, in which a CT cell feeds back (directly or indirectly) to a single permanent layer 4 "partner" cell, and to the incipient and actual relay inputs to that 4-cell. This seems to be the standard form of proofreading that cortex does, because typically CT cells share the receptive field properties of neurons in the overlying column, rather than of the relevant relays, and there is evidence for early developmental sibling cortical pairing[8], which provides a simple partnering mechanism. It is also probably easiest to wire up because the required "partnering" would be difficult to coordinate between separate structures like cortex and thalamus, but could be easily achieved by an early developmental clonal mechanism involving daughters of radial glial cells. The other extreme version, postsynaptic unitary distributed ("poud") proofreading (not shown), would involve CT feedback to a single permanent relay partner, and to all the midlayer targets of that relay. Intermediate (nonunitary) cases are also possible but would be difficult to coordinate. "poud" seems unlikely because it predicts that the RFs of CT cells would resemble those of relays, and there are more than an order of magnitude more CT cells than relays[9]. However, it is possible that in primate visual cortex, where CT cells are relatively less numerous, and 4C neurons are extremely numerous, poud proofreading is done (particularly for 4C alpha neurons, which in any case have RFs resembling those of relays).

Background brainstem inputs (not shown) also influence the relay mode, and TRN state, but this action is more global and depends on sleep/wake state, and perhaps the waking but alert/inattentive states. Specifically in the waking and dream states cholinergic input depolarizes relays and hyperpolarizes TRN cells, priming the latter to fire bursts in response to CT input. In slow wave sleep, ACh action largely stops, depolarizing TRN and hyperpolarizing relays, which favors intrinsic thalamic rythmicity (e.g. spindles).

Also not shown is a collateral of relay axons that goes to TRN[4]. This collateral connection seems weak and, except perhaps in SWS, probably does not by itself fire TRN cells. However, it does depolarize them; if this depolarisation is fairly persistent (~10 msec) it would "prime" a TRN cell so it is more likely to fire in response to an arriving CT spike. This mechanism would tend to screen out TRN responses triggered by CT spikes that resulted from proofreading errors, improving proofreading accuracy.

This scheme places most of the burden of selective enabling of those TC connections which have just experienced a coincident spike pair on selective innervation and plasticity-disabling of potentially connected ("incipient") relays, rather than selective enabling of currently-connected relays. Nevertheless, the functional outcome is selective enabling, since although all the connected or connectable relays receive hyperpolarizing TRN input, this is selectively countermanded by depolarization of a subset (those receiving direct red reverse arrows in the above diagram). Note that these direct CT-relay synapses are located on the distal dendrites, because the control of plasticity operates on a somewhat slower time scale (~10-100 msec) than the control of relay spiking by driving, proximally-located synapses (~0.1–1 msec). Mode-switching would be particularly effective if the main (persistent) hyperpolarizing TRN  influence were via GABA-B receptors since this involves a minimal conductance increase (since these receptors inwardly-rectify). This "global/selective" approach, while apparently less specific  than a purely selective approach (with TRN cells currently targeting subsets of relays that do not currently receive direct CT input), has several advantages. First, it places the burden of selective layer 6 innervation of relays on a single mechanism. The direct connections are made by 'drumstick" synapses which are well suited to the task of continuous rewiring of the CT connections to keep pace with changes in feedforward connectivity during ongoing learning. Possibly the selectivity of this process is enhanced by the "drumsticks" which act as presynaptic spines[10], especially since in this pathway ltp seems to be presynaptic[11]. Second, much of the relevant circuitry can be established by early developmental mechanisms, without needing ongoing detailed TRN-relay wiring to keep pace with thalamocortical learning. Indeed, it seems unlikely that plasticity of TRN-relay synapses could be sufficiently selective, since these synapses do not involve spines or drumsticks, and ltp at inhibitory synapses is not synapse-selective[12]. Specifically, if topographic TRN-relay connections are established in a early fixed pattern, an arriving CT axon, after waiting in the perireticular nucleus[13], needs only to find the particular TRN neurons that already innervate the relevant, topographically-grouped relays. Since the targeting of both TC axons to layer 4, and TRN cells to relays, is largely topographic 4 this initial CT-TRN selection would only require minor adjustments from the initial topographic patterns established in subplate and perireticular nucleus. In the simplest case, only a single TRN cell needs be targeted (since the set of potential connections in layer 4 is heavily influenced  by initial topographic wiring: a relay cannot connect to any layer 4 cell, but only to those whose mature dendrites ramify within the relay's axonal arbor). The initial process of the layer 6 CT axon finding the relevant group of TRN cells is probably accomplished in the transient perireticular nucleus, in concert with the topographic sorting that is simultaneously taking place in the subplate[13]. Then, after TC axons starts to arborise in layer 4, where target cells dendrites are also developing, the CT axons can start to find the appropriate set of TRN cells. This process may generate the complex axonal interweavings that mark the entry to TRN[13].

In essence this topographic strategy deals with the limitation (imposed by crosstalk even after proofreading) that a cortical neuron can only learn from ~ 1000 inputs, by learning only "local" features. This strategy is viable because of the "smoothness" or "chunkability" of the world; provided progressively higher-order correlations are more "coarse-grained" this still allows powerful learning in a hierarchical manner.

Third, it is an efficient strategy because it is likely that connected relays outnumber incipient relays (especially later in maturation as learning converges), so it makes sense to use the permanent default TRN connections to target all relays, and then "pick out" the smaller subset of unconnected relays by a selective learning process. At first glance this seems to be contradicted by evidence that cortical functional synaptic connectivity, while extensive, usually lies in the 10-25 % range . However, this assessment ignores the likelihood that many of the connections are electrically "silent" (lacking AMPARs), and thus overlooked in typical electrophysiological experiments. There is evidence that many TC connections are electrically silent, at least early in development[14,15]. Between neighboring layer 5 cells the average number of touchpoints is 6[16]. If this applies to TC connections, and 10% of these touchpoints are occupied by synapses (filling factor = $0.1^{[17-19]}$), the typical fraction of available relays that are unconnected at any one moment would therefore be $(1-0.1)^6$ or ~ 53%. So the figure should actually show many more blue ovals; however, although most of the blue synapses are probably very weak, they may not all be completely silent.

This diagram, and Fig. 3b, sketches the basic TC connections underlying orientation tuning[108,109]. However, it omits the on relays, which also contribute to the RF of the layer 4 cell show and its layer 6 partner. Both on and off relays would be targeted (directly or indirectly) by the layer 6 CT cell, which proofreads all these connections in distributed fashion. The direct connections go to incipient on relays. The reader should be able to deduce the appropriate connections. As in Fig. 3b, the blue oval represents a currently connected relay with all-silent TC synapses. This neuron does not currently contribute to the RF of the layer 4 cell, but must be proofread because it could strengthen as a result of spike-coincidence. Such coincidences would typically be rare because this neuron does not contribute to the current RF (i.e. the layer 4 cell's "model" of its world), but if that model changes, these coincidences would become more frequent, and "suspicious", thus driving RF adjustment. In this sense silent connections constitute a "recessive" pool of anatomical mutations that can be recruited if the environment changes.

The diagram only shows one of the many CT cells that converge onto particular relays. These relays also participate in the RFs of many other layer 4 cells. Thus they receive indirect enabling input (via TRN) from the 6-CT cells corresponding to those 4-cells, which provide disabling direct input to the appropriate "complementary" set of relays that do not currently innervate those 4-cells. Thus every relay will get a combination of both direct epsps and indirect ipsps from layer 6, as typically observed[7]. However, in some cases there will be an imbalance in the ratio, and stimulating some small clusters of CT cells could produce net inhibition or excitation. This probably explains the offset examples[7].

**Supplementary Figure Legends.** These legends repeat some material contained in the regular figure legends, but include additional explanations and details, including some material that provides general background.

**Fig. 1 Supplementary Legend.**

**a:** the 1-unit ICA-with-error model. The top row shows 3 "sources", zero-mean scalar signals $s$ that fluctuate, from pattern to pattern, with equal variance according to a nonGaussian (typically Laplacian) pdf (left source, solid circle marked with L) or a Gaussian pdf (middle and right sources, dashed circles, marked with Gs). The Laplacian distribution, which decays exponentially towards zero probability symmetrically for both positive and negative signal values, is "superGaussian", with "fat tails" (exp-s decays less rapidly than $\exp(-s)^2$ ), and has kurtosis 3. We also tested "logistic" distributions, which have the logistic function $1/(1+e^{-ks})$ as cdf; this is also superGaussian but has smaller kurtosis (1.2), and does not allow learning (i.e. convergence) for as wide a range of almost-white inputs as does the Laplacian. For a given $\mathbf{M_o}$, the threshold error was about halved using logistic sources. We also tested cases with n=2 or >3, with similar results. However n=3 is the most transparent, since for n=2 it is theoretically possible that the Gaussian IC could be stably learned (some ICA learning rules allows 1 Gaussian source, though perhaps only multiunit rules where the antiredundancy term alone might find the last IC), while for n>3 the number of possible equilibria (ICs and PCs) rapidly grows. Furthermore, we find learning collapses when 1, 2 or 3 sources are nonGaussian, but the 1 nonGaussian case is particularly transparent because there is only 1 stable IC. With 2 or 3 nonGaussian sources learning also collapses close to approximate PCs, but since one of the stable ICs is often quite close to a possible PC, it is more difficult to exclude the possibility that learning simply switches to another IC. This ambiguity gets worse with even higher n. In the all Laplace n=5 case, 3 of the ICs destabilised at low error, and moved close to the approximate least PC. For n=10 all Laplace, 5 collapsed at low error, but we did not test whether the collapse was to approximate PCs.

The source signals are linearly combined in the next row to form the mix signals $\mathbf{x}$. The mixing coefficients (i.e. the values of the fixed weights from $\mathbf{s}$ to $\mathbf{x}$, shown as small colored dots, whose sizes reflect the coefficients) make up the elements of the quasiorthogonal mixing matrix $\mathbf{M_O}$; the red mixing weights form the first column of $\mathbf{M_O}$, the blue weights the second column and the green weights the last column. In simple terms each mix unit $\mathbf{x}$ gets a little bit of each source, in proportions determined by the relevant column of $\mathbf{M_O}$. Mathematically this means that the random vector $\mathbf{s}$ (with elements $\mathbf{s}$) is premultiplied by the matrix $\mathbf{M_O}$ to give the new random "mix" vector $\mathbf{x}$: $\mathbf{x} = \mathbf{M_O s}$. We show both $\mathbf{M_O}$ and its inverse $\mathbf{M_O}^{-1}$ (first 2 matrices in Fig. 1c). If $\mathbf{M_O}$ were perfectly orthogonal, then columns of $\mathbf{M_O}$ would become rows of $\mathbf{M_O}^{-1}$, which is why we have colored the first row of $\mathbf{M_O}^{-1}$ red. However, because $\mathbf{M_O}$ is not exactly orthogonal, the red row of $\mathbf{M_O}^{-1}$ (the target of learning) is not exactly the same as the red column of $\mathbf{M_O}$. $\mathbf{M_O}$ is calculated as described in part c of this legend, and the obtained mix signals, the elements of the vector $\mathbf{x}$, are sufficiently white (i.e. $\mathbf{M_O}$ is sufficiently orthogonal) that in the absence of error the nonlinear rule can learn the IC. We justify this "offwhitening", rather than the almost perfect whitening usually used in ICA, on the grounds that the brain cannot perfectly whiten, in large part because learning whitening filters also uses slightly inaccurate Hebbian rules.

The set of mix signals, $\mathbf{x}$, then provides input to the model neuron (y) via modifiable "synaptic" weights w, generating the scalar output y as a response to the current input pattern (and thus to the current, but hidden, source pattern). The weights are now very slightly (learning rate typically 0.002) adjusted using the nonlinear learning rule. Then a new set of s values is selected, generating a new $\mathbf{x}$ "input pattern" and the process repeated many times (or "epochs"; we typically used 50,000 to several million epochs depending on how long the weights, averaged over suitable windows, took to reach a steady state). The initial weight vector was randomly chosen, and the resulting trajectory is shown schematically in Fig. 1b (though actual runs used many more patters than show in the sketch). The weights on the output neuron are shown as appropriately-sized red dots to indicate that after convergence of the weight vector $\mathbf{w}$ to the IC, the weights are very close to the red mixing coefficients (first column of $\mathbf{M_O}$) and on average exactly equal to the red row of $\mathbf{M_O}^{-1}$). The output $y$, caused by the current input pattern $\mathbf{x}$ is given by the dot product of $\mathbf{x}$ and the current $\mathbf{w}$, $\mathbf{x^T w}$, i.e. the "projection" or "noon shadow" of $\mathbf{x}$ on $\mathbf{w}$). The yellow curved arrow shows that part of the update (*not* part of the activity!) from the first input "leaks" to the third connection; all synaptic weights leak symmetrically to each other, according to the error matrix $\mathbf{E}$, but only one leak is shown here.

Fig. 1b sketches the behaviour of the connection weights in typical numerical experiments. Normalisation confines the weights to the surface of a unit ball (i.e. to a sphere; only the hemisphere facing the reader is shown; if the weights start in the other hemisphere, they end up at antipodean locations on that hemisphere)). We used "brute force" normalization: all the weights are multiplied by the factor that restores the weight vector's length to its original "unit" value. While this is unbiological, various biologically plausible schemes are available; exact normalization is not essential, and the norm of the weight vector does not have to be the same in the presence or absence of error, though of course it is using brute normalization. A black dot has been placed at the centre of the ball to represent the origin of the coordinate system defined by the weights or the sources or input patterns, though only the weights are shown. It's important to realize that while the weights are confined to the surface, the source patterns or input patterns can be represented as points that can occur anywhere within or beyond the ball; this "cloud" defines the input joint distribution (i.e. the conditional distributions along all the possible directions). In the case of the sources, which are independent, the joint distribution is defined by the product of the marginal distributions. Specifically, with Laplace sources, the distribution of **s** in any particular direction (i.e. the pdf of variously scaled versions of a particular example of **s**) is itself Laplacian, though narrower than any of the source Laplacians. For the input patterns **x** the joint distribution "cloud" is not given by the product of the marginal distributions, since the elements of **x** are not independent.

Note that straight lines represent unnormalised vectors; other lines represent successions of points on the weight sphere (i.e. weight vectors normalized to unit norm). The weight sphere has been rotated so the red vector (which corresponds to the first row of $\mathbf{M_O}^{-1}$ and unmixes the nonGaussian source) points to the reader, instead of the third weight pointing to the reader. (Thus the marginal source distributions are defined along lines that all tilt from the reader). During learning, the first input pattern $\mathbf{x}^1$ (not shown but can be thought of as an **x**-point anywhere within or beyond the ball) causes a randomly chosen initial weight vector (star) to move slightly off the weight surface (e.g. into the ball) parallel to the direction of that pattern by an amount that is proportional to the length of the pattern, the learning constant k and the nonlinearly transformed output f($y$); normalization then drops the new weight vector back to the sphere; the net movement is show as a "zig" on the sphere and eventually, with no crosstalk, the zigzag path reaches the IC vector at the red dot. The detailed trajectory, but not the destination, depends on the form of the nonlinearity. For the illustration zigzagging has been exaggerated in size and reduced in number; however the movement to the IC is typically quite noisy except at very small k (see error bars in Fig. 2).

During our numerical experiments thousands or millions of **x**-points are generated and the density of these points in the 3 dimensional input space (of which the ball is the inner core), representing their joint distribution, is rather subtly nonGaussian, and can be thought of as battered soccerball "cloud" (because $\mathbf{M_O}$ is roughly orthogonal, the cloud is roughly round; its irregularities reflect the slightly nonGaussian statistics of the mix vectors **x**, and its roundness the approximate whitening or "sphering"; it is cloudy because the points  lie at variable distances from the origin, though because of the zero mean source assumption they cluster near the origin; if we had used **M** not $\mathbf{M_O}$ to do the mixing, the cloud would be a battered rugby ball; however, usually this does not allow convergence to the IC).  However, any particular subset of these points (eg the "batch" we use to generate $\mathbf{C_S}$) will tend to scatter around the underlying joint distribution because of sampling errors.

 The other 2 weight vectors corresponding  to rows of $\mathbf{M_O}$ are shown as blue and green dots; the corresponding unnormalised vectors are shown as broken lines: since they correspond to Gaussian sources they are not stable ICs (though they are equilibria of the rule, since now the output exactly tracks the corresponding source). Subthreshold crosstalk only produces tiny movements of the average weight vector away from the true IC (within the red dot), but at the threshold the weight vector zigzags abruptly (as a function of error, though the movement is quite slow in terms of numbers of epochs) away from the approximate IC to the least eigenvector of $\mathbf{EC_O}$ (square). Note that with all Gaussian sources (or insufficiently white $\mathbf{M_O}$), the error-free nonlinear antiHebb rule would typically (but this is not shown) zigzag directly from any start position to the purple spot, the intersection of the smallest Principal Component (the least eigenvector of $\mathbf{C_O}$, shown as a purple line) with the sphere. The yellow line (dotted and solid) shows (schematically) the loci of the normalised least eigenvectors of **EC** (learned when all sources are Gaussian) on the sphere for various crosstalk values increasing from zero (purple dot) to the "trivial" zero specificity value where all Hebbian specificity is lost. The square shows the locus at the crosstalk threshold. The yellow loci (dotted) between the purple dot and square are hypothetical unstable equilibria of the erroneous rule if one source (at least) is nonGaussian; they are apparently stable for all-Gaussian sources. The yellow loci beyond the square (solid) are stable for the 1 nonGauss case. Thus at the error threshold learning switches (dotted zigzags) from complex (red dot, driven by hocs) to simple (yellow solid line) driven by socs, though it can take millions of epochs to complete the switch. We do not show the time course of this switch in Fig. 2 but it is appears to have 2 components, one quite rapid and comparable to the initial fast rate of convergence to the IC, and one much slower. Indeed, fast and slow components are also visible following step changes in error even for all Gauss inputs. We always waited until the weights showed no further trend before evaluating the average weights. Note that the red, blue, green and purple unnormalised vectors all originate at the black dot, the origin of the coordinate system. The red, blue and green vectors are approximately orthogonal, with the red vector pointing towards the reader. If they were exactly orthogonal then only the component of an input vector that was due to a single source would be "seen" by the neuron's output, since the projection of the other components on the IC weight vector would be zero. Thus all patterns would, on average, only shift the weight vector in the direction it already points; this shows that when the weight vector reaches a red, blue or green dot it is at an equilibrium, however only the red equilibium is stable because it corresponds to the nonGaussian source.

Fig. 1c The mixing matrix $\mathbf{M_O}$, its inverse $\mathbf{M_O}^{-1}$, the large-sample covariance matrix $\mathbf{C_L}$, and error matrix $\mathbf{E}$ are shown corresponding to Fig. 2 (values correspond to the threshold "per connection" error b= 0.045 in Fig. 2; see legend to Fig. 2). The original random $\mathbf{M}$ (entries uniformly distributed between 0 and 1) used to generate $\mathbf{M_O}$, via $\mathbf{C_S}$, was

$$\mathbf{M} = \begin{bmatrix} 0.7318 & 0.0731 & 0.305 \\ 0.4191 & 0.0827 & 0.0169 \\ 0.8225 & 0.3565 & 0.5619 \end{bmatrix}$$

The small sample covariance matrix $\mathbf{C_S}$ used to generate $\mathbf{M_O}$ is shown in Supplementary legend 2. Usually all the offdiagonal elements of $\mathbf{E}$ were set equal, as shown in Fig. 1c. The columns of $\mathbf{M_O}$ and the rows of $\mathbf{M_O}^{-1}$ are colored using the scheme of Fig. 1a. The rows and columns of $\mathbf{M_O}$ and $\mathbf{M_O}^{-1}$ are of approximately equal lengths and directions since $\mathbf{M_O}$ is close to orthogonal. In different tests different $\mathbf{M_O}$ s (generated from different $\mathbf{M}$ s) were used. $\mathbf{C_L}$ (fig 1c) was computed using a batch of 100,000 x-vectors generated using $\mathbf{M_O}$. If $\mathbf{M_O}$ were exactly orthogonal $\mathbf{C_L}$ would be diagonal with equal entries. Its eigenvectors, which would have equal eigenvalues, could not be learned, either linearly, or nonlinearly using all-Gauss sources, because although they are equilibria they are not hyperbolically stable; stability requires distinct eigenvalues; indeed all vectors are eigenvectors of $\mathbf{I}$). But since $\mathbf{M_O}$ is only approximately orthogonal, the offdiagonal elements are nonzero but small, and the diagonal elements are almost , but not exactly equal. These 2 factors allow small amounts of crosstalk to stabilize learning of an eigenvector (now of $\mathbf{EC}$). The orthogonality of $\mathbf{M_O}$ was characterized using the Frobenius norm (square root of the summed squares of the matrix elements) of the matrix $(\mathbf{I} - \mathbf{M_O}\,\mathbf{M_O}^T)$, which equals 0 for an orthogonal matrix. We call this measure of orthogonality the Orthogonality Factor (OF).

**Fig. 2 Supplementary legend.**

1 Laplace source and 2 Gaussian sources (i.e. n=3) all of unit variance and centered on zero were used. A learning rate of 0.002 was used. The original Matlab seed for the $\mathbf{M}$ used here was 23. The small sample covariance matrix $\mathbf{C_S}$ that was calculated using a batch of 1000 unwhitened vectors mixed using $\mathbf{M}$ for the calculations in Fig. 2 is as follows:

$$\mathbf{C_S} = \begin{bmatrix} 1.32 & 0.65 & 1.65 \\ 0.65 & 0.37 & 0.78 \\ 1.65 & 0.78 & 2.29 \end{bmatrix}$$

It can be seen that this deviates much more from $\mathbf{I}$ than does $\mathbf{C_L}$ (Fig. 1c) calculated using a much larger sample (100,000) of offwhite input vectors. $\mathbf{C_S}$ was then used to "offwhiten" the mix vectors used for learning, i.e. these were calculated using $\mathbf{x} = \mathbf{M_O}\,\mathbf{s}$ where
$\mathbf{M_O} = \mathbf{C_S}^{-1/2}\,\mathbf{M}$. The OF for $\mathbf{M_O}$ was 0.0807 and for $\mathbf{M}$ it was 1.609. Thus $\mathbf{M_O}$ (the decorrelated mixing matrix) was approximately 20 times more orthogonal than $\mathbf{M}$ (the original mixing matrix).

The average value of the weights were typically calculated over windows of 100,000 patterns or more. Large windows were needed, especially in the presence of crosstalk, because the weights become quite noisy (which accounts for the quite large standard deviation bars in Fig. 2a), either using partly nonGauss or all Gauss sources. This noise may reflect an approach to instability; nevertheless, in both cases the average weight vector closely matched the least eigenvector of $\mathbf{EC}$. Because long averaging windows had to be used it was not possible to study the kinetics of the IC-PC transition, but it appeared to occur over a time scale similar to a slow component in the weight noise, requiring 100,000 patterns or more at k = 0.02. Similar slow kinetic were also see in the all-Gauss case. A very similar IC-PC transition at a similar crosstalk threshold was also seen with k = 0.004. Since the PC is stable at just suprathreshold error, and the IC at just below threshold error, it seems that we are seeing a bifurcation between 2 stable equilibria, not an onset of instability (which would occur at error levels much closer to trivial). This is supported by the observed lack of dependence on k. We also tested the function $f(y) = y^3$; for this function one takes the positive sign in the learning rule. Since this rule is Hebbian, one expects to obtain the leading eigenvector of $\mathbf{EC}$ beyond the error threshold, as we observed.

Only 2 of the weights (Fig. 2a) are shown: the third weight is completely determined by the first 2 weights and the explicit normalization which was applied after each successive input vector. Fig. 2b shows the dot products of the average stable weight vector with either the first row of $M_O^{-1}$ (red points and red "L" line, or the least eigenvector of $EC$ (blue points and blue "L vs G" line. These vectors are all normalized so these dot products give the cosines of the angles between the vectors.

The crosstalk parameter used ("error") was b, which discounts the contribution of n to the quality factor Q, assuming that weights change in analog fashion[20], using $E = nb/(1+nb)$ where $E = 1-Q$ (Q is the diagonal element in the error matrix; see Fig. 1c). For n = 3 the "trivial" value of b for which update specificity is completely lost is 0.66 Thus the "error threshold" value of b, $b_c$, is about 5% of the trivial value in this case (close to our average value using 4 different $M$s).

In the absence of crosstalk the weights stabilize very close to the first row of $M_O$ (Fig. 2b red line). The slight discrepancy reflects the finite learning rate and averaging window. We checked using a larger averaging window that the stable weight vector is actually closer to the appropriate row of $M_O$ rather than the appropriate column (see Fig. 1c; the difference arises because $M_O$ is not exactly orthogonal due to the partial whitening procedure). As crosstalk increases there is at first a slight gradual movement of the stable learned weights away from the only stable IC, corresponding to the Laplace source (see red line in Fig. 2b). This means that the output no longer exactly tracks the Laplace source. However, beyond the threshold error $b_c = 0.0425$ (equivalent to a quality factor Q of 0.88) the weights move sharply (though slowly!) away from the IC, to a new stable values that coincides with the stable weights that are obtained using the same $M_O$ and all-Gaussian sources (black points in Fig. 2a; blues points in Fig. 2b), which corresponds closely to the least eigenvector of $EC_L$ (black theoretical lines in Fig. 2a, see also blue points and line in Fig. 2b). The red lines in Fig. 2a just link the points and have no theoretical significance. Using $f(y) = y^3$, a similar result is obtained, except that beyond the threshold the weight vector aligns with the largest eigenvector of $EC_L$, because this rule is Hebbian (not shown).

## Fig. 3. Supplementary Legend.

a. Dedicated proofreading. Plasticity of the I-J connection is enabled when they fire highly-correlated spikes, as judged by the K cell. The strength change is only made as a result of the combination of the Hebbian coincidence detected by NMDARs at the J-I synapses and the delayed plasticity-enabling "handshake", "approval" or "confirmation" signals generated by the K cell and delivered to both the J and I cells. The K cell and the J-I synapses both act as coincidence detectors, and they both have to be triggered for changes in the strengths of the J-I synapses. The handshake must arrive and be implemented before the next coincidence, although the expression of plasticity can be delayed, and accumulated as a batch.

The diagram assumes the 'default' mode (set by diffuse brainstem input) is plasticity-disabled, and that coincidence-detection by K enables plasticity. There are a number of important issues. 1: How does a K cell detect coincidence? 2: How does its coincidence detection appropriately enable the plasticity of the J-I connection? 3: How do the necessary specific connections develop and then keep track with ongoing learning at feedforward connections? 4: Are K-cell signals used to adjust processes other than plasticity? We have studied these important questions over a number of years[21,22] and have found that known or plausible processes can answer all of them; we summarize here likely answers to the first 2 only. However, some of the most interesting matches between our model and known facts come as answers to the last 2.

1) Our original suggestion for the basis of neuronal coincidence detection was a combination of 2 known mechanisms: passive cable delays combined with spike-thresholding, and NMDARs. In the first idea, a presynaptic spike that arrives early on distal dendrites would trigger a somatic epsp that coincides with a proximal epsp generated by a postsynaptic spike. In the second idea, the somatic depolarization produced by the first spike would favor NMDAR-based epsps. A third possible mechanism for coincidence detection emerged with the work of Larkum et al.[23] on the apical tufts of layer 5, 2 and 3 pyramidal cells: they discovered "BAC-firing", in which a single backpropagating dendritic Na spike combined with a suitably timed subthreshold tuft synaptic input could trigger an all-or-none tuft calcium-based response that in turn triggers a burst of Na –spikes in the initial segment. Accurate coicidence detection based on dendritic location is already known in auditory neurons[24]. Recently, a similar BAC-coincidence mechanism has been revealed in the distal dendrites of layer 6 cells, which ramify in layer 4 (Larkum, personal communication).

Some anatomical evidence appears to favour this view: thalamocortical axons typically arborize richly in layer 4, and they may synapse with the distal apical dendrites of layer 6 CT cells in layer 4 (ref 26 but see ref 151). Also, layer 4 spiny stellate cells typically send a vertical descending early-developing collateral into layer 6 [25]; this axon could be seeking out a permanent layer 6 partner. Recent developmental data support this possibility[8]. However, there is almost no physiological or anatomical evidence for or against the postulated "basal" 4-6 monosynaptic connection .

The opposite arrangement is also possible: TC axons ramify in layer 6, and they make significant numbers of large synapses directly on layer 6 CT cells[151] . A TC axon makes many more synapses on layer 4 cells, so these rather weak layer 6 inputs are unlikely to suffice to drive the layer 6 cells (in agreement with the proofreading model, which requires that the thalamic input to CT cells be only "semidriving", in the sense that only combined thalamic and layer 4 input would cause CT cells to fire). Also, layer 4 axons ramify extensively in layer 4, and could, and do[26], make synapses on distal apical 6 dendrites (however, some at least of the CT cells in that study were layer 5 cells). Of course all 3 possible coincidence mechanisms could work in both types of scenario (T-input proximal and 4-input distal; T-input distal and 4 input proximal; a combination of both). It's possible that both T and 4 inputs are each located both proximally and distally, perhaps with the early-established basal connection providing an electrical clue that tells an individual layer 4 axon which particular single distal dendrites (those belonging to its layer 6 "partner") it should selectively target. This issue is closely related to the question of what inputs create the receptive field properties of layer 6 cells. It is also related to the theoretical question of the fundamental operation that neurons perform on their inputs: addition or multiplication[27]. At the least there are both experimental and theoretical supports for the notion of layer 6 CT cell coincidence detection, but no direct proof.

2) The problem of selectively ratifying, or enabling, the plasticity of the connection which has just experienced a coincidence has 3 main parts: spatial, (delivering the approval signal to the right connection), temporal (getting it there before the next coincidence) and mechanistic. We discuss all 3 in the Supplemental Note[DD]. The key to the first is to supply the approval selectively to both the pre- and postsynaptic neurons contributing to the connection, and requiring that both handshakes be present (as shown in Fig. 3a); the key to the second is mainly getting the presynaptic approval quickly down the axon to the synapse; this requires some sort of multiplexing on the spikes themselves. In its simplest form, the presynaptic component of approval enables the connection plasticity retroactively, by rapidly switching the relay to burst firing mode, and using whatever driver-triggered (e.g. sensory) spike comes next to carry the signal. This inevitably means that approval is always "1-step behind", much in the way that if DNA polymerase proofreading detects an error, the additional exonuclease operation produces a "hiccup" in smooth replication. The key to the third is to hold the synaptic-coincidence initiated strengthening in "draft" form pending confirmation arrival – essentially the layer 6 signal enables the transition from "induction" (NMDAR-based) to expression (e.g. complete dodecamer autophosphorylation). A possible mechanism would be that the confirmation prolongs the intermediate Ca-calmodulin-induced partial phosphorylation lifetime, perhaps by inhibiting phosphatases, so the "approved" coincidence can effectively contribute to the "ramping-up" of autophosphorylation to the point where it triggers all-or-none strengthening. In other words, strengthening would actually be triggered by a batch of approved coincidences. It possible that for TC connections the enabled integration time of the kinase is quite long (hours?).

In distributed proofreading, a single layer 6 cell "guards" many different connections. This is possible because if coincidences anywhere in the relevant set of neurons are rare, and the proofreading "handshake" plasticity-approving signals act very quickly and are typically over by the time the next coincidence occurs, only the relevant synaptically-detected coincidences are approved. There are 2 simple extreme versions of distributed proofreading. In the version shown here, each proofreader has its own permanent unique layer 4 partner, hardwired early in development, and feeds back to only that partner and the set of relay cells that currently innervate both the partner and itself. This version would mean that the layer 6 cell shares the receptive field (RF) properties of its layer 4 partner, as found experimentally. The alternative extreme version would require that a CT cell be hardwired to a unique permanent relay partner, and would feedback to that partner and to all the layer 4 cells it currently innervates. This would endow the CT cell with center-surround RF properties, which is typically not observed. Furthermore this long-distance hardwiring might be more difficult to achieve developmentally. Note that an inhibitory TRN neuron which may be intercalated between the CT feedback and its target relays is not shown in the diagram (but see the Supplementary Figure). This TRN cell would hyperpolarize the relevant relays, shifting them from tonic to burst, which would enable plasticity presynaptically (represented by a forward arrow). Relays that are potentially (but not currently) connected to the layer 4 cell (open ovals) would receive direct depolarization (reverse arrows), a 'pull' mechanism that helps ensure plasticity is only enabled as a result of spike coincidence. In this case, the 'default", disenabled mode is defined by absence of all CT firing, coupled with diffuse Ach action. Usually connections receive a balance of "push" (plasticity enabling via TRN hyperpolarization) and "pull" (disenabling via direct depolarization, plus background brainstem cholinergic action), though at any one time the state is either enabled (burst relay firing) or disabled (tonic). The postsynaptic plasticity enabling signals (red forward arrows on layer 4) are delivered by modulatory "drumstick" synapses on dendritic shafts.

## Section 3

**Nonmathematical Guide to PCA, ICA and figures 1,2.**

Here we explain Fig. 1, and associated concepts, for the complete nonmathematician. For convenient exposition there is some overlap with other material in the paper and Supplement. There are no references.

Fig. 1 a shows our model neuron, at the bottom, which generates an output signals y in response to various inputs signals $x_1$, $x_2$…. which are passed through corresponding synaptic weights $w_1$, $w_2$… The output is simply a weighted sum of the inputs: $y = w_1x_1 + w_2x_2$….This is the simplest possible way to represent the input-output relation.

The inputs x are generated in turn from source signals s in the same fashion. Thus $x_1 = m_{1,1}s_1 + m_{1,2}s_2$…where $m_{1,1}$ is the mixing coefficient from source 1 to input 1 etc. The complete set of mixing coefficients are listed in a matrix **M (Bold** is used for vectors and matrices). The whole process is conveniently represented by the equation $\mathbf{x} = \mathbf{Ms}$ where the vector **s** is the set of s and the vector **x** (i.e. the $x_1$,$x_2$…. values) is the set of inputs x. These vectors are just ordered lists of the current values of the sources and inputs.

Thus the basic structure of the model is very simple: the output is generated by linear mixing of inputs and the inputs are generated by linear mixing of sources. We assume that the sources are statistically independent: if they have even deeper causes, they must each have their own separate causes, with no intermingling. If the output of the neuron can learn to track the fluctuations in one of the sources, by suitably adjusting the weights, it has implicitly "understood" how that source contributed to the fluctuating input signals: the sources themselves cannot be further analysed as being due to combinations of even deeper causes: one has reached the end of the line. This loosely resembles a simplified version of "object recognition": one can recognize several objects in a cluttered scene even though the objects can be in various orientations and position: these fluctuations in object presentation are analogous to the fluctuations in the sources, but there are nevertheless independent objects and independent sources. But in the ICA model the "objects" are single points, and the variation in presentation generates input fluctuations in the simplest possible way, linear mixing of objects. Thus the complicated nonlinear laws of occlusion, foreshortening, shadowing etc have been simplified to simple linear mixing by the unknown matrix **M**. We assume that if synaptic crosstalk prevents the brain from learning to solve even this simple problem, crosstalk would usually be an even more severe problem for more complicated problems. This is the reason why we study ICA learning, not because it is exactly how the world or the brain operate. We suspect that the much more difficult problem of visual object recognition is solved by a multistep learning process, involving several successive cortical areas, each step of which uses a nonlinear Hebbian component that is sensitive to hocs. This "hierarchical" learning process would require that, preparatory to each step, cortical information be rerouted via thalamus, so proofreading could be done in the crucial first stage of the next step. In the simplest case, each step would involve proofread nonlinear Hebbian learning (stage 1; thalamus to layer 4), application of a nonlinearity to create new hocs (stage 2; layer 4 to 2/3) and rewhitening (2/3 to 5, and thence to a higher-order thalamic nucleus for new proofreading for the next step in a new cortical area, as well as distribution of the whitened intermediate data to subthalamic "motor" structures). In the case of classical ICA, only 1 step is required.

The way the neuron solves the problem of identifying a source is by learning, i.e. systematic adjustments (following a suitable rule) of the modifiable synaptic weights **w**, based on the current inputs and output, in the hope that if the rule governing the adjustments is appropriate, eventually the weights will converge onto approximately fixed values that allow the neuron's output to equal the current value of a particular source. In our simulations (and in reality) the weights continue to fluctuate even after "convergence" because of the finite learning rate. In principle one could reduce the learning rate to very low values and get less weight fluctuation, but we found this did not change the average equilibrium values, with or without crosstalk. The continuing minor weight fluctuation after learning has stabilized is a necessary evil: it allows learning to readjust if the input statistics change. After the end of a cortical "critical period", learning becomes very slow, minimizing weight fluctuations, but this slowness would partly reflect increased proofreading accuracy, rather than a general decrease in plasticity. During the critical period, proofreading neurons would have many inputs and make many mistakes, falsely enabling plasticity much of the time and allowing rapid learning if input statistics are favourable. Later in development, and in maturity, the numbers of silently connected inputs to layer 4 cells would decrease, increasing proofreading accuracy by making it less "distributed", and reducing the overall enablement level.

  Converging to weights that allow a source to be tracked corresponds to interpreting the various possible views of the objects on a table as transformations of an underlying set of "causes", the objects themselves.  If one develops a set of neurons that can "see" these objects despite the way they shift around on the table, or the viewer moves, one has accomplished object recognition. More generally, the goal of neocortical learning would be to discover the "causes" of one's experiences, since this is the best guide to productive behaviour.

Another way to approach the problem would be to have a "teacher" that can instruct the neuron when, and possibly how much, the output deviated from what it would be if it were successfully identifying the object. This is supervised learning, but we regard having a teacher as already having solved most of the problem: after all the teacher knows the solution, and the neuron merely has to learn the teacher's solution. In the ICA model, learning is unsupervised, and must bootstrap itself into understanding the problem simply by studying the available clues, the various instances of $\mathbf{x}$ generated by the unknown mixing process, or the various possible views of the table. Most of the learning the neocortex does must be unsupervised because it is rare one is given suitable labeled inputs.

Before proceeding further, it is useful to remind oneself of the difficulty of the table task confronting a naïve viewer with no previous visual experience (e.g. a baby). All the baby sees are patches of light that seem to change unpredictably from one moment to the next as the baby moves. The only clues she can use are the various relationships, or correlations, between the different views (or, more precisely, of the pixels that make up the views). Fortunately there are spatiotemporal correlations: sometimes the baby's eye and head movements are relatively smooth, which can help enormously. However, this means the learning rule must itself be sensitive to time, i.e. STDP. In basic ICA, as in most connectionist models, one ignores time, so the input data correspond to randomly oriented snapshots, rather than a continuous movie. This simplifies the mathematics, although it may complicate the learning problem. The view we take here is that while time may make the learning task easier in some ways (by providing additional clues), it makes it harder in others: first, one has to learn to unscramble a spatiotemporal mixing process (basically, a time-dependent matrix); second, it will be more difficult to implement a time-dependent learning rule; in particular crosstalk is more likely to be important at later times. In essence, connectionist models simplify the problem by reducing spatiotemporal correlations to spatial correlations, so they can be treated in a single simple framework.

Because it is so simple, the ICA problem can be "solved" very straightforwardly: if the set of input synaptic weights, the "weight vector" $\mathbf{w}$ of the neuron, is fixed to be equal to a row of the "unmixing matrix" $\mathbf{M^{-1}}$ then the output will track one of the sources as it fluctuates. This is because $\mathbf{M^{-1}}$ is *defined* to be the set of coefficients that, by a second linear transformation, exactly undoes the mixing. There are mathematical procedures for calculating the inverse transform $\mathbf{M^{-1}}$ given $\mathbf{M}$ but they cannot be used in this case because $\mathbf{M}$ is not explicitly given to the neuron: all it sees is many examples of the results of the initial linear transform, the successive vectors $\mathbf{x}^1$, $\mathbf{x}^2$, etc. generated from the successive values at times 1, 2 etc. of the fluctuating sources $\mathbf{s}^1$,$\mathbf{s}^2$ etc. Clearly the relationships between these different vectors (ignoring the fact they happen to occur in a particular temporal sequence because this, given that the sources are random signals, is irrelevant) are the only possible clues the neuron can use in this unsupervised case. But it is not intuitively obvious that the relationships, or correlations, do contain the type of information that is necessary to solve the problem. In fact it is fairly easy to see that in one special, but typical, case, the correlations are **not** adequate: this is the case where the sources have a Gaussian distribution. In this case the joint distribution of the inputs $p(\mathbf{x})$ will be a multivariate Gaussian distribution, which is defined by $p(\mathbf{x}) = \mathbf{M}p(\mathbf{s})$ with $\mathbf{s}$ Gaussian: the probability of seeing different linearly scaled versions of a particular combination of x-variables (the complete set of these defines the "joint" distribution) has itself an exactly Gaussian distribution. One can calculate[BB] that under these conditions all the even higher-order moments, such as $<x_1x_2x_3x_4>$ (i.e. the average value of the quadruple product), reduce to combinations of the pairwise moments (covariances or "socs") $<x_1x_2>$ etc: there is no real higher-order information. Now the distribution of $\mathbf{x}$ (and hence the set of its socs) depends on 2 unknowns, $\mathbf{s}$ and $\mathbf{M}$, and there is no way to calculate each separately from that distribution (or equivalently, from the socs): the problem is ill-posed.

This limitation disappears when the sources are nonGaussian, because now the joint distribution of $\mathbf{x}$ is not defined by $\mathbf{M}p(\mathbf{s})$ (even though the x values themselves *are* defined by $\mathbf{x} = \mathbf{M}\mathbf{s}$). Now the hocs do provide lots of additional clues, although still not enough to completely identify $\mathbf{M}^{-1}$: all one can get are weight vectors that are proportional to rows of $\mathbf{M}^{-1}$, because one cannot dissociate the ambiguity between the magnitudes of rows of $\mathbf{M}$ and entries of $\mathbf{s}$ (they are both unknown and so their product cannot be uniquely factorised  - unless there is a restriction to integers; weight digitization is another interesting issue not explored here).  In summary, if the sources are nonGaussian, the hocs present in a large enough ensemble of input vectors could in principle allow identification of rows of $\mathbf{M}^{-1}$, to within some arbitrary scaling factor, and to a precision which is set by the ensemble size (or learning rate). In figures 1,2 one of the sources has a "Laplace" distribution: the probability of finding the source in a particular, small, range decreases exponentially as that range moves away from zero, symmetrically in both positive and negative directions. Since the tails of the Gaussian decay as exp-$x^2$, the Laplace tails are "fatter", if both distributions have the same variance, as we always imposed. The deviation from Gaussian is usually measured using the "kurtosis" which is the difference of the fourth order moment $<s^4>$ for a Laplace distribution and an equivariant Gaussian variable. (All odd moments are zero for symmetric distributions).

It's important to note that there are 2 related but separate issues here: how best to encode the input vectors (so they can be sent to other destinations without being corrupted by transmission noise) and how best to solve a specific problem (in this case, recover the sources). If there are as many output neurons as there are inputs, then almost any set of weights will effectively encode the inputs, though some weights will encode better than others in the presence of noise. In general the mixing process guarantees, by the central limit theorem, that the inputs will be close to Gaussian, and in this case the optimal encoding is to use sets of weights which are "eigenvectors" of the covariance matrix $\mathbf{C}$ (which simply systematically lists all the socs). An eigenvector of a matrix is a vector whose direction (in n-dimensional space) is unchanged by the linear transformation represented by that matrix. Fig. 1b shows an example: the purple line going from the center of the ball (black dot) to the purple point on the surface of the ball is an eigenvector of the $\mathbf{C}$ for a large set of input. In this case we used 100,000 inputs so the large sample covariance matrix estimate $\mathbf{C_L}$ is quite close to true $\mathbf{C}$ for an infinitely large sample (i.e. the expectation of $\mathbf{C}$). Fig. 1b actually shows the weights, not the input patterns, but for every weight there is a corresponding current input value, so we can use the same coordinate system (in this case a 3D system because there are only 3 inputs and weights) for both. To a first approximation the input patterns would show up as a diffuse ellipsoidal cloud of points, and the eigenvectors of $\mathbf{C}$, such as the purple line, correspond to the axes of this ellipsoid. However, the input distribution is not *exactly* Gaussian in all directions, because the sources are nonGaussian, and sample-limited; these small deviations from perfectly ellipsoidal equiprobable contours reflect the presence of hocs (and one could usefully regard the learning task as finding the overall shape of these deviations). Each eigenvector has an associated eigenvalue, a number which represents the input scatter along that direction (and corresponds to the variance of the neuron's output when its weight vector points in this direction). For more than 3 inputs the input patterns distribute in a higher-dimensional space which cannot be easily visualized, but the same concepts (generalized to hyperellipsoids etc) still apply. It may help the reader to visualize the input patterns in Fig. 1b as a roughly ellipsoidal cloud of points which is thinnest in the approximate direction of the purple line (because the antiHebb (tanh) rule we use typically learns the *least* eigenvector of $\mathbf{C}$).

By using this type of Principle Component representation one is essentially fitting the data with a multidimensional least squares linear encoding: the axes of the ellipsoids are the fitted lines, and instead of sending the original multidimensional data, one is sending the location on that line of the "proxy" point on the line that is closest to the datum, which is a single number. By using a set of mutually orthogonal lines one can reduce the approximation involved (that the true data point coincides with its proxy) to any desired degree. In particular, by knowing the "compressed" activity of the million retinal ganglion cells one can know the activity of the 100 million photoreceptors almost exactly. Thus Principle Component Analysis is a form of linear regression. The largest eigenvector (first PC) is the best fitting line and the least eigenvector (least PC) the worst-fitting. One could view ICA as a type of nonlinear regression.

The goal here is not merely to compute the average correlation between 2 inputs, but to discover a characteristic pattern underlying the entire ensemble of socs. Dedicated neurons could multiply 2 activities, but that is only part of the overall computation done by a neuron equipped with Hebbian weights. What is special here is not sensitivity to an individual soc (or hoc), which by itself is practically useless, but the ability to find a pattern underlying the socs (or hocs). This is why many input have to come together at 1 neuron, not an inability of neurons to multiply.

This discussion shows that learning the largest eigenvector of $\mathbf{C}$ would be a useful if the goal is merely to transmit information elsewhere in the brain, and to a first approximation this is what the retina does: the center-surround ganglion cell RFs correspond to the principle eigenvector or Principle Component (corresponding to the longest axis of the hyperellipsoid) of a small image patch centered on the ganglion cell. Fortunately the retinal socs (averaged over all natural images, and possibly all possible images transformed by the eye's optics) quickly decay to negligible values away from the center, so this "local" form of soc removal works fine (but it means that the retinal locations of the ganglion cell RFs must somehow be accurately known to the target structures, such as thalamus: the mapping must be "topographic"). Of course some retinal ganglion cells do already solve problems (e.g. they may act as hardwired "bug" detectors) but much of the problem solving is deferred to central structures like the tectum or neocortex (since the tectum does not have proofreading, it cannot systematically exploit hocs, but could act as a large and useful learned "look-up" table that maps image to action; retinal soc removal, by decorrelating patterns at second-order, will help learning such a table).

Encoding images using neurons whose input weights are unit-length eigenvectors of the **C** for those images results in a joint output distribution which is close to hyperspherical, so neurons are uncorrelated at second order. This is called decorrelation, "whitening" or, perhaps, "unsocking". Not only is it useful for sending the signals over noisy channels, but it is extremely helpful for further analysis: distortions in a hypersphere are easier to detect than in a hyperellipsoid. The retina does a "local" decorrelation, as noted above, but the outcome is the same: the activities of ganglion cells averaged over a complete set of images is decorrelated. This does not mean that as the image of a soaring bird moves over the retina the ganglion cell activities are uncorrelated, though it does mean that as one views an album of snapshots (or indeed as the eye dithers over a single image) the activities are largely uncorrelated.

A linear Hebbian rule is well suited to learning such eigenvectors: the weight change is driven by the pairwise correlation between an input and the current output (i.e. by the pairing of pre- and postsynaptic spikes), and since in a linear neuron the output depends linearly on all the inputs, the growth of a weight depends on all the input socs, and is fastest in the direction of the largest eigenvector of **C** (after all, the scatter in the points, which is what causes weight change, is greatest along that axis). Because growth in the various possible directions is exponential (because they are scaled by the output), and can be completely described by combinations of growth along any set of mutually orthogonal axes, multiplicative normalization of the weights leads to competition and selection of the fastest growing direction. The reasoning here is the same as spotted by Malthus and Darwin: if different types of a population all grow exponentially, subject to the constraint that the whole population remains constant, the fastest growing type eventually wins. This would not necessarily be true for other types of growth: for example superexponential growth (e.g. following $\exp(at^2)$), types that happen to be initially more numerous could win even if they do not have high a values); subexponential growth (e.g. following $\exp(at^{0.5})$) leads to incomplete victory.

Doubtless in the retina ganglion cells do not learn their entire RF: since it is always true that socs decay to zero over short retinal distances (in a way that mainly reflects the optics of the lens), it makes sense to hardwire in an initial strong "arbor" constraint, so a ganglion cell only sees a tiny patch of visual space. But one could say that evolution "learned" this arbor function, and that the entire RF is "learned". This means that retinal Hebbian learning only has a small set of available inputs to deal with, and can be quite accurate even in the presence of crosstalk. Unfortunately there will be some re-correlation in the thalamus, because there is convergence and divergence there (much of which may be due to inaccurate thalamic Hebbian learning[19]).

The process of running through a large set of inputs, slowly adjusting (the allowable maximum learning rate depends on how much variation there is in the patterns) and renormalizing the weights, is a type of averaging operation. The result is not an explicit listing of socs, but even more useful: it reveals the essential patterns of the socs, the characteristic or "eigen" vectors of $C$ (its "soul"). If one knows all the eigenvectors one can recreate $C$, but this is not necessary or useful: instead one wants to exploit the structure of $C$ to generate representations that allow efficient information transmission and subsequent hoc learning. In particular, one want to use learned eigenvectors of C to pairwise decorrelate signals, and make them have equal variance (so they use the full bandwidth of the available transmission channels, and in preparation for subsequent nonlinear learning). The process of decorrelating and matching the power of signals is called "whitening", since such signals have equal power at all frequencies, like white light.

Now we consider briefly what would happen using a linear but inaccurate Hebb rule. We will then return to the ICA problem. Our definition of inaccuracy is very basic: we simply assume that when a bunch of spike coincidences occurs at a connection, the ensuing weight change does not occur exclusively at that connection, but "spills over" to other connections. 2 obvious ways this could occur are either that some of the NMDAR-mediated calcium influx into the spine head, or some of the newly-inserted perisynaptic AMPARs, might reach other spine heads. Because we assume the Hebb rule is linear, it would be likely that all steps in strengthening are linear too, so the exact nature of the leaking agents(s) would not matter (although we strongly believe that the first molecule, calcium, is both the most critical and the most difficult to control). For a first analysis we ignore the (very likely) case that the amount of crosstalk will depend on the weights, and list the diffusional coupling between any 2 connections as entries in an "error matrix" $E$. This means that the actual pattern of updates produced by an input pattern can be calculated from the vector of weight changes that would occur in the completely specific case (i.e. using the "correct" learning rule) by transformation (in this case, "premultiplication") by $E$. Essentially this means that the incoming patterns are linearly distorted by the crosstalk: each pattern looks like $Fx$ instead of $x$, where $F = E^{1/2}$. (The reason why the input distortion is by $F$ rather than $E$ is because linear Hebbian learning is driven by covariances, and $E$ acts on the updates, so it is equivalent to $F$ acting on the patterns themselves). This leads immediately to the conclusion that the inaccurate linear Hebb rule will lead to learning of the principle eigenvector of $EC$; in antiHebbian learning one typically (but not always[BB]) learns instead the least eigenvector of $EC$. In both cases, as crosstalk increases, the learned weights shift increasingly away from the error-free values, so typically the encoding gets worse (in the Hebbian case), but for slight error the change in encoding efficiency is very small (and essentially negligible; the real problem arises in the next stage). As a further initial approximation, we ignore the complicated physical relationships between the synapses that comprise connections, and assume that, because each has many synapses, that the synapse physical locations change over time, and the intersynaptic messengers (such as calcium) diffuse further than the typical intersynaptic distance, all weights influence each other equally. At very high error (beyond the biological range) all weights tend to change by the same amount; combined with normalization this means the weights hardly change at all. This type of learning "failure" is not just unbiological, but also completely trivial.

The yellow line in Fig. 1b sketches how the weights corresponding to the least eigenvector of $\mathbf{EC}$ gradually move from the principle component of $\mathbf{C}$ (purple dot) as crosstalk increases. This is the line that would describe the stable weights, as a function of error, if one used a linear learning rule, but because we used a nonlinear rule, the weights change in a more complex manner, discussed below: above a crosstalk threshold they jump to the solid yellow line.

One further important detail is the way that increasing the number of inputs $n$ will increase crosstalk. In the simplest case, the additional inputs must be accommodated on the existing dendrites, so synapses get closer together in proportion to $n$. This is why the fraction of weight update that leaks from a given connection (which we call E = 1-Q, where Q, the diagonal element of E, is the "quality" of the updating process) should increase with $n$. If the updates are analog rather than digital[19], this leads to E = nb/(1-nb) where b is a "per connection" error rate that reflects all the factors influencing crosstalk (such as fractional escape from the spine, synapses per connection and dendritic length/diffusion space constant ratio). It is this factor b that we plot in Fig. 2; note that b is usually a very small number because calcium isolation by spines is excellent, but since $n$ is typically quite large (even for a tiny 10X10 image patch), Q can be significantly less than 1. Of course the error problem would disappear if dendrites could be made indefinitely long as inputs are added, but this is impossible, because of cable properties; the only way to overcome cable properties is to introduce strong dendritic nonlinearities, but now learning only occurs in short dendritic segments, again greatly restricting the number of inputs that can be learned from. Thus the outcome of crosstalk in linear learning is inadequate whitening, which while unimportant for signal transmission can be fatal for subsequent learning. We now describe how the nonlinear rule allows one to solve the ICA problem, and why it fails when crosstalk exceeds a threshold.

Solving the ICA problem requires using input hocs and not just input socs. If there are no inputs hocs, i.e. the input distribution is multivariate Gaussian (bell-shaped in every direction) , there is no way to solve the ICA problem because the ambiguity between $\mathbf{M}$ and $\mathbf{s}$ cannot be resolved. A nonlinear rule responds to hocs because the higher order terms in the Tatlor expansion of the nonlinearity make the update depend on powers of the output (and hence on powers of the other inputs), so averaging over input patterns yields weights that reflect hocs. More specifically, they reflect higher order moments of the input distribution. However, although these moments do also include socs, the soc contribution can be eliminated by first "whitening" the inputs, by passing them through a bank of (learned) decorrelating filters, as reviewed above. These preliminary filters will not be perfect because they cannot be perfectly learned even given unlimited time.

Although the mixing process does tend to make the inputs more Gaussian than the original sources, it also introduces socs and hocs (mixing by definition creates dependencies); the hocs can then be exploited to learn how to unmix the inputs. Indeed, mixing creates a specific pattern of hocs, which can then be used to infer a suitable unmixing process. However, the ambiguity that is fatal for Gaussian sources does not completely disappear: if all the sources are nonGaussian, the one the neuron chooses to track is largely arbitrary. This is why in our study we make all the sources but one Gaussian: this means the neuron can only learn one particular row of $\mathbf{M^{-1}}$. Since we know which row (i.e. target set of weights) it is supposed to learn, we can easily check if crosstalk prevents it from finding that row. Of course, it might prevent it in a graded, "graceful" manner (as in the linear case) or it might do it in a "catastrophic" manner, so it can no longer learn from hocs (but could learn from residual socs, if the prewhitening were imperfect). Our key result, around which our whole edifice of reasoning and speculation is built, is that the failure is catastrophic. Since the ICA model is the simplest, most robust, way to learn hocs, and yet still fails catastrophically if there is crosstalk, we suspect this would hold more generally, even when inputs are generated in more complicated ways that linear mixing.

So what sort of nonlinear learning rule can learn to do ICA? There have been various approaches to this problem, though they are all deeply connected. The most powerful strategy would be to use a rule that makes the outputs of a set of neurons as independent as possible; since the sources are independent, and the mapping from sources to inputs to outputs is deterministic, it follows that only when the neurons exactly unmix their inputs can the outputs be independent. This strategy boils down to learning weights such that the joint distribution of the outputs (the probabilities of all possible combinations) is the same as the product of their marginal distributions (the distributions of individual output neurons). This is known as a "factorial" representation, and is the holy grail of theoretical neuroscience, since it would provide a complete answer to object recognition and even deeper problems: the outputs (in particular, the product of their firing rates) would represent the probability that a set of objects, or other cause, is present. However, this approach is too ambitious: there is no way, for large n, to estimate the joint distribution from even a very large sample of observations. So the various approaches use heuristics that should provide robust approximations. A rather direct approach is to try to minimize the output interdependencies using a precisely "tailored" nonlinearity. It can be shown that the ideal nonlinearity corresponds to the cumulative density function of the sources, but since this is unknown one has to use approximations. The source cdf is generally sigmoidal, and this has lead to the widespread use of sigmoid nonlinearities such as the logistic (Boltzmann) and tanh functions, as heuristic approximations to the true but unknowable nonlinearities. Before going further, it is useful to sketch the 2 main mathematical issues at stake.

First, having devised a nonlinear learning rule based on a promising heuristic, one must show that the equilibria of the rule (in the slow learning limit) correspond to the desired solutions (permutations and scalings of $\mathbf{M^{-1}}$). Second, one must show these equilibria are *stable*: that small displacements away from equilibrium (due to the arrival of a new pattern, generated by the same underlying process) do not grow away from the equilibrium, but die back towards it as new patterns continue to arrive.

Here we consider only the simple case studied in our paper: that of a single output neuron. To derive a suitable learning rule in this case Hyvarinen and Oja used a powerful heuristic: they maximise the deviation of the output distribution from the Gaussian distribution, based on the idea that the Gaussian has the largest entropy (unpredictability) of any unbounded distribution (the entropy of the uniform distribution has maximum entropy in the bounded case, which is why we toss dice). It turns out, rather nicely, that provided the inputs are prewhitened (to remove the socs and thus make moments into cumulants[BB]), almost any nonlinear function will do this trick, so a row of $\mathbf{M_O^{-1}}$ is an equilibrium; for example, a cubic function will maximise the *kurtosis* (fourth order cumulant) of the output (kurtosis is zero in the Gauss case). However, the cubic is unbounded, so occasional large outlying inputs signals have a disproportionate effect, and it is better to use a bounded nonlinearity, such as tanh. It's relatively easy to see why a row of $\mathbf{M_O^{-1}}$ is an equilibrium of (almost any) nonlinear rule: the key step is the whitening, which converts the original $\mathbf{M}$ to the orthogonal matrix $\mathbf{M_O}$. An orthogonal matrix is one where all the rows and columns are mutually orthogonal and of equal length: they are vectors that point in mutually perpendicular directions. It has the interesting feature that its inverse is simply its transpose (i.e. a version in which corresponding rows and columns are interchanged). Now the eigenvectors of $\mathbf{C}$, which are used to whiten, are mutually perpendicular (they are the axes of the hyperellipsoid which has been fitted to contours of the joint distribution), and in order to whiten the mixed vectors coming out of $\mathbf{M}$ one must multiply them by a close relative of $\mathbf{C}$, its inverse square root (see the above discussion of $\mathbf{F}$ and $\mathbf{E}$). The result is that the sources are transformed in 2 linear steps to generate input signals: first mixing by $\mathbf{M}$ and then further mixing by the whitening process; the overall transformation is y = $\mathbf{M_o s}$ where $\mathbf{M_O}$ is the product of $\mathbf{M}$ and the whitening matrix. Now if the weights come to lie exactly in the direction of a row of $\mathbf{M_O^{-1}}$, which is orthogonal, then they must lie exactly in the direction of the corresponding column of $\mathbf{M_O}$ itself, which means that the output must exactly track the corresponding source (the column of $\mathbf{M_O}$ is just the set of coefficients leading from that source to the set of inputs). This means the neuron only sees the component of the input signals that is due to that source, which is by definition exactly proportional to those mixing coefficients; therefore any adjustment in the weights must keep the weights pointing exactly where they already are, which is the definition of an equilibrium.

However, to test the stability of this equilibrium one should realize that the above argument breaks down if the weights deviate even slightly from equilibrium (due to the arrival of a new pattern, which will slightly push the weights in its own direction, away from the equilibrium). Now the component of the input that is not due to the source in question is no longer exactly "blanked out", and it could push the weights further away from the equilibrium, towards any other available, and perhaps stable, equilibrium. Here the argument gets more involved, but clearly will depend on the nonlinearity itself, since the rate of growth in the various possible directions varies nonlinearly with the output (unlike the situation for Principle Components). The key quantity that determines the stability turns out to be the sign of the difference between 2 nonlinear quantities averaged over all the patterns: the nonlinear moment of the relevant source, and the corresponding nonlinear moment of a Gaussian variable that has the same variance as the source. This difference is essentially akin to a generalised cumulant, and reduces to the source kurtosis in the cubic case. This is why the rule has to be correctly signed: this corresponds to ensuring that the sign of the generalized cumulant is correct. Note that the tanh rule, unlike the cubic rule using $y^3$, would still work fine if the sources had zero kurtosis, provided they have even higher cumulants.

In summary, the 1-unit rule is biologically attractive because the nonlinearity does not have to be accurately implemented. Furthermore, the stability does not depend on the particular $\mathbf{M}$ used, so the rule is "universal" – but only if whitening is perfect, which we argue is biologically impossible. Of course, one does have the further complication that one needs a bunch of neurons, possibly (although not necessarily) finding different ICs, which may involve some additional means of coordination. However, that is a rather separate problem not treated here.

The obvious problem is that biologically it is impossible to exactly whiten the inputs in the way that seems to be required for the 1-unit rule to work. This is because, as explained above, crosstalk prevents one from learning precisely the appropriate eigenvector filters. (An "innatist" would argue that one should evolve, instead of learn, accurate whitening filters, but this ignores 2 facts: individuals, not populations, must learn, and evolution is subject to the same error catastrophe as learning). Fortunately, the 1-unit rule shows some robustness in this regard: typically the inputs do not have to be *exactly* white. This is basically because, if the rule is stable for white inputs, it will also be stable for small deviations from whiteness. The price one pays for this flexibility is some sacrifice in the generality: the amount of whitening that is required depends on the details of $\mathbf{M}$ (our results using a multi-unit rule suggests that one of these details is how close to orthogonality $\mathbf{M}$ is initially). Whitening is golden, but crosstalk alloys it with silver.

Crosstalk also affects the nonlinear rule itself, as we show in this paper. Indeed it is the combined effect of crosstalk on linear followed by nonlinear learning that can be catastrophic. Crosstalk can have 2, related, effects on nonlinear learning, one gradual, the other fatal. First, it can slightly affect the apparent distribution of the input vectors, in much the same way we described for the linear case. To a first approximation, one is learning, below the error threshold, source vectors mixed by $\mathbf{E} \, \mathbf{M_O}$ not $\mathbf{M_O}$, so one does not exactly earn a row of $\mathbf{M_O}^{-1}$ or retrieve the source at the equilibrium of the modified rule. This can be seen in the way the average equilbrium weights slightly change as one starts to increase crosstalk (Fig. 2). We have not checked whether the new weights agree exactly with a row of $\mathbf{E} \, \mathbf{M_O}$, because this would require extremely long runs at low learning rates. Now a key ingredient of the stability analysis (that the equilibrium is a row of $\mathbf{M_O}^{-1}$) breaks down, but because there is a zone of "structural stability" the breakdown does not occur until crosstalk exceeds some sufficiently small value (which clearly depends on $\mathbf{M_O}$, as one would expect). This breakdown is the second aspect of crosstalk.

However, thinking of error as modifying the input vectors is, in the nonlinear case, only a crude approximation. If one looks carefully at the (error-free) nonlinear rule, one sees the term $\mathbf{x}f(y)$ is actually $\mathbf{x}f(\mathbf{x}^T\mathbf{w})\mathbf{x}$, where $\mathbf{x}^T$ means "the input vector written as a column vector" ; this is just notation that says that one scales the input vector by the "nonlinearly transformed weighted sum of inputs" to get the weight change. Now if the effect of error were to change the input vector, we would again use $\mathbf{Fx}$ and $\mathbf{x}^T\mathbf{F}$ as in the linear case. The outcome of ICA learning would be a row of $(\mathbf{E}\,\mathbf{M_O})^{-1}$, not of $\mathbf{M_O}^{-1}$, since all one has done is change the overall mixing process. In general one would not expect an error catastrophe (though this may not be true in the offwhite case: the modified mixing matrix may become impossible to learn for a given degree of offwhitening even though the unmodified matrix is, since the stability proof does not extend to this case). But this is **not** what we think error does (although it would be if the errors reflected transmitter spillover between synapses). Instead, we think that AMPARs and/or calcium ions (as well as intermediates) diffuse, which can be represented by $\mathbf{E}\mathbf{x}f(y)$ (which is what we present here) or by $\mathbf{x}f(\mathbf{E}y)$ (which also shows an error catastrophe). One might argue that all 3 approaches are closely related, increasingly so as the nonlinearity weakens, and the real reason learning collapses is that one cannot sufficiently whiten. In our view, whitening and crosstalk are inextricably linked: better whitening reduces the impact of crosstalk, and less crosstalk yields better whitening. Furthermore, the best way to whiten is to reduce crosstalk (which requires proofreading) and the best way to reduce the impact of imperfect whitening on inaccurate nonlinear learning is also proofreading. It probably doesn't matter which way one applies proofreading, but it seems as if historically the second approach was adopted by the mammalian brain. Even though nonlinear rule may learn $(\mathbf{EM})^{-1}$ this does not mean that all one has to do is use an estimate of $\mathbf{E}^{-1}$ to slightly correct the final answer: for offwhite inputs $(\mathbf{EM})^{-1}$ may be *impossible* to learn, in which case there is nothing to correct. And minor correction of a row of $(\mathbf{EM})^{-1}$ is probably pointless because at subthreshold error the answer is almost correct anyway.

One needs a much more radical approach: proofreading. Dedicated proofreading largely solves the problem but is impractical. Distributed proofreading uses an approximation: that provided there are not too many current anatomical inputs (silent or not), a coincidence recently detected by a layer 6 cell belongs to the connection to its partner layer 4 cell across which that coincidence recently occurred. Provided the "confirmation" signal from the layer 6 cell gets back to the entire set of connections on that layer 4 cell before the next expected coincidence, it will approve the appropriate expression of the induced draft plasticity change (if it acts solely retroactively). This would probably not work if all the available connections were anatomically present: not only would this greatly increase the synapse density (and thus the crosstalk), but there would be mistakes: some incorrect approvals would occur (because new coincidences could occur before the approval had time to act). A more efficient strategy is to reduce the anatomical connectivity, so that weak connections spend some, or perhaps most, of their time anatomically disconnected. These incipient connections do not have to be proofread. However, this strategy does require continual, sleep-like, updating of the proofreading connections so that if a new, silent, connection happens to form as a result of a new spine forming, it can be enabled if it experiences a recent coincidence. Clearly, by judicious choice of the new spine formation rate, one can adjust the load of anatomical connections that must be proofread so the expected total coincidence rate, and thus proofreading inaccuracy, can be kept low. As learning advances, this gets easier, since there will be fewer coincidences. Of course even when learning is complete, a low level of coincidence, approval and strengthening continues, but with no net change in weights because of normalization, and as the population of incipient and silent connections swaps around, some ongoing updating of corticothalamic feedback would be needed.

Interestingly the sleep-like updating of the CT connections does not have to learn a strict generative model (inverse of the feedforward weights). It only has to learn a black/white version thereof (i.e. to disconnect layer 6-relay direct connections that correspond to current feedforward weights). This could be done by reverse-correlation and antiHebb learning. Because these weights are all-or-none, this learning should be very error-resistant, reducing the "quis custodiet custodes" problem.

The other key result of this paper is that at the failure point another equilibrium, one that is *always* stable for correlated Gaussian inputs (if the learning rate is appropriate), suddenly becomes stable. In a general sense this is not too surprising: the one exception to the linear learning outcome we described above is when the inputs are *perfectly white*. Under these conditions $C$ is just $\mathbf{I}$, the identity matrix (a diagonal of ones), the eigenvectors are 1,0,0..; 0,1,0,0… etc, but none of them are stable: they all have equal (or almost equal) eigenvalues, and stability requires the eigenvalues be significantly different. Furthermore, even if $\mathbf{C}$ deviates slightly from $\mathbf{I}$ round-off errors will inevitably interfere with the nominal stability, unless one imposed weight (and input) digitization. Now we have the reverse situation from ICA: addition of a small amount of crosstalk *stabilizes* linear learning!

With all this in mind, we turn one more to Fig. 1b (and Fig. 2; see also Supplementary Figure legends). Fig. 1b illustrates the 3 input case; since the weights are automatically normalised they always lie on the surface of the unit ball, and we can track their path over the surface of the ball as successive patterns arrive and learning proceeds. It illustrates schematically weights initialized at an arbitrary location, and changing in little steps as each new input pattern arrives. (We exaggerated the steps for clarity; normally only a very low learning rate is used, so each step is tiny). The patterns themselves are not shown but they form a complex 3D cloud, at various distances from the center, whose shape represents the underlying joint distribution (or, equivalently, the pattern correlations at all orders). An arbitrary pattern drawn from the cloud is presented as input, in the absence of crosstalk: this moves the weight by a tiny amount in the direction of the pattern (i.e. parallel to the direction in which the pattern points, not literally towards the pattern); the amount depends on the nonlinearly transformed output produced by the pattern, which is just the "projection", or shadow, of the pattern on the *current* weight vector. If the amount was just proportional to the projection, then the average outcome would be an eigenvector of $\mathbf{C}$, but the nonlinearity destroys the stability of this outcome. The little movement parallel to the input pattern moves the weights slightly off the surface of the ball, but the brute normalization step moves the weights vertically and exactly back to the surface. The outcome is a slight displacement on the surface (the first zig). Subsequent patterns cause further zigzagging, but these get smaller (because the patterns tend to cluster in the direction of the final weight vector), until the weights, averaged over a suitable time window, reach the stable equilibrium of the rule (the red point). They do continue to show further small movements near this equilibrium because of the finite learning rate, but these are not shown (they can be thought of as occurring mostly within the diameter of the red dot). This "red" direction corresponds to the row of $\mathbf{M}^{-1}$ that is most similar to the column of $\mathbf{M}$ that emanates from the Laplace source (see Fig. 1c). The coordinates for plotting the weights have been rotated so that the red column of $\mathbf{M}$ points exactly to the reader. The prewhitening was not perfect (we used a relatively small batch of 1000 vectors to "learn" the whitening matrix), as shown by the fact that the covariance matrix $\mathbf{C_L}$ (Fig. 1c) of a *large* sample of whitened input vectors is not exactly $2\mathbf{I}$; the factor 2 comes in because we want to scale the input vectors to match the tanh nonlinearity, which runs from -1 to 1). This is why $\mathbf{M_O}$ is not exactly orthogonal, and the rows of $\mathbf{M_O}^{-1}$ (the red, blue and green lines) are not exactly mutually perpendicular.

Thus we see that under the action of the nonlinear rule the weights move to an average position that corresponds to a row of $\mathbf{M_O}^{-1}$, so the neuron's output almost exactly tracks the Laplacian source: it sees the "object" hidden in the successive patterns.

Now we add crosstalk. For very small crosstalk, the average weights stay very close to the red dot, though they do perhaps wander somewhat more. But suddenly, at a typically quite low threshold crosstalk level (i.e. far below the "trivial" value at which all update specificity is lost), they start a rather rapid new zigzag march to an entirely new location, which lies on the yellow line corresponding to the least eigenvector of $\mathbf{EC}$ (i.e. what a linear rule would learn from the same inputs, or what the nonlinear rule learns when using all-Gaussian sources). Of course at first they jump to the (average) position that corresponds to the threshold value of Q. Further increases in crosstalk push the weights further along the yellow line. The dotted part of the yellow line corresponds to the various eigenvectors of $\mathbf{EC}$ at various *subthreshold* errors. In Fig. 2a the jump to the eigenvector of $\mathbf{EC}$ (which corresponds to linear learning, or to nonlinear learning from correlated Gaussian vectors) is shown by a sudden large movement of one of the weights (red points and line); the other 2 weights show smaller jumps (though in other cases we studied all the weights made medium jumps). In Fig. 2b, we see that the angle between the average stable weight vector and the appropriate row of $\mathbf{M_O}^{-1}$ is very close to zero up to the threshold level of crosstalk, when it jumps far away; conversely, at the threshold the angle between the weight vectors learned using either all 3 Gauss or 2 Gauss plus one Laplace source (blue line) jumps to zero at the threshold. We recommend that the reader go back and forth between figures 1 and 2, until their relationship is clear.

Thus hoc learning collapses to soc learning if the update rule is not highly specific. We also looked at all Laplace cases (n=3,5 and 10) but these are more difficult to understand, because there are now n possible stable ICs: we found that all would destabilize, but at different crosstalk thresholds, so the weight vector moves around in a complicated way between different ICs and the least PC as error increases. We believe that the 2 Gauss 1 Laplace case is the simplest to study and understand, although it may be easiest to get analytic results in the 1 Laplace, many Gauss, case, since as the vectors get increasingly high-dimensional, the results should get closer to an "average" case, with an $\mathbf{M}$-independent crosstalk threshold. Of course, one can achieve much the same result by studying many individual examples of low-dimensional cases; it is therefore likely that the "average" error threshold, for biologically reasonable degrees of whitening, for large numbers of random $\mathbf{M}$, is quite close to that we determined ($b_c = 0.04$). This value probably underestimates the true "self-consistent" error sensitivity, in the sense that if the whitening matrix had been learned using a rule with crosstalk, a much lower error threshold ($b_c \sim 0.01$) would have been found. However, both these values seem to be close to biologically plausible lower limits to Hebbian accuracy[BB]. Note also that when we used logistically-distributed sources (kurtosis 1.2), $b_c$ was about halved. Proximal causes underlying sensory signals may typically be quite close to Gaussian, making nonlinear learning difficult and extremely error-sensitive.

Some last general points. In these unsupervised models (PCA and ICA, and also more general models that exploit hocs that are generated in more complex ways than in ICA), learning is ongoing, on-line and open-ended. The rules not only try to find the current "best weights", but they update their estimates gradually and continuously, so as to track any change in the underlying statistical model, or any changes in upstream networks (for example due to neuron death). The downside of this flexibility is that if the input statistics become less favorable, then all previous learning is lost: a true catastrophe! For example, if a unit learns, at a suitably low level of crosstalk, to find the weights corresponding to a Laplacian source, but then that source shifts to a logistic distribution, the neuron will completely unlearn what it knew! If higher-level nets have learned using the inputs from this now useless unit, they will also come tumbling down, so the whole neocortex returns to a "blank slate", or demented, condition. Unfortunately, this learning collapse cannot be prevented, except by bringing 'clarity' back to the input statistics. There is no way the neuron can "know" that at some point it has reached the "right" weights, and stop there – the only "template" it can use is the input statistics themselves.

We focus on PCA and ICA not because these techniques are powerful general techniques that the brain uses to understand the world (they are not) but because they allow a quantitative discussion of the key problem the brain faces in implementing any connectionist-style strategy: making the connection updates sufficiently specific. Our view is that as the learning problems get harder, the specificity problem will only get worse. However, if proofreading is, by necessity, built into the very core of the hierarchical learning the cortex does, it may very well be used in additional ways (some of which we have hinted at in prior publications) as part of the overall strategy. Also, if there are several ways to skin the complex learning cat, ways that place more emphasis on precise, proofread, feedforward learning may work best.

## Section 4

## Supplementary Notes

**A** The "mind" is usually viewed as a manifestation of the coordinated interaction of many cortical areas: "modules"[28,29,30]. Since these areas differ in extent, number and function between mammals, this view is probably incompatible with the possibility that all mammals have "minds". Our viewpoint is that it is the type of processing/learning/something else that is common to all cortical areas that enables "mind"; in this view a core cortical "columnar" microcircuit would, uniquely, allow complex learning (which we define as learning from higher-order correlations). The possibility of hierarchically linking together multiple cortical areas would then stem from their ability to do complex learning, the inverse of the traditional view. Complex learning is inexhaustible (since hocs are essentially infinite), and must be done gradually, in many passes, whereas simple learning (from socs) can be done in a single pass. We have previously argued that neocortical "proofreading" confers the ability to systematically learn from hocs[31,32]; of course non-cortical structures could also, in favourable circumstances, also learn from hocs, but not systematically (because they frequently would suffer error catastrophes). We call this the "hocus-socus" viewpoint. Cortical areas could be viewed as providing higher-order corrections to actions initiated by subcortical structures, consistent with the observation that cortical afferents are merely collaterals of axons that supply subcortical motor structures that also receive collaterals from layer 5 "output" corticothalamic axons[33,34].

This view parallels the modern view of "life', as being not a collection of complex functions (respiration, digestion etc) but as the molecular machinery which evolves these functions. Thus the key to life is the ability to evolve genes, and the key to mind the ability for neurons to learn from hocs. Of course one needs many genes and many neurons, but the problem of coordinating them is a separate one not tackled here or in basic accounts of the molecular basis of life.

Our approach focuses on the fundamental operation of connectionist brain models, the activity-dependent adjustment of individual synaptic weights. It has been previously recognized that this operation should be connection-specific (indeed, this is the central, though usually unstated, assumption of connectionism) and that biological adjustments are indeed highly specific. However, previous work has not addressed the central issue: how specific must the adjustments be for sophisticated learning to be possible? Here we take "sophisticated " to correspond to "hoc-driven".

**B** The most familiar form of correlation is simple pairwise correlation between 2 random variables (which throughout we assume have zero mean), measured by a covariance (average value of product of 2 variables $x_1, x_2 = \langle x_1 x_2 \rangle$ over the entire set of observations). These "socs" make up the covariance matrix $\mathbf{C}$ (Fig. 1c), and linear Hebbian learning finds a set of characteristic synaptic weights, eigenvectors of that matrix. $\mathbf{C}$ (or equivalently its eigenvectors) completely characterizes the joint distribution of multivariate Gaussian random variables. From joint distributions one can calculate how probable it is one is observing a specific cause given an observation. A multivariate Gaussian distribution results from linearly combining independent Gaussian variables, using a mixing matrix $\mathbf{A}$, with $\mathbf{C} = \mathbf{A}\mathbf{A}^{\mathbf{T}}$. However, in this Gaussian case $\mathbf{C}$ does not allow identification because one cannot get $\mathbf{A}$ from $\mathbf{C}$. If the underlying causes are nonGaussian $\mathbf{C}$ alone does not allow the causes to be inferred nor the mixing process identified. While pairwise decorrelation ("whitening"; see below) is a useful preprocessing step (e.g. before sending signals over long, costly axons) it cannot uncover probable causes. The real clues about the causes of sensory data (which aid survival in a complex world) are contained in hocs: average values of products of more than 2 measurements. When we refer to "hocs" we are actually referring to "crosscumulants" (i.e. the joint cumulant of several random variables $x_1, ..., x_n$), which are moments corrected for the contributions equi-covariant Gaussian signals would make (i.e. all cumulants are zero for Gaussian signals; the full set of cumulants are equivalent to the variables' joint distribution). A specific example of a fourth order crosscumulant would be

$$cum(x_1 x_2 x_3 x_4) = <x_1 x_2 x_3 x_4> - <x_1 x_2> <x_3 x_4> - <x_1 x_3> <x_2 x_4> - <x_2 x_3> <x_1 x_4>$$

It's worth underlining the key property of hocs: while $(x_1 x_2)(x_3 x_4) = (x_1 x_2 x_3 x_4)$ so that one can do the overall multiplication by separately multiplying pairs of variables, and then multiply the products (e.g. using 3 separate neurons), this is NOT true for moments, since $<x_1 x_2 x_3 x_4>$ is NOT equal to $<x_1 x_2><x_3 x_4>$ UNLESS the variables are statistically independent. This means that hoc learning *must* be done by single neurons: in this case many hands do not make light work. In particular, a Hebbian rule at a single synapse can, in principle, do an analog high-order moment calculation based on coincidence detection. In principle one could use networks of neurons to multiply signals, for example using recurrent connections. But this is a hopeless approach to learning, since there are many more synapses than neurons. (Our "proofreading" idea does use neurons to multiply but they are used to *supplement* learning by synapses, not replace it; we suspect that any scheme that uses neurons to "boost" Hebbian multiplication will prove equivalent to our proofreading scheme.)

The whole point of our paper is that in practice the coincidence-detection has to be so extraordinarily accurate that it cannot be achieved at typical observed synapse densities given the known, likely or conceivable biophysics of Hebbian synapses. Therefore either the neocortex does not need to learn high-order relationships (for example because they have already been discovered by genetic evolution, and hardwired in), or it does learn such relations and therefore must have overcome this biophysical limit (for example, by proofreading). Since the same error catastrophe limits genetic evolution (to the observed ~ 1 billion bytes), the former is actually not a possible solution, despite widespread "innatist" belief, and since we show the neocortex has the circuitry and physiology required to proofread, our explanation is the most parsimonious. It does however run counter to the prevalent belief that the brain can be noisy and sloppy and still do extraordinary things. Experimental neuroscientists know that synapses cannot be accurate but tend to think in terms of pairwise correlations, for which accuracy is less important; conversely, neural theorists with a machine learning perspective know that hocs are important, but tend to assume that synapses are highly accurate. Our paper tries to unite these viewpoints. Interestingly some psychologists have realised that computation requires impossibly accurate synapses[37,38] but have then wrongly concluded that computation must be done nonsynaptically, by some as yet undiscovered neural mechanism that possesses the extreme accuracy of DNA! However, we take a less radical tack: we suggest that much of the mysterious circuitry of the neocortex endows synapses with the necessary (albeit not complete) accuracy. The defining feature of neocortex would thus be its ability to systematically learn hocs and thus partly understand the world, which we take to be the defining property of minds. This view is both radical and prosaic, since it equates mind to the repetitive operation of a seemingly trivial, but highly accurate, synaptic pairing rule. Similarly, modern biology equates life to the repetitive operation of a seemingly trivial, but highly accurate, molecular pairing rule.

The type of unsupervised learning we consider in this paper is quite different from memorization, which involves storing explicit templates (an observed activity pattern). Instead we consider learning of implicit templates: the input patterns are created from hidden templates by a particular, not completely known, "generative" rule, and the learning task is to infer the templates from the observed input patterns ("objects from pixels"). We study here a particularly simple model: the "generative" rule is linear deterministic mixing of independently fluctuating sources, at least one of which is nonGaussian, by a square matrix of coefficients $\mathbf{M}$, and the goal is, roughly speaking (for quasiorthogonal mixing), to infer a column of the matrix (i.e. the set of mixing coefficients emanating from a particular source). Even this type of learning is difficult because it requires sensitivity to hocs, but it is otherwise the least difficult type of learning (and our main result, that it is impossible when there is too much crosstalk, therefore likely generalizes to more difficult cases). Note that even in this simple case, an enormous range of possible problems, of varying difficulties, emerges, depending on the choice of $\mathbf{M}$ and the nonGaussian source distribution. If the generative rule is less favorable, then simple feedforward learning may not converge, because unique weight vectors are not constrained by the observed input. If it does converge, the outputs may now not be completely independent (although more so than the inputs), and their residual hocs could in principle drive further useful learning (but to do so the outputs would have to be first nonlinearly transformed and rewhitened[Y]).

To test whether crosstalk can prevent feedforward learning in a wide variety of circumstances, one must have a model that can, in the absence of crosstalk, reliably learns in a variety of circumstances; the ICA model fits this requirement. Furthermore, if there are only feedforward connections, neurons must learn individually. This leads to the 1-unit ICA model. However, the 1-unit rule can find any of a set of nonGaussian sources (and any of the rows of $\mathbf{M}^{-1}$). This creates a complication: since our criterion of hoc learning failure is that any stable weight vector must correspond to an eigenvector of $\mathbf{EC}$, one has to be able to clearly distinguish these eigenvectors from any of the possible rows of $\mathbf{M}^{-1}$ in inevitably noisy (finite learning rate) numerical experiments. Even for n=2 or 3 this is often not easy. Therefore we further simplified the model by making at least 2 of the sources Gaussian (ICA can learn an IC corresponding to a single Gaussian). This is why we focused on the n=3, 2-Gauss case. However, we found that even in the n=3 all nonGauss case the learned vector above the threshold was indistinguishable from the least eigenvector of $\mathbf{EC}$ (though it could also be close to an IC).

A Hebbian rule, where the weight adjustment depends on (some function of) input and output activities (this definition includes supervised delta-type rules), directly responds to input-output correlation, and indirectly to input-input correlation (since the output depends, via the weights, on all the inputs). If the rule is linear, then for sufficiently small learning rates (in any case necessary for weight convergence) all the hocs drop out; a nonlinear rule, if all the terms in its Taylor expansion are nonzero, is in principle sensitive to all the hocs. However, in the paper we show that a nonlinear rule loses its hoc sensitivity if it is insufficiently accurate (which in turn means that when the brain learns to whiten, it does not need an exactly linear rule, which would be difficult to engineer). In a recent paper[20] we showed that an inaccurate linear rule learns to decorrelate (albeit imperfectly).

**C** These "molecular accidents", i.e. mutations, are mistakes in copying DNA, which are amplified by successive rounds of replication. Although mutation is often viewed as the driving motor of evolution, the fact that life is possible only because mutations are exceedingly rare[40-45] suggests that the real motor is accurate amplification; mistakes are unavoidable, and usually bad, accompaniments. (Interestingly, Eigen speculates that similar ideas might apply to the brain[44]). There are 2 slightly different proposals for the synaptic equivalent of "mutation", both based on the digital nature of synaptic adjustments[35]. (1) mutations could correspond to formation of new connections (by the growth of a spine at close axodendritic approaches to form a silent synapse[46,47,C]. (2) mutations could correspond to Hebbian errors[48], for example due to intersynapse calcium diffusion. Although Hebbian errors are rare[36], they do occur[102]. In our early papers[21,22] these 2 ideas were combined. Although they are mechanistically distinct, they are closely related. We have distinguished[20] 3 types of Hebbian error: Type 0 (activity noise, which merely alters pattern statistics and does not otherwise affect learning); type 1 (synaptic updates noisy); type 2 (synaptic updates do not occur independently, primarily due to chemical diffusion; these errors require "proofreading"). Although silent synapses do not affect the current weight vector, they are not a free lunch: they are targets for type 2 error. The greater the number of connections (silent or not), the lower the accuracy of Hebbian amplification[20]. One could regard a silent synapse that forms a new connection as a recessive mutation. Diploidy (and the possibility of recessiveness) and sex[HH] are quite different though they often occur together.

**D** The spine neck is long and narrow and carries calcium pumps[49] that prevent almost all the NMDAR-mediated Ca signal from reaching the shaft [50-57]. However the exact degree of escape has been controversial, largely because measurements are noisy and obscured by "piggybacking" on added diffusible dye[58,59]. The neck cannot be too long and narrow, because this would electrically uncouple the synapse (even silent synapses have to "see" the backpropagating spike; strong synapses must have neck conductances higher than the synaptic conductance[27]. This issue is also controversial[60-63]; the double controversy indicates that synapses cannot unequivocally be both chemically isolated and electrically coupled, despite widespread hope. Indeed, spines on olfactory granule neurons are long enough to provide complete chemical isolation, but they are also electrically isolated[160].

**E** If n-length binary strings are copied with a per-base error rate e, and one particular "master" sequence replicates fastest, then since the probability of incorrect sequence copying is $(1-e)^n$, in a fixed population the survival of the master requires $1/e > n/2$. This argument neglects backmutation; the corrected formula[40,42] is $1/e > n/(\ln s)$ where s measures the "superiority" of the master. In the simplest case, where all nonmaster sequences replicate at the same rate, s is the ratio of master to nonmaster replication rates[42]. Thus as more information about the environment is packed into a string (n increasing), e must decrease. Since mutation usually produces only small changes in fitness, a good rule of thumb, simple but profound, is that e must be less than 1/n, i.e. the accuracy of the elementary step (base copying) must be greater than the string is long. More generally, if there are m bits stored per position (in polynucleotides m=2), the accuracy must be greater than 1/nm. The fact that evolution fails abruptly for e > (ln s)/nm is known as the Eigen "error catastrophe", and it critical for origin of life theories: life originated by the chance, purely chemical, formation of an RNA that could act as a replicase that is more accurate than the reciprocal of its own length[64]. In this paper we argue that as more environmental information is packed into a weight vector (n increasing), by selective adjustment of connections ("learning"), the more specific the weight adjustments must be. This does not imply a direct equivalence between synapses and bases, though in each case a specific pairing (of spikes or bases) operation underlies the learning. Since as inputs are added synapses must get closer, one expects a similar dependence of (1-Q) on n in both the evolution and neural learning cases[20]. If one instead were to lengthen the dendrites to accommodate the extra inputs, one must counter the ensuing cable attenuation by adding more synapses to each connection, in a self-defeating manner. In essence we equate the origin of mind to the origin of life: both produce structures whose complexity hinges on the accuracy of the (molecular or synaptic) elemental, structure-producing, step.

The Eigen model, on which 1/e < n/ln(s) is based, assigns a fitness value to every sequence; this assignment implicitly specifies the "environment" in which the polynucleotide population evolves, but the environment itself is not otherwise specified. One could regard the (fluctuating) environment "seen" by the population as an n-dimensional random vector whose hocs define the fitness-sequence mapping; in this sense the Eigen model is very closely linked to hoc learning – indeed, the world seen by the evolving genetic population and by the brain is *the* world itself.

**F** Our original proposal was that special neurons measure coincidence across connections and control their plasticity by a purely presynaptic mechanism[21,22]. We then realized that greater accuracy, and even better agreement with observed circuitry, could be achieved by controlling plasticity both pre- and postsynaptically; we also realized that this mechanism is a type of "proofreading"[31]. Our original model incorporated a type of pushpull control, since it compared coincidence across existing and incipient connections. However, the current pushpull model (Fig. 3b) is simpler and agrees well with available data.

**G** In general the spread of plasticity signals from connection to connection ("crosstalk") will depend on the current geometry (dendritic locations of constituent synapses) and current strength of those connections. We first make the simplification that these 2 factors are independent. Second, we assume the spread is weight-independent. This is unlikely to be exactly true since the spine neck electrical (and diffusive[57]) conductance probably adjusts to comfortably exceed the synapse conductance[63], in an effort to minimize diffusive coupling and maximise electrical coupling. However, there must be some minimum level of neck conductance, probably greater than 1 nS (calculated assuming 10 open NMDARs), even in a silent synapse, since the bAP must efficiently invade the spine head to unblock the NMDARs. Thus our weight-independence assumption is highly conservative. Additional weight-dependent crosstalk will further distort learning. Finally, we usually assumed that all weights give and receive crosstalk equally (i.e. $\mathbf{E}$ has uniform diagonal elements Q and uniform offdiagonal elements), despite the fact that crosstalk between individual synapses is highly local. We make this drastic simplifying assumption based on 3 arguments[20]: first, cortical connections are multisynapse[50,65,66]; second, synapses form and disappear over time[6,16-19,46,47], even in the adult; in both cases it is likely that the relative locations of excitatory, spinous, synapses is largely random (at least within a given dendritic zone, e.g. basal dendrites), based on haphazard axodendritic close approaches[16,17,18,19]. Third, calcium spreads over dendritic distances ( several microns[57]) greater than typical intersynaptic distances (in cortex 0.5 $\mu m$[67]; on Purkinje cells down to 0.06 $\mu m$[68]). The combination of these 3 factors should result in an average error matrix that has almost uniform diagonal or offdiagonal entries, at least to a first approximation. This assumption also corresponds to the usual connectionist assumption that all weights of a given type follow the same learning rule, and to the assumption usually made in molecular evolution models of position-independent mutation rates. However, we found that the results described in this paper were not qualitatively (and only slightly quantitatively) affected by randomly perturbing the error matrix entries (by mean factors up to 30%) from uniformity ("hotspots").

The assumption of a fixed error matrix is a "mean-field" (i.e. neglecting fluctuations) simplification of the real situation where the crosstalk pattern evolves unpredictably as learning takes place. If crosstalk is weight-dependent, this may be an oversimplification, and it will be important to check whether this approach is valid. Even more importantly, this means the brain cannot "know", *a priori*, the current crosstalk distribution (though it can know the current connection pattern), and therefore cannot "deconvolve" crosstalk-induced blurring. In other words, an error is not an error if it can be corrected. Proofreading does not "correct" errors – it merely lowers the effective error rate. Mismatch repair does correct errors, using the information in the parent strand; this approach is not available for neural learning because the world keeps changing.

**H** We varied the task in several ways: varying n, $\mathbf{M}$, the source statistics, the proportion of Gaussian sources or the degree of whitening. However we expect that the main biological effect of varying n would arise because as inputs are added, synapses must move closer together, increasing crosstalk[20]. Therefore we mostly focused on the n = 3 case, which is easiest to analyse. All these factors will affect the relative amount of hocs and socs in the inputs (mixing Gaussian sources only generates socs).

**I** Hoc learning must occur in a single neuron because this is the largest neural unit where multiple inputs converge. It has been proposed[69-72] that neural subunits (e.g. excitable dendritic segments) could also learn from hocs, but necessarily this would greatly restrict the richness of the available hocs. A subunit could learn a useful representation based on this hoc subset, and these representations could then be combined in a further learning process, but the underlying higher-dimension hocs would have been discarded. In this sense composing a neuron out of excitable subunits does *not* increase the neuron's "computational power", even though the neuron becomes formally equivalent to a network of "subneurons"[71]. We believe that since complex learning is the key to intelligence, this difficulty is fatal. Of course this limitation applies to whole neurons too; it seems likely that the world's statistical structure is in some sense sufficiently smooth (i.e. the joint distribution of all brain inputs is, in our world, somewhat redundant) that simultaneous access to all possible hocs, by a single "superneuron", is not required. This is related to the problem of whether the world is entirely comprehensible. More prosaically, the cortex evolved to provide sufficient understanding that mammals can usually survive. Connectionist models of hoc learning (e.g. Boltzmann and Helmholtz machines[73-75]) ignore the problem of crosstalk because learning rules can be implemented almost exactly on digital computers, but this suffers from the reverse flaw: it requires serial processing. The brain uses massive parallelism and in our view the most significant parallelism is hoc learning by single neurons. This process appears to be quasidigital: it involves detection of individual spike pairs and their encoding by multibit synapses. Proofreading makes this process even more accurate. It is a way to make sloppy analog neural computation more digital without sacrificing massive parallelism.

In the brain different neurons learn different aspects of hocs (e.g. different ICs) using additional sets of connections (e.g. enforcing "antiredundancy"[76]). If some neurons have successfully learned, these connections could provide additional clues to other neurons, but the antiredundancy connections or rules would themselves be subject to crosstalk. More generally, neurons receive inputs generated in a more complex "generative" manner than square deterministic linear mixing, and simple feedforward learning may fail to converge to stable weight vectors that reflect aspects of the underlying generative process; finding networks architectures and rules that allow useful learning is a focus of much current work[73-75]. Nevertheless, given appropriate and appropriately-learned additional inputs from other neurons (e.g. lateral or top-down), feedforward learning by single neurons must converge to reflect aspects of the hocs present in the feedforward inputs if the neuron is to be learn. In the single unit ICA neuron studied in this paper[77-80], no additional inputs are needed. We conjecture that since crosstalk would also interfere with learning appropriate "additional inputs", top-down or lateral input cannot overcome the catastrophic effects of crosstalk between feedforward inputs, unless such input is formally equivalent to the proofreading scheme outlined here.

**J** ICA learns sets of weights that maximize the independence of the output neurons; if the inputs to these neurons are generated by linearly mixing independent nonGaussian "sources", this learning procedure culminates in weights that "unmix" the inputs, so the outputs match the sources (exactly in the slow learning limit). In one-unit ICA, instead of maximizing output independence, one maximizes output nonGaussianity, subject to a normalization constraint and to prewhitening, and learns just one row of $\mathbf{M}^{-1}$. If the linear mixing assumption is not met, ICA learning still tries to maximize output independence (or, in the 1-unit case, nonGaussianity). If it succeeds in finding stable weights, these will either be "pseudoICs" (e.g. unmixing rows for noisy ICA), which do not allow accurate source tracking) or, if whitening is insufficient, least or greatest PCs. Since the sources are not recovered, the outputs will contain residual hocs even after learning stabilizes. Additional information about the sources might be needed to exploit these hocs (for example in further cortical areas).

**K**      **M** is the matrix of mixing coefficients; unmixing corresponds to finding the "inverse" matrix $\mathbf{M}^{-1}$ that "undoes" the effect of $\mathbf{M}$. We generate $\mathbf{M}$ by picking numbers randomly in the range 0 to 1. However, the 1-unit rule usually cannot learn to find a row of $\mathbf{M}^{-1}$, basically because there are strong socs which interfere with hoc learning. (Instead, it typically finds an eigenvector of $\mathbf{C}$). This problem can be solved by "whitening" the inputs: using preprocessing (by an auxiliary neural net) which removes the socs. Indeed, the standard 1 unit rule assumes this whitening can be done perfectly[77-79]. As explained in the paper this process converts $\mathbf{M}$ to a new matrix $\mathbf{M_o} = (\mathbf{C_s}^{1/2})^{-1} \mathbf{M}$, which, when $\mathbf{C_s}$ approaches the true $\mathbf{C}$, is "orthogonal": its rows and columns all have equal length and lie orthogonal to each other. In principle linear Hebbian learning can generate neural whitening filters. However, we do not completely whiten the inputs (i.e. we use the small sample $\mathbf{C_s}$, not the true $\mathbf{C}$), since we have argued previously that crosstalk makes this neurally impossible[20]. Thus our $\mathbf{M_O}$ is not completely orthogonal, though the lengths of the rows and columns are roughly the same, as are their directions (see Fig. 1c). Fortunately (for both the authors and the brain) this is usually adequate to allow learning; in the paper we show this is unfortunately no longer true in the brain when there is low, but significant, crosstalk.

The 1-unit rule is based on an approximation to "negentropy" as a measure of nonGaussianity[79] (rather than on information maximization as in the original BS rule[76]). By adjusting the weights to maximize the "nonGaussianity" of the source estimates, one can find ICs. This approach builds on the fact that the Gaussian distribution has the highest entropy of all unbounded distributions. Estimating output entropy requires knowing the full output probability distribution for all possible weights vectors, which is infeasible. However, by using approximations to entropy, one can obtain practical online learning rules. The simplest approximation is kurtosis, but this leads to a cubic learning rule, which (being unbounded) is outlier-sensitive. Use of bounded nonlinearities such as the tanh function used here gives more robust learning[79]. One must expect that because practical ICA rules involve approximations, they will fail in certain situations (for example when $\mathbf{M}$ is "illconditioned", or close to noninvertible, and especially when the assumptions are violated). One could regard crosstalk as narrowing the range of conditions in which the necessary approximations/assumptions are valid, and proofreading as widening this range. This suggests proofreading could also be regarded as a technique for learning stable weight vectors in situations where the ICA assumptions are violated, rather than just combating crosstalk. In this paper we do not attempt to outline the full set of conditions under which proofreading should work; in particular we do not discuss how the necessary proofreading circuitry can be learned[35,36,81,82]. It turns out that adjusting distributed proofreading circuits so as to keep pace with feedforward learning requires learning a simplified form of internal generative model. This provides links from our work to other ideas in unsupervised learning[74,75]. Since crosstalk and other deviations from the ICA assumptions are interlinked, it is not possible to completely disentangle these different aspects of proofreading. If there were no crosstalk, proofreading could still be useful; if there were no deviations from the 1-unit ICA assumptions (in particular, perfect whitening), one would not need proofreading to combat inevitable crosstalk.

The proof that the 1-unit rule converges[78] assumes that $\mathbf{M_O}$ is exactly orthogonal. The underlying reason for this restriction is that socs interfere with learning hocs, and when $\mathbf{M_O}$ is exactly orthogonal $\mathbf{C} = \mathbf{I}$ for the mix vectors (perfect whitening; no socs). Furthermore, the stability condition involves only nonlinear moments of s (and of a Gaussian variable having the same variance as s; Eq. 8.35 in reference 79). However, the Hartman-Grobman theorem[83] suggests that even if $\mathbf{M_O}$ is not exactly orthogonal the rule should still be stable, since "structural stability" applies not just to small perturbations of variables, but also of parameters. Indeed, we find that provided $\mathbf{M_O}$ is sufficiently orthogonal (either because a fairly large batch number is used to estimate $\mathbf{C}$, or, biologically, because the learning rate and crosstalk are both small enough), the 1-unit error-free rule reliably converges to the IC, to a degree that reflects the learning rate. When $\mathbf{M_O}$ is not exactly orthogonal the rule converges to a row of $\mathbf{M_O}^{-1}$ not a column of $\mathbf{M_O}$; see Fig. 1c; of course for $\mathbf{M_O}$ exactly orthogonal the distinction vanishes. We measured the orthogonality of $\mathbf{M_O}$ using the Frobenius norm of $(\mathbf{I} - \mathbf{M_O M^T})$ where $\mathbf{I}$ is the identity matrix (see Supplementary Legend Fig 2). For the case shown in Fig. 2, the partial whitening made $\mathbf{M_O}$ twenty times more orthogonal than $\mathbf{M}$. Interestingly, we found that when $\mathbf{M_o}$ was not sufficiently orthogonal, the error-free 1-unit rule converged to the least PC, even with nonGaussian inputs.

The appropriate choice of sign in the nonlinear learning rule reflects the nonlinearity used[79]. For the $f(y) = y^3$, one must use + but for $f(y) = \tanh(y)$ one uses - . We refer to the former as a Hebb rule and the latter as an antiHebb rule.

**L** Most biological weight adjustments probably use spike-timing-dependent-plasticity (STDP[84-88]), rather than mean rate rules. Simple STDP rules, combined with reasonable assumptions about the way spikes trigger spikes via epsps, lead to correlational learning[88]. The introduction of time greatly expands the statistical richness of underlying generative models and complicates (but can aid in) deciphering "causes". Now learning must be sensitive to the temporal sequencing of input/output spike pairs and thus of input patterns. This places additional demands on synaptic wetware: specifically, even in the simplest case, the size of adjustments must reflect the time intervals between input/output spike pairs (or even higher-order spike relations). At minimum this means the synaptic machinery must carry some "memory" of the spike arrival time(s), which means that some chemicals must persist. The longer a chemical signal persists, the further it, or its products, can diffuse. Thus STDP rules are likely to be more crosstalk-prone than rate-based rules. Chemical can be anchored, but some local diffusion within the synapse must take place, since these chemicals are triggers for downstream expression mechanisms. For example, one could anchor calmodulin in the postsynaptic density, but at the price of not capturing all the incoming calcium; alternately, one could sense more of the calcium using mobile calmodulin, which it would diffuse in all directions. The spine neck exists to restrict such diffusion, but cannot be totally effective without isolating the spine electrically.
One would incorporate errors into STDP rules by using a time-dependent error matrix. One can also make errors and STDP weight-dependent.

More specifically, it seems likely that a trigger for STDP is brief (10 to 100 msec) NMDAR-mediated calcium influx; the size of this calcium signal, which persists about 1 sec, reads out the time of arrival of a spine-invading bAP relative to the opening of NMDARs, which is triggered by a (presynaptic) forward spike[89,90,91,161]. However, this graded calcium signal, via calmodulin, is then locally[92-96] "stored" (in hippocampus, for about 1 minute) on dodecameric CamKinase by phosphorylation of a graded number of subunits; when sufficient subunits have been so phosphorylated, autophosphorylation of the entire dodecamer then ensues, which triggers a unitary strength change[35,92-97]. Obviously small (and possibly experimentally undetectable) fractions of the Ca, calmodulin or kinase (but see Lee et al.[98]) could diffuse to neighboring spines over these time windows (although the fully-phosphorylated form is likely to be firmly anchored). Thus the synaptic adjustments triggered by many variably-delayed spike pairs are expressed in "batch" mode, rather than completely "online", presumably because single AMPARs cannot be reliably delivered to the synapse, nor can spine-head calcium signals be noise-free. The physical arrangement ("towers"[95]) of dodecamers is clearly designed to "trap" as much incoming calcium as possible, but it cannot be 100% efficient. The current paper shows that 99% efficiency is often not enough.

These principles are likely to apply, with modifications, to any physical device that both stores and transmits information, and the approach of our paper boils down to asserting that learning must obey the laws of physics.

**M**   It is useful to spell out why spike coincidence detection leads, in a simple scenario, to a multiplicative rate-based Hebb rule. Assume spikes occur randomly (Poisson-distributed) at mean rates x (presynaptic) and y (postsynaptic). Assuming that a small standard increase in strength occurs whenever pre- and post- spikes occur in some fixed window (e.g. 10 msec), the rate of increase in synaptic strength is proportional to xy. Specifically, if NMDARs open for 10 msec, and always lead to just enough calcium influx to trigger an increase, the multiplicative Hebb rule is obeyed, and should be crosstalk-immune (because any extra Ca arriving from other synapses becomes irrelevent if a coincidence does occur, and any calcium arriving when there is no coincidence cannot trigger strengthening). However, a variety of deviations from this picture make crosstalk inevitable, particularly, as noted above, the time-dependence of the update rule. In particular, calcium that diffuses late from other synapses can add to small local calcium signals to increase the strengthening probability, interfering with the synapse specificity of STDP. We use an "error matrix" to represent, in the simplest possible form, diffusional coupling between synapses. A more sophisticated model could use weight- and time dependent errors, but these would merely add extra amounts of crosstalk and would not ameliorate the core problem studied here. One must remember that the bAP also triggers calcium increases in (presumably all) spine heads, and the specific coincidence-triggered increases ride on top of this; since it affects all synapses equally, it is presumably cancelled out by renormalisation, but it would make it very difficult to use a nonlinear calcium-ltp relation to discriminate against stray calcium.

**N**   The simple connectionist model we use allows negative weights and firing rates, and these are presumably implemented biologically by using combinations of "on" and "off", and excitatory and inhibitory, neurons. If sources are subGaussian inputs, then a Hebbian rule should be used, or equivalently one could simply invert the sign of all input neurons; the outcome would be that a stable IC would be learned, but would have reversed sign. It thus appears that using separate on and off neurons to represent the positive and negative signals means that either Hebbian or antiHebbian learning can be done, irrespective of the source distribution. Thus in the brain provided inputs do not keep shifting from sub to super and back, stable learning should be possible using fixed wetware.

**O** Averaging the weight changes over all the input patterns we have $<\mathbf{x}f(y)>$; expanding the nonlinearity around zero gives $<\mathbf{x}(k_0+k_1y+k_2y^2+k^3y^3+k_4y^4....)>$. The first term just produces endless weight fluctuations; the second term corresponds to the linear Hebb rule[99]; the third term is zero because the sources are symmetric; the remaining terms express all the higher order moments; these moments reflect the crosscumulants plus terms which are combinations of covariances[B]; the covariances become negligible if the sources are sufficiently nonGaussian and the inputs are sufficiently white. Note that even when the nonlinearity is a pure cubic, the rule is still covariance-sensitive (i.e. for nonwhite input).

**P** In our model imperfect whitening is necessary to reveal the effect of crosstalk. We usually achieved this by using a limited batch number (typically 1000), so the small sample covariance matrix $\mathbf{C_S}$ deviates from the true $\mathbf{C}$ (and from its better estimate $\mathbf{C_L}$). This means that the calculated $\mathbf{M_O}$ or $\mathbf{M_O}^{-1}$, is not exactly orthogonal (see Fig. 1c) and the input vectors are not completely white. However, if the input vectors are insufficiently white, the 1-unit rule converges to the least PC, not the IC. Crosstalk essentially acts to make the vectors appear less white than they are, preventing convergence to the IC, as though pairwise correlated "virtual" noise were added to them. The amount of this added virtual noise depends in turn on how white the vectors were to begin with. Note that to calculate the eigenvector of $\mathbf{EC}$ (Fig. 2a black theoretical curve) we used a much larger batch (100,000) of the approximately whitened input vectors to calculate $\mathbf{C_S}$. In our paper on the effect of crosstalk on linear Hebbian PCA learning[20] we show that weights converge to the leading eigenvector of $\mathbf{EC}$, but for antiHebb learning we expect convergence to the least eigenvector of $\mathbf{EC}$. Although in general it is not possible to analytically calculate how much the learned eigenvector moves away from the PC as error increases, we provided bounds in a number of interesting cases, and generally observe a sigmoidal relation between the cosine of the angle between the learned eigenvector and the PC; except in pathological cases or very close to trivial error there are no bifurcations.

We found empirically that using brute force normalisation and a small learning rate (<= 0.02) an erroneous linear Hebb rule converged to the leading eigenvector of $\mathbf{EC}$, while the linear antiHebb rule converged to the least eigenvector, for all tested source distributions. Furthermore, the erroneous nonlinear rule also converged to the leading (or, for antiHebb, least) eigenvector of $\mathbf{EC}$ when all the sources are Gaussian. If one or more source was Laplacian, the erroneous nonlinear Hebb rule converged to the leading eigenvector of $\mathbf{EC}$ above the error threshold, while the nonlinear antiHebb rule converged to the least eigenvector. We previously showed that with Oja-style normalisation the erroneous linear Hebb rule converges generically to the leading eigenvector of $\mathbf{EC}$ for sufficiently small learning rates [20]. Convergence of the erroneous antiHebb rule (for explicit or implicit normalization) will be analysed elsewhere. In the paper we avoid the important question of the biological implementation of normalisation by assuming explicit normalisation.

**Q** It seems likely that sensory signals presented to the cortex by thalamic relays cells have indeed been efficiently though not perfectly decorrelated[100,101]. The imperfection reflects at least 3 biological difficulties: the learning rate has not been completely annealed, so decorrelating weight vectors continue to fluctuate (equivalent to a decreased batch number); the learning rule is subject to crosstalk[102,103,20,39]; sensory information has been greatly compressed, throwing away all minor components (e.g. in retina the compression factor may approach 100). This suggests that an alternative to proofreading a nonlinear learning rule (as suggested here) would be instead to proofread the preprocessing learning. This would require proofreading circuits to be distributed all over the CNS; it is more efficient and evolutionarily feasible to consolidate them centrally, as in thalamus and the input layers of neocortex.

**R** If $\mathbf{M_O}$ is orthogonal then $\mathbf{M_O}^{-1} = \mathbf{M_O}^T$, i.e. swapped columns and rows. Thus a row of $\mathbf{M_O}^{-1}$ is a column of $\mathbf{M_O}$, and the IC just corresponds to the mixing coefficients from a source (e.g. the red coefficients in Fig. 1a top row). If the weights converge to exactly match these coefficients, the neuron's output becomes insensitive to the other sources, which can no longer make contributions to the average change in the weight vector; furthermore, changes in the weights due to the matched source simply drives them in the direction they are already in; thus this IC is a stable equilibrium of the rule. The IC remains an equilibrium even when $\mathbf{M_O}$ is not orthogonal, but now any small deviation of the weights from the IC allows the averaged weight changes caused by the other sources to increasingly drive the weight further from the IC, which is no longer stable. However, the structural stability of the hyperbolic fixed point (row of $\mathbf{M_O}^{-1}$) means there will be a small zone where the rule remains stable, as we observe.

**S** There are 2 extreme cases of our approach[32]. In the case studied here, we assume that crosstalk affects the weight adjustments themselves ("error last"); thus the matrix $\mathbf{E}$ premultiplies the vector of weight changes $\mathbf{x}f(y)$ to generate a corrupted set of weight changes. This could correspond to a few newly added perisynaptic AMPARs escaping to nearby synapses. The other extreme case would be "error first" i.e. delta $\mathbf{w} = \mathbf{x}f(E\mathbf{x}^T\mathbf{w})$: the primary triggering molecule calcium would diffuse to other synapses, where it would trigger small weight changes or, more likely and equivalently, lower thresholds for larger, discrete weight changes induced by activity at those synapses. If the nonlinearity is generated by steps (such as spike-generation) that precede calcium diffusion, and intraspine mechanisms then "count" the total number of calcium-pulses, the overall effect is error-last, even though the underlying cause is calcium, not AMPAR, diffusion. There are also many intermediate possibilities, corresponding to intersynapse diffusion of intermediate signals (e.g. Ca-calmodulin, Ras etc.[104]). While we think it likely that calcium diffusion is the worst culprit (because the calcium signal could trigger ltp-induced spine neck 'tightening'[62], choking off diffusion of downstream intermediates), all these possibilities may occur. More generally it is impossible that any natural process can be completely accurate.  We find that both extreme cases produce very similar results, so here we focus on the case we studied in detail, error last.

T   The quality factor Q in the error matrix reflects 2 different factors: the number of inputs n and a "per connection" quality q ($<=1$) that reflects parameters such as spine neck and dendritic shaft messenger attenuation factors, dendritic length and the average number of synapses that making up a connection "weight"[20,]. These factors are all interconnected, implying that q cannot approach 1 extremely closely, because improving one harms another. The best solution is to optimize all factors simultaneously, without making any perfect. Thus if a strong connection were only made of 1 synapse that synapse would have to have a high conductance, which would entail a low spine neck resistance and impaired compartmentation. Conversely, if a strong synapse is made up of many weak synapses, the synapse density increases. If density is decreased by lengthening dendrites, synapses need to be stronger to counter cable attenuation.  We considered 2 possible relations between q and $Q^{20}$. In one model the weights adjust discretely and we get $Q = q^n$. This is similar to the case for DNA replication, where q is the probability of correct copying of a single base, and $q^n$ the probability that a whole sequence is correctly copied. Alternatively, if the weights adjust continuously, one expects $Q = 1/(1+nb)$ where b is a "per connection" error rate. In Fig. 2 we use the "per connection" error rate b based on the analog model. It is useful to define a "trivial" $b_t$ value for which update specificity is completely lost and no effective learning is possible. For n = 3 as in Fig. 3 this occurs at Q=1/3 and $b_t$ = 0.66. The observed threshold $b_c$ = 0.0425 thus corresponds to a situation where crosstalk is only 6%  of the completely unselective value (i.e. 94% accuracy).

However, there are good reasons for thinking that the situation is much worse. For the **M** used in Fig. 2, the whitening (using a batch of 1000 vectors) was probably much better than could be achieved biologically because biological decorrelation filters must be learned using rules which are also subject to crosstalk. Furthermore, although 1000 is a relatively small sample, it seems unlikely that a biological online stochastic gradient ascent rule with a reasonable learning rate could produce whitening filters that are as good as used in Fig. 2. A much more self-consistent approach would be to take the $\mathbf{C_s}$ determined using 1000 sample vectors, and then further degrade it by premultiplying it by an error matrix corresponding to b = 0.0425. The whitening now gets much worse; in fact the rule will no longer converge even without crosstalk! We then progressively lowered the b value (used both for the inaccurate PCA preprocessing and the inaccurate 1 unit rule) until stable IC learning was just achieved.  We define this as a "selfconsistent" $b_c$ value. For the **M** in Fig. 2 the selfconsistent error threshold was only 0.015, about 2% of trivial.

In both models Q approaches 1 at low n and 0 at large n (since q is expected to be close to 1 since messengers are well compartmented). These models capture the vital point that significant crosstalk can occur even with excellent synapse compartmentation if synapse density is very high. It is currently impossible to estimate q precisely from either models or experimental data. The calcium attenuation between neighboring thin spines seems to approach  $0.01^{51,}$ but could be much worse in thicker spines [54.] In many cases the synapse density (e.g. 16 μm$^{-1}$ in Purkinje cells[67]) is much greater than the length constant for calcium spread along the shaft (~ 4 μm[57]). This would yield a Q value worse than 0.64. Interestingly Purkinje cell spines spiral around the dendritic shaft, perhaps in an effort to maximise the inter-insertion distance and minimize calcium coupling[105].

It's also important that our use of a fixed known error matrix **E** to represent crosstalk radically simplifies the true situation. While this is a reasonable "mean-field" approximation for an initial study, further exploration is required. The key point is that the true error matrix $E$ which describes the "instantaneous" distribution of errors that reflect the current geometrical synaptic configurations and proximity relations is unknown and ever-changing. We assume[20] that on average the distribution of errors conforms to **E**: **E** = <*E*>. But this does not imply that one can use the known **E** to "unscramble" the effect of errors. There are 2 reasons why such a strategy is doomed. First, consider the linear learning case, which finds an eigenvector of **EC**. Writing **EC** = **FCF** with **F** = **E$^{1/2}$** (for a justification see[20]), if one could first postmultiply the input data by **F**, it appears one could overcome crosstalk. However, this is not feasible: one would have to use $E^{-½}$ instead, which is unknowable. Second, for the nonlinear rule this preprocessing, even if it could work in the linear case, would introduce new hocs, so one would learn **(FM)$^{-1}$**, which is no good. See also note AA.

Although Q can be used as an index of weight adjustment accuracy, it is perhaps better to regard the trivial value that Q takes when all weights change equally in response to any pattern (e.g. because calcium spreads throughout the dendritic tree) as "completely inaccurate". As Q decreases beyond this trivial value, accuracy would increase again, although this situation is unbiological. It corresponds to the fact that for per-base error < 0.5, copying of binary strings generates erroneous direct copies, while for error > 0.5 it generates erroneous complementary copies (and exact complements for error = 1). Therefore when we refer to "accuracy" or "inaccuracy" we mean "relative to the trivial value".

**U**  The Laplace distribution decays exponentially from the peak probability density at zero for both positive and negative values; the decay constant equals the variance. The logistic distribution has a cdf given by the function 1/(1+exp-x) where x is normalized with respect to the variance. All sources were set to have unit variance; the factor of 2 in the covariance matrix used for prewhitening (Fig. 1c) matches the range (-1 to +1) of the tanh function[76].

**V** Although the averaged weight vector moves from the IC to the least eigenvector of **EC** with only a slight increase in error, the weights fluctuate, making it difficult to trace the detailed shape of the curve in the transition region. Although the weights show a large change over a small error range near threshold, they move quite slowly and noisily to their new equilibrium values, both for nonGauss and all Gauss sources.   See Supplementary Legend to Fig. 2

**W** Although Gaussian variables do exhibit higher-order moments, these moments are completely defined by the covariance matrix; in this sense we say that Gauss variables have no hocs, but technically we mean they have no high-order cumulants. As noted above[B], a nonlinear rule is sensitive to higher-order moments but for Gaussian inputs these terms are nevertheless all covariance-driven.

**X** We found that if $C_O$ is prepared using large batch numbers (~100,000), even high levels of crosstalk (fairly close to the trivial value) often do not prevent IC learning (depending on the particular $M$ and input distribution used). Conversely, if $M_O$ is insufficiently orthogonal, the error-free rule converges to the appropriate PC, rather than an IC. This is because whitening does roughly "half" of ICA[79], since if $M_O$ is orthogonal only a rotation (i.e. orthogonal) matrix has to be learned, which has half as many degrees of freedom as a random matrix. This significantly simplifies learning. This means that the effects of crosstalk we describe are only significant in a narrow range of conditions. However, since the cortex must learn from hocs, and since biological preprocessing is necessarily imperfect, this range of conditions is the one of most biological interest.

**Y** Bell and Sejnowski[115] reported that the "IC"s of natural image patches resemble oriented edge-detectors, raising the possibility that such receptive fields (RFs) are formed by an ICA-like learning process. In parallel, Olshausen and Field[159] proposed a related, but more complex, learning rule for learning overcomplete sparse-coding filters, which are also typically edge-detectors. Their rule reduces to the Bell-Sejnowski rule in the complete case. However, the issue of to what extent simple cell RFs are established by learning from natural images is controversial, with much evidence that the basic form emerges before eye opening, with subsequent refinement. Both ICA and sparse coding learning rules are sensitive to hocs. Our arguments are independent of whether visual cortex uses such rules, and we do not use natural images as input. However, imagining that the early visual system learns edge filters by using ICA-like rules is a useful way of making hoc-learning more concrete.

Although ICA of natural images finds edges and the first neurons in visual cortex also find edges, there is no proof this is more than coincidental. Indeed, a recent study [106] found that using ICA-based filters provided only weak improvements in representational power over PCA or even randomly-generated filters, despite the fact that the ICA filters are tailored to represent hocs. The authors conclude that, since hocs are vital, perhaps one learns nonlinear filters. However, simple cells are basically linear[107-109]. We draw a different conclusion: because ICA only provides a marginal representational improvement (judged using "multi-information") over PCA in the case of natural images, learning ICA filters might be quite difficult for the cortex. This would in turn make it very crosstalk-sensitive, and would demand good proofreading. In that study, ICA filters were learned by a rather sophisticated 3-step process (prewhitening, batch-based fast ICA using deflation to force different neurons to find different ICs, and finally a search in the space of possible orthogonal matrices). Cortex presumably learns ICA filters by a slightly different, but equally sophisticated process: proofreading. The key point in the study is that ICA did provide significantly better, though slight, improvement in representational efficiency as judged by a multi-information criterion. Provided that each pass through the cortex provides slight improvements, nonlinear learning of linear representations could be useful, especially if it can be done using pure feedforward online learning, and if it can achieve results that are not easy done with simple specialised hardware.

The result that ICA, which exploits hocs, only provides marginal improvements in capturing the regularities of natural scenes, is in line with other theoretical work that suggests that connectionist networks have great difficulty in discovering the higher-order relations that are key to solving interesting problems[73,75,110,111,112], basically because the needle of high-order relations is swamped in the haystack of low-order relations. Swamping also underlies the error catastrophe identified in this paper (since it can be avoided by removing all socs). In principle a hierarchy of nets can learn progressively higher-order relations, over increasingly larger patches, because the recoding at each layer reduces swamping in the next layer, but the remaining implicit hocs require more samples, so learning gets progressively slower, and stops when the animal, or his brain cells, dies. The key to solving interesting problems is thus reducing crosstalk, by proofreading, to a level where the necessary high-order relations can, eventually, be discovered (given that a sufficiently large number of examples is available). This is why we suggest that the key to the neocortex's ability to learn solutions to difficult problems (i.e. and e.g. "mind") is its proofreading ability, which enables a massively hierarchical approach. It is interesting to note that a useful hierarchy of linear rerepresentations requires the introduction of nonlinear transformations between each level, followed by re-whitening (to remove socs that reappear as a result of the nonlinearities). Perhaps the required 3-layer structure at each level (rewhitening; linear transformation based on proofread hoc-learning; nonlinear transform;) corresponds to the canonical 3 layers of cortical processing (2/3 to 5; 5 to 4; 4 to 2/3). It is not clear whether the nonlinearity should be point-wise, or would need nonlinear mixing (for example, a "max" function [110]). Either way, this fits nicely with the idea[4,114] that the main route for information flow between cortical areas is via thalamus, which is essential for proofreading, and with other suggestions about cortical learning[110]. Direct intercortical (and intracortical recurrent) connection strengths would be set by soc learning, which does not require proofreading.

The other side of the coin is that while ICA of natural image patches only provides marginally improved representations, it does at least work, despite the fact that natural scenes are not generated by linear mixing of independent sources. This seems to arise because the patches are small and rather coarsely digitized, so the number of possible hocs is limited. Indeed, one might conjecture that as long as patches are small and coarse one can always find "ICs"[115] no matter how complex the underlying generative model. Obviously, to completely (albeit implicitly) represent the input, one might need very many such IC-like units, and one has to keep track of which patch they represent (which boils down to between-area topographic mapping, presumably done in the subplate[5]). An expanding hierarchy of such patchy representations might in principle assemble a complete "recognition" model that would allow "objects" and even deeper "causes" and "explanations" to be recognised. Real-world "smoothness" would allow the hierarchy not to expand indefinitely. Such a hierarchical model would appear to be accomplished solely by feedforward learning, and not to require accompanying learning of internal "generative" models. However, this is not quite true, since it does require accurate proofreading, and updating the proofreading circuitry is tantamount to learning a simplified generative model[AA].

The ultimate success of such an "expanding patch" feedforward approach obviously hinges on the assumption that the world itself has a "hierarchical" or "modular" structure[28-30], (local smoothness of the joint distribution), so that "bottlenecks" (representational compression, reduced numbers of neurons) does not discard significant hocs. This assumption may be necessary for both "nature" and "nurture" approaches. In the primate visual system, bottlenecks occur above striate cortex.

**Z** These nonlinear learning rules can be given a broader interpretation when the underlying generative model deviates from simple square linear mixing: they still tend to make the outputs as independent as possible (although, in these more complex, biologically realistic situations, no longer completely independent). Thus one can view these rules as a powerful general strategy for finding useful provisional representations of complex input data. Subsequent stages of learning could further increase independence, though it would require additional data. Our results suggest that such a general hierarchical strategy would fail in the presence of limited crosstalk. In the case of 1-unit learning, nonlinear rules tend (assuming adequate prewhitening) to make the output more nonGaussian (by maximizing a measure of negentropy[79]). This is also a useful goal, since nonGaussian often means "interesting". This strategy underlies the notion of "projection pursuit", which provides a theoretical basis for the BCM family of rules[116], which would also presumably fail if there were too much crosstalk.

**AA** As noted above, much current work in neurally-inspired unsupervised machine learning focuses on devising algorithms that can learn from inputs created by more complicated, and powerful, generative models than that assumed for ICA. Typically these approaches involve additional "top-down" feedback to guide feedforward learning, as well as stochastic neurons and recurrent connections. A typical example is the Helmholtz machine[74] which uses layers of stochastic neurons with feedforward and feedback connections between layers, sometimes supplemented by within-layer connections. The principle involved is to try to learn, during a "wake" phase an internal, multilayer generative model whose output exactly matches the hocs present in the input layers. This can then be used to learn, during a "sleep" phase, an appropriate recognition model that recovers the sources. In the wake phase the recognition weights are fixed, and supply input to the top layer, which fires stochastically; if the recognition weights recover the sources, the top layer recovers the sources (on average) and can be used to learn the generative weights using a Hebb-like "delta" rule, which minimizes the mismatch between the generative "prediction" of the inputs, and the actual inputs. Conversely, during the sleep phase, the top neurons are spontaneously active, "dreaming" pseudo-inputs via the now fixed generative weights. These fantasized input activities would, if the generative model were correct, have exactly the same joint distribution as real inputs. Recognition weights can now be learned during sleep using a delta rule, by comparing the internal "sources", the top activities, with the predicted output of the recognition model. Note that while this learning is stochastic/delta, it achieves the same result (more slowly!) than deterministic ICA-type learning would achieve: good recognition weights.

Indeed, one could either view stochastic binary learning as an approximation to continuous deterministic learning, or the biological implementation of continuous deterministic learning as being stochastic binary (i.e. slow stochastic multiplication by coincidence detection). If these "sleep/wake" phases are alternated, the network eventually converges on correct multilayer recognition and generative models. The generative model is not used to drive behavior, but it is necessary for learning the recognition weights that do drive behavior. It seems likely that crosstalk could prevent learning both the recognition and generative model. Consider the case where (1) inputs are generated by linear square mixing (2) the machine has learned a correct internal generative model (3) the machine tries to learn the recognition model using fantasized vectors generated by topdown stochastic activity. Now, the stochastic/delta learning of recognition weights will gradually learn the correct recognition weights (it is just inverting the internal generative model, a la ICA), and this recognition learning is equivalent to deterministic feedforward learning using a nonlinear Hebb rule (i.e. the averaged behavior is the same in both cases; indeed, in biologically implementing a deterministic multiplicative Hebbian scheme, one actually resorts to a stochastic coincidence scheme[M]). Since the latter cannot work using a sufficiently inspecific rule, neither could the former. So even in the most favorable case the Helmholtz machine with crosstalk might not work, unless it uses proofreading. This would be true *a fortiori* for more complex external generative models, and in the multilayer case.

There is an intriguing connection between this approach and "proofreading". One of the reasons why distributed proofreading works is that a layer 6 neuron does not have to monitor coincidences across all possible connections onto its layer 4 partner, but only the subset of currently existing connections. However, as learning proceeds, some existing connections will disappear (because acausal post-pre activity drives all the comprising synapses to silence, which then have only a limited lifetime) and other incipient connections are made actual (because new, silent synapses that happen to arise at empty touchpoints become strengthened by "causal" pre-post co-activity). This rewiring process requires that proofreading neurons also be continuously re-wired so they continue to monitor and regulate the appropriate set of connections. We have sketched the sleep-like processes that could accomplish this elsewhere[81,82]; a key step (accomplished by internal "reverse correlation") is that the layer 6 cell must learn to contact the set of relays that do not currently feedforward to the layer 4 partner of the layer 6 cell. However, in essence this means that to appropriately update proofreading circuitry, a type of "internal generative model" must be learned (the transpose of the current "recognition" connection matrix). Fortunately this type of learning does not require proofreading since it can be done using only socs (though the transposition will be slightly inaccurate, and proofreading efficiency slightly lowered). Since only the internal recognition model requires proofreading, one must regard it as the core step. If this reasoning is correct, it would imply that a pure feedforward net equipped with *dedicated* proofreading would not need any form of internal generative model, but of course it is no longer really "connectionist" because it implicitly uses very extensive (exponential in n) multiplication operations.

**BB**

An alternative, related, view of the failure is that crosstalk introduces a form of input noise: the crosstalk causes weight changes that are slightly incorrect, and resembling those that would be caused by noisy input vectors. This crosstalk "noise" is correlated, because connections affect each other, but in ways that are, moment–to-moment, unknown (since they reflect the happenstance of current synaptic geometries). Noisy ICA[79] is more difficult that plain ICA, and in general even if the mixing matrix can be learned, the sources cannot be exactly recovered. A powerful general approach to noisy ICA is "bias removal" and requires that the inputs be "quasiwhitened" using

$(\mathbf{C}-\mathbf{N})^{-1/2}$ where $\mathbf{C}$ is the input covariance matrix and $\mathbf{N}$ is the (known or assumed) covariance matrix of the noise which has been added to the inputs. This might suggest that similar suitable quasiwhitening could combat crosstalk, but this is unlikely to work, for several reasons. First, clearly the best preprocessing is accurate whitening using $\mathbf{C}$ derived from a large sample of actual inputs: in the large sample limit this provides perfect crosstalk-protection. Second, we use $\mathbf{E}$ as a "meanfield" approximation to the true situation where an unknown and fluctuating $\boldsymbol{E}$ describes the true error distribution; true error prevention would require $\boldsymbol{E}$ to be known[20]. Third, neither "error first" nor "error last" corresponds exactly to modification of the input vectors[S], because of the nonlinearity. In error first, $\mathbf{E}$ acts on $\mathbf{x}^{\mathrm{T}}$ *inside* the nonlinearity (e.g. calcium diffusion). In error last, $\mathbf{E}$ acts on the vector of weight updates (i.e. on $\mathbf{x}$, a row vector; e.g. AMPAR diffusion). Both these cases are distinct from $\mathbf{F} = \mathbf{E^{1/2}}$ acting on both versions of $\mathbf{x}$. Obviously all 3 cases become more similar as the nonlinearity weakens. Fourth, even in the case where crosstalk is equivalent to added noise, the brain does not know how to appropriately quasiwhiten.

**CC**   Crosstalk will depend on 2 factors: the extent to which 2 spinous synapses that occur very close together on a dendrite are chemically isolated from each other, and the extent to which separation along the dendrite provides additional attenuation. The former factor can be regarded as expressing synapse biophysics (including properties of the spine neck) and the latter as expressing dendrite properties together with synapse density. We proposed a very simple picture of how these 2 sets of factors combine[39,20] resulting in the equations:

$Q = 1/(nb+1)$  or  $Q = (1-b)^n$

$b = (\lambda_c \, \alpha \, a)/ \, L$

where Q is the Quality (i.e. the diagonal entry of $\mathbf{E}$; see Fig. 1c), n is the number of inputs, and b is a "per connection" error rate (see Fig. 2 legend) that combines all factors other than n. $\lambda_c$ is the space constant for chemical attenuation along dendrites, $\alpha$ is the average number of (actual, anatomically existing) synapses made by a feedforward connection, and L is the dendritic length (or, more accurately, the total length in the dendritic zone that is targeted by the particular set of feedforward axons under consideration). The 2 alternative equations for Q reflect 2 different assumptions about the way that synapses change strength: the first assumes the strength change is analog and the second, digital. However, since typically Q is presumably close to 1, both assumptions yield similar dependencies on n and b (i.e. b must be smaller than 1/n). In the following we take the specific case that the diffusing chemical is calcium; however the diffusion of other relevant chemicals such as ras[104] will make crosstalk even worse, without otherwise affecting the main arguments.

Unfortunately even for calcium none of the 6 parameters in these equations is precisely known, and may vary between species, cortical areas, developmental status etc, and current estimates are typically order of magnitude. However, since we do not know how "difficult" the learning problems that the cortex solves are, this uncertainty essentially means that we do not know exactly when to expect biological learning to fail. A similar difficulty attends the Eigen molecular evolution problem: the strength of selection, which determines the exact threshold error value at which evolution fails, is unclear. Nevertheless, there is some direct experimental evidence that the mutational "error catastrophe" does occur [132]. In any case, on-one doubts the necessity to copy bases at accuracies approaching the reciprocal genome length.

Reasonable ball park figures for the above parameters might be 1000, 2 $\mu$m, 10, $10^{-2}$, and 2 mm, yielding a Q of 0.91, close to the threshold $Q_t$ in Figs. 1,2 (which in turn is close to our average value for Laplacian sources). However, as noted above[0], this $Q_t$ value is not "selfconsistent", so learning is probably more error-sensitive than this indicates. This would mean that if in some sense the "Laplacian unmixing problem" is in some sense "typical", learning would almost never succeed. Of course if spine neck attenuation were much better than the assumed value, learning could succeed more often even without proofreading. Three arguments mitigate against this possibility. First, there is no direct experimental evidence that attenuation exceeds 99%. Second, one must expect that substantial improvements beyond this value will produce some degree of spine head electrical isolation (spine neck resistance are already close to values impeding electrical coupling[63]). Third, there have been repeated claims that calcium isolation is much worse than 99%[50,57,54,52]; while in some of these experiments calcium "dye piggybacking" may have been a problem, in several cases explicit precautions against such an effect were taken.

Another possibility is that the strength changes are not linear in the calcium concentration, so that $a$ should be raised to a power $h^{20}$, discriminating against "stray" calcium signals. An extreme version of such a nonlinearity would be an ltp/ltd "threshold", as used in the BCM model[116]: calcium signals straying beyond an active synapse would decrease neighboring weights, apparently "sharpening" the specificity of synaptic strengthening. The problem with these ideas is that in our model the learning equation is *already* nonlinear but instead of "solving" the crosstalk problem, the nonlinearity *creates* it. For example, we find that using a cubic nonlinearity (essentially h = 3) allows ICs to be learned in the zero-crosstalk case, but learning fails at a low, threshold, crosstalk level. The problem is of course that if one postulates a nonlinear relation between calcium and strengthening in neighbors of synapses, one also has to postulate the same relation in the synapse itself: calcium is calcium, no matter where it comes from. In addition, stochastic effects in small spines will tend to linearise the response, and coincidence-specific small calcium signals will ride on larger background calcium signals due to the bAP itself, also linearising their contribution.

It seems to us that for rate-induced plasticity, the task of the synapse is already quite difficult if it is reduced to its barest, but already challenging, requirement: the detection of spike coincidences within a narrow (~10 msec) window; in this scenario the synapse's task is simply to decide whether or not a significant calcium elevation occurred; if it did, this should be registered as a "coincidence". If the synapse has to also measure the size of that signal, the task becomes intractable. This is also why we prefer the "error last" formulation of the nonlinear crosstalk model. There are several reasons why this "minimal task" is difficult. The average number of free calcium ions in an average spine is on the order of 1; since the diffusional equilibration time is on the order of 100 msec[51], then in the course of a day about 4 pseudocoincidences (defined here as spontaneous tenfold elevations in calcium) will occur (calculated assuming Poisson statistics). It could be argued that this is still much less than the expected number of true coincidences, but the relevant variable here is the net number of real coincidences (i.e. the difference in the mean numbers of pre-post and post-pre real coincidences). After learning stabilizes there will be no net real coincidences (by definition), and 4 pseudocoincidences would be a problem: weights will all ramp up. This could be avoided additional (nonSTDP) forms of normalization, since on average it will affect all weights equally. However, the real problem is that there will be considerable scatter in the daily synapse to synapse numbers of pseudocoincidence, constituting an unavoidable source of plasticity noise (which we have previously called "Type 1 errors"; these cannot be handled by proofreading and must be kept well below the expected proofreading failure rate). Pseudocoincidences can be overcome by increasing the threshold calcium level that marks a spike coincidence, but at considerable expense: the larger calcium signals required are more likely to trigger crosstalk (Type 2 errors), since they are more likely to saturate neck and shaft pumps. The problem can also be overcome by having more synapses per connection, but this again increases type 2 errors. The problem boils down to the difficulty that as learning stabilizes suspicious coincidences become extremely rare, so the background noise has to be correspondingly low. From this point of view a synapse detecting coincidences is in the same boat as a rod detecting photons. Multisynapse weights help, like pooling rod signals in starlight, but at the expense of reducing spatial resolution (in this case, the number of learnable inputs). We expect that the level of Type 1 errors is kept comparable to the level of Type 2 errors (in the presence of inevitably faulty crosstalk), since there is a diminishing evolutionary advantage to reducing them further. This may play an important role in setting the sizes of synapses: they are large enough that type 1 errors become rare compared to type 2 errors (after including crosstalk). Fluctuations will also affect the sizes of induced calcium signals, but this will probably interfere mostly with temporal aspects of plasticity.

Another way that these type 1 errors, reflecting calcium fluctuations, could be reduced is by "batching": accumulating coincidences (i.e. suprathreshold calcium signals) on a register, and using the accumulation to trigger strengthening. If the register "forgets" with a time constant much shorter than a day (e.g. 1 minute), then a large decrease in the number of pseudostrengthenings can be achieved. Batching has other advantages: it makes the learning rule less stochastic (it is typically used in Fast ICA), and is thus equivalent to a lower learning rate. It may also help decrease the effective synaptic bit resolution: if 10 coincidences have to be accumulated to trigger a unitary strength change equal to adding 10 AMPARs[35,97], it becomes equivalent to the ultimate achievable resolution: 1 AMPAR molecule added per spike coincidence. Given that biology often operates near biophysical limits (e.g. rods detect single photons), it may be that learning cannot be improved (except by social mechanisms).

It seems likely that the batching machinery is the dodecameric CaMKinase molecule: sporadic individual calcium elevations, triggered mostly by true pre-post spike coincidences, reach a calcium threshold and trigger monomer phosphorylation, which persists for about a minute. If a "batch" of coincidences occurs within that minute, the multiple phosphorylations may trigger autophosphorylation and insertion of a dollop of AMPARs.

Recent experiments have provided strong direct evidence against the view that the synapse can "read out" the magnitude of the spine head calcium signal and decide whether to trigger LTP or LTD, or make even finer-grained adjustments of the amount of strengthening [89]. These new experiments favor the "2-sensor" view of LTP/LTD[133,134]: one sensor triggers all-or-nothing LTP, and another sensor all-or nothing LTD. The overall STDP curve would be an expression of these 2 perhaps independent processes. In this view adjustments are stochastic and all or none, and the attained peak calcium level (and consequently the time integral of the calcium signal) probably adjusts the probability of the unitary increases. Decreases would be triggered differently, for example by retrograde endocannabinoid signaling[135,136]. In an STDP model, one can think of LTP as underlying the Hebbian rule, and LTD as underlying normalization (which, since LTP increases weights, must always decrease weights). The specificity requirements for LTD are far less severe than for LTP, since it is not necessary to exactly normalise the weights. Also, the error matrix for LTD would be quite different if it is accomplished by retrograde signaling[39]. Thus insofar as calcium levels determine the extent of LTP, they endow it with timing sensitivity, rather than making it nonlinear. The necessary Hebbian nonlinearity would therefore presumably reflect the nonlinear dot-product/firing relation (as implicitly assumed in the original Bell-Sejnowski formalism), rather than a nonlinear calcium/strengthening relation. In this view the plasticity machinery simply counts individual spike coincidences, as in the basic rate-Hebb view; in models that explicitly incorporate time, the counting is done using appropriate temporal weighting. Of course this has the consequence that the output neurons do not "recover the source", and spike coding is less sparse but there is no loss of information, more channel capacity is used and the same learning is done. Indeed it seems that layer 4 neurons do not code sparsely[137].

If inputs are added to a fixed dendritic tree, synapses become more crowded and crosstalk worsens (typically linearly with n[39,20]). It is not possible to overcome this difficulty by adding dendrite so as to keep synapse spacing constant, because cable attenuation gets worse, requiring either that synapses be stronger (which requires that spine necks be shortened or widened) or more numerous (which restores crowding). Thus adding dendrite is a self-defeating strategy. Dendrite can be added in 2 ways: simple elongation, or adding branches. The former leads to classical electrotonic attenuation; the latter, though requiring less elongation, increases local input resistance and pushes the synapses closer to electrical saturation. The root cause of crosstalk is the incompatibility of complete chemical isolation and complete electrical coupling. In particular, in simplified biophysical models[138] the ratio of the electrical and "calcium" space constants is less than 1000. This cannot be substantially improved by increasing pump density or efficiency once pumps are shoulder to shoulder and narrow necks limit energy supplies.

It is interesting to note that CT cells fire only rarely even in awake behaving animals[117]. Our model predicts this finding, since these cells would only fire when appropriate spike coincidences occur. Furthermore, if learning is antiHebbian, it acts to reduce the numbers of spike coincidence. Indeed, we would predict that once stable learning has been achieved spike coincidences (and thus CT firing) would mostly occur by chance (i.e. at the level predicted on the assumption that spikes occur independently), and would not be triggered by patterned input. If inputs and outputs are firing merrily (say at 10/sec), then a CT cell using a 10 msec coincidence window would only fire at 10% of the rate of a layer 4 cell. We predict that CT cells would fire more often in inexperienced animals or in higher-order thalamic nuclei. This may account for the increased burst-firing in higher-order relays[139-141])

Another approach to estimating biological-plausible levels of Hebbian inspecificity is from the crosstalk experiments themselves. Harvey and Svoboda[103] reported that induction of ltp at a single spiny synapse lowered the threshold at about 20 neighboring synapses, although it did not cause ltp or maintained spine enlargement at any of those neighbors (confirming ref 36). The experiments used 2 ltp induction protocols, a 4 msec uncaging "suprathreshold" protocol at one synapse (which "reliably" triggers ltp) and a 1 msec "subthreshold" protocol, at one of the neighbours, which by itself "reliably" fails to trigger ltp, but which becomes suprathreshold in the neighborhood of a synapse subjected to a suprathreshold protocol. We place "reliably" in quotes for 2 related reasons. First, there is considerable variation in both upscs and spine head volume indices for ltp, which means that one cannot completely reliably decide whether a given protocol never or always produces ltp. Second, other work has shown that ltp at single synapses is all-or-none and unpredictable, with stronger stimuli being more likely to trigger ltp[97,35]. Ideally, Harvey et al. should have determined the quantitative relationship (presumably a sigmoid curve) between ltp probability and uncaging duration, which could also be interpreted as a relation between average degree of ltp and uncaging duration. Initially we will ignore this problem and simply assume that because neighboring synapses, tested with the subthreshold protocol, "see" the induced ltp at a synapse subject to the suprathreshold protocol, they have undergone some form of ltp.

Thus 20 incorrect synapses would be strengthened for each correctly strengthened one out of a total of around 10,000 made by CA3 axons on a typical CA1 neuron ( in radiatum and oriens combined), giving a Q value of 0.998 if each CA3 axon makes 1 synapse per CA1 neuron. However, very likely each axon makes 3-10 synapses [147] so the Q value could drop to 0.980, comparable to the average "selfconsistent" threshold value that causes the hoc collapse.

Several factors could affect this value. First, in these experiments synapses were selected that were well isolated from the dendrite and from other synapses, which would lead to an overestimate of Q. On the other hand, the fact that single-synapse ltp only affects the threshold at neighbors, and does not itself cause frank ltp, implies that assuming 20 incorrect synapses would be an overestimate. Here the key question is the shape of the ltp-probability/uncaging curve. This curve must be quite steep, because in another paper[104] the authors found that a "reduced uncaging" protocol, with a 2 msec pulse, produced about 80% of the ltp that the the suprathreshold protocol produced  (and about 60% of the NMDA response). It seems to us that in a more realistic situation, with complex patterns of activity at many synapses, one should evaluate crosstalk in the linear region of the crosstalk curve (i.e. not make a correction for thresholding). The principle here is the same as translating synaptic input to firing activity: the fact the spike has a sharp threshold does not mean that a weak input will have no effect, one should consider the effect of that input in the context of other ongoing inputs, and to a first approximation assume a linear firing/current relation.

The similarity between the Q estimates by 2 different routes (assuming calcium is the crosstalk culprit, or using the direct crosstalk data) suggests that perhaps calcium leakage is the main cause of crosstalk. One likely scenario is that in the crosstalk experiments the suprathreshold protocol pulses each produce a small (~1%) transient (~ 1 sec) increase in calcium at neighboring spines. This leads to an accumulating, but typically still subthreshold, increase in CaMkinase phosporylation (interpulse accumulation, over a ~ 1 minute period must occur because ltp induction requires many repeated uncagings). The application of the subthreshold protocol now produces additional calcium entry (approximately ¼ of the suprathreshold entry and ½ of the 'reduced uncaging" entry) which suffices to trigger the CaMKinase switch and induce ltp. In this scenario, one predicts that use of an intermediate subthreshold duration (e.g. 1.5 msec) would by itself trigger ltp in a significant fraction of trials, and that this fraction would be greatly increased by a suprathreshold protocol applied to a neighboring spine. Understandably, Harvey and Svoboda preferred to use a cleaner, simpler protocol but this makes it more difficult to quantitatively evaluate crosstalk.

**DD**  The important issue of the timing of proofreading can be separated into 2 aspects: problems that would arise even if proofreading could be done in dedicated fashion (e.g. Fig. 3a) and additional problems that arise for distributed proofreading (e.g. Fig. 3b).
Given that a dedicated proofreading neuron could quickly and reliably detect a coincidence occurring at the connection it monitors, how could this swiftly "enable the plasticity" of that connection so that the confirmed spike coincidence can lead to synaptic strengthening? Here we simplify the discussion by assuming that the synaptic coincidence induces some local change in the synapse that prepares the synapse to express ltp, but that the expression will only occur if 2 "enabling" signals, one postsynaptic and one presynaptic, arrive sufficiently quickly. Note that the timing of the expression is not the issue, instead it is the timing of the *decision* to express. In conventional ltp models, "induction" corresponds to "decision", but here we are adding an extra step (or conjunction of steps): detected coincidences lead to induction (generally by Ca entry through unblocked NMDARs) and are then *confirmed*.
It is likely (see above) that induction is itself a series of steps: calcium entry leads to stepwise increases in phosphorylation of CamKinase dodecamers, which ultimately triggers complete switchlike autophosphorylation and, by a further chain of events, expression[92-96]. In principle it might be possible that only the final steps after the switch are confirmed (i.e. that proofreading operates in batch mode), but this would clearly weaken the underlying principle of double coincidence detection. So here we assume that each coincidence is confirmed. A simple mechanism would be that if confirmation does not arrive, there is a rapid "default" CamKinase dephosphorylation that erases the "memory" of the preceding coincidence.

An alternative scenario is that the signals arriving at the synapse, via both pre- and postsynaptic routes, "disconfirm" coincidences, for example if the default duration of the Ca-induced CaMKinase phosphorylation is long but is truncated by the arrival of "disconfirmation" (specifically, a presynaptic burst accompanied by postsynaptic mGluR activation). This scenario is not very attractive, because it means that CT spikes would be caused by lack of coincidence; since most arriving spikes presumably do not trigger layer 4 cell firing, there would be a lot of disconfirmation going on. It seems more plausible that coincidences are rare (indeed, that is the result of antiHebb learning) and that CT cells therefore fire rarely (as observed[117,118]). The implication is that most arriving TC spikes fail to directly fire most of the layer 4 cells, because they are not sufficiently synchronous[119,120] so that the cells that do spike convey a "sparse code". Recurrent amplification would complicate the simple feedforward view: indeed, proofreading can also be viewed as a mechanism for sorting out which are the relevant, direct, causal, "feedforward" spikes, as opposed to irrelevant (from the feedforward learning perspective) recurrent spikes.

The postsynaptic aspect of confirmation is relatively straightforward: CT axons collateralise heavily in layer 4 and make many "drumstick" synapses on the shafts of thalamorecipient spiny stellate cells[121]. These synapses, which resemble layer 6 CT synapses on relays, presumably activate mGluRs which are often necessary for plasticity[122, 123] and there seems no reason why this action should not be reasonably fast, and somehow prolong (or permit the retention of) the CaMKinase "memory" of a previous synaptic coincidence. For dedicated proofreading, one requires that the action be fast compared to the mean time between coincidences *at a specific firing connection*. If the mean relay and layer 4 spike rates are 10/sec, and the coincidence window is 10 msec, this intercoincidence time is 100 seconds. This gives lots of leeway for using distributed proofreading (but means that the number of current feedforward connections to a layer 4 cell not exceed ~100; this is probably partly why many connections are merely incipient).

The key to efficient proofreading is that the relay spikes themselves must carry the enabling confirmation since no other message can travel from the relay somata to the TC terminals in time. We suggest that the relay burst mode (which is controlled by CT feedback[124,125,126]) carries the signal. There are several components to the delay. First, the CT conduction time. In the case of the fastest (probably magnocellular) feedback[127-129] to LGN this takes 5 msec (compared to 4 msec for the TC path). Next, the TRN mediated hyperpolarisation must last long enough to remove enough of $I_{Ca,t}$ inactivation to allow bursting[130]. Lastly, the burst (which replaces the next tonic spike) must reach the thalamocortical terminals, and somehow affect the spine head expression machinery. Possibly the burst leads to multivesicular release which swamps subsynaptic receptors and engages specialized perisynaptic receptors[131]. The exact machinery for presynaptic enablement is not the issue here; what it important is that known processes could implement presynaptic enabling of postsynaptic expression within the necessary 100 msec window. If the machinery is inefficient, it will lower proofreading accuracy, but any level of accuracy at all buys major increases in learning power.

**EE**   Here we summarise the landmark experiments of Wang et al.[1]  and explain how they relate to Fig. 3b and Supp. Fig. 1. These experiments were a *tour de force*, and for the first time allow the pattern of corticothalamic feedback to be related to the feedforward thalamocortical pattern.

Wang et al. recorded simultaneously from a layer 6 CT neuron and several topographically corresponding relays. Because the relays and the CT cell responded to visual stimuli in the same region of visual space, the relays correspond mostly to the undotted "available" relays shown in Fig. 3b. Their results, though very clear, are not easy to follow, and their paper repays careful study (and their Supplementary Figure 3 is slightly misleading, as outlined below).

Wang et al. determined the RFs of the recorded CT neuron and relays. As expected, the relay RFs were center-surround, and the CT RFs were composed (typically) of 2 flanking on and off elongated and oriented lobes. Note that although there would be corresponding oriented layer 4 cells in the overlying column, these were not recorded from, but may contribute to the RF of the recorded layer 6 cell. However, only the layer 6 CT cell feeds back to LGN.  The relays fell into 2 classes: "overlapping" and "nonoverlapping". The RFs of the former class overlapped with the CT RF, while the latter class did not. The overlapping relays could be further divided into 2 subclasses: "matching", where the sign of the relay center coincided with the sign of the lobe with which it overlapped (so these relays might, and probably did, contribute to the CT neuron's RF), and "nonmatching", where the signs disagreed. Nonmatching cells do not contribute to the cortical RF, despite overlapping them, and very likely they are not connected to the cortical cell, even weakly or silently (see below).

Wang et al. then determined that visually-stimulated (dark or bright spots) layer 6 cell firing was enhanced by local application of a GABA-B receptor antagonist to the CT neuron). Only low drug levels, inducing minimal enhancement, was used, in an attempt to ensure that only increased firing of the recorded CT cell was affecting LGN. They asked whether this increased firing was associated with a change in the overall firing mode of the simultaneously recorded relays, either from tonic to burst (T-B, presumably mediated by disynaptic hyperpolarisation via TRN), or from burst to tonic (B-T, presumably mediated by direct monosynaptic depolarization). Any such shift would be evidence for functional feedback. The drug caused no change in the firing rate of the CT cell in the absence of visual stimulation. For nonmatching relays, the stimulating spots presumably fell in the relay surround, and therefore stimulated the center of other relays that did match, and therefore could influence the firing of the CT neuron. For matching relays the direction of the change was almost always (one exception only) to relatively more bursting (i.e. corresponding to plasticity-enabling red arrows in Fig. 3b). For nonmatching relays the increased CT firing always caused a shift to relatively more tonic (corresponding to disabling reverse red arrows going to the pink "nonmatching" off-row in Fig. 3b). In all cases the effect was on the burst/tonic ratio rather than complete switches in mode. Unfortunately in the crucial experimental figures 2a and 3b it is not initially made clear which of the 2 recorded relays in Wang et al. Fig. 2a correspond to which data in Wang et al. Fig. 3b but this can be inferred from the subsequent presentation (the left "matching" relay in 2a corresponds to the bottom (T-B) plots in 3b). It's important to realize that the B/T mode ratio of the relay in the absence of drug application is itself set, *inter alia*, by ongoing CT feedback; the drug-induced increased visual firing allows one to deduce the functional polarity of the feedback from the recorded CT cell to the recorded relay.

Thus CT feedback onto overlapping matching relays was "phase reversed", in the sense that a relay whose firing putatively contributed to the CT neuron's firing (on-center relay contributing to the on-lobe or off-center relay contributing to the off-lobe), presumably by depolarising it directly (and/or possibly via a corresponding layer 4 cell), would be shifted to more burst firing by CT firing (i.e hyperpolarized disynaptically, via TRN). However, overlapping nonmatching relays would shift to a more tonic mode, presumably by directly depolarization by that CT neuron. Note the firing of a nonmatching off-center relay is caused by central darkness so illumination of its RF center would, if it made a direct excitatory connection to that layer 6 cell, "hyperpolarize" the CT cell; in this "fictive" sense nonmatching relays also receive phase-reversed CT feedback. Thus the feedback, instead of amplifying the RF of the CT and presumably the corresponding layer 4 cell, tends to annul it (except of course that CT firing does not actually drive spikes in TC relays – it is modulatory). This is unexpected and, as the authors note, counterintuitive, since previous papers (including some from the Sillito lab[142]) had suggested that cortical feedback might amplify feedforward responses.
Wang et al. suggest that this effect could contribute to the phase reversing spatiotemporal RF properties of relays; however, this is already seen in retinal ganglion cells, which do not get CT feedback.

Note that all overlapping relays significantly influence the CT firing mode, half one way (matching, T-B, phase reversed) and half the other (nonmatching, B-T, fictively phase-reversed).

It was also found that about half the nonoverlapping relays' firing mode were significantly affected by CT visually-induced firing, but in this case it is obviously not meaningful to separate them into matching and nonmatching subclasses. These influences were equally divided between B-T and T-B. Note that nonoverlapping relays either have their RF centers on a line drawn from the CT RF center that is nonparallel to that CT RF, or are parallel to it but beyond the ends of the RF (see Wang et al. Fig. 6). Since these "parallel but nonoverlapping" relays must be further topographically from the center of the relevant visual patch, they would be more likely to correspond to "unavailable", dotted relays in our Fig. 3b (see below for further discussion of this point), and we would not expect that they receive feedback.

We now explain how these results map onto proofreading (Fig. 3b, Supp. Fig. 1).

First, one must realize that the simplified schemes we show (Fig. 3b and Supp. Fig. 1), if interpreted as applying to lgn and striate cortex, only indicate off-center relays (there is an equal population of on-center relays, not shown). The reader should examine Wang et al.'s Supp. Fig. 1, and, with caution, Supp. Fig. 3, which shows the 2 sets of off and on relays, which each tile visual space. The green relays correspond to the "matching" set which drive the off-lobe of the CT neuron (show as a green bar inside the layer 4 and 6 cells). Thus a dark bar centered on the green relays causes these relays, and hence also the CT cell, to fire. Spot darkening over a single green relay would also fire the CT cell, but less strongly. The pink row of off-relays above this overlaps with the on-lobe of the CT neuron. The on-lobe (represented by a yellow bar inside the layer 4 and 6 cells) is driven by "yellow" on-relays, which however are not shown. These pink off-relays do not drive the CT cell, and probably do not connect to it even by silent synapses[N], though of course they do contribute to the off-lobes of other layer 4 and 6 cells (not shown). We interpret this non-connection as the outcome of the fact that if the CT cell is consistently driven by light falling on the set of "on" relays, these off-relays will be silent, and this anticorrelation with the CT cell, will, for Hebbian learning, lead to disconnection. (For antiHebbian learning, the argument is reversed, but has the same outcome[N]) Proofreading requires that all connected relays be enableable (red arrows Fig. 3b; black TRN arrows in Supp. Fig. 1) and all unconnected (but available) relays be disenableable (reverse red arrows Fig. 3b and Supp. Fig. 1). This exactly matches the Wang et al. results for overlapping relays, since the green off relays are presumably connected and the pink off relays are not.

Consider now the bottom row of (nonoverlapping) relays. These relays do not contribute to the RF of the CT cell, but they could nevertheless be connected by silent or very weak synapses. The firing of these cells is, on average, neither correlated not anticorrelated with the CT firing, so one expects that silent synapses, if they form, are neither strengthened nor quickly eliminated[N]. An example of a nonoverlapping relay that has formed a silent connection (open dot) is shown in blue. Since this is an existing connection, it must also be enableable (red arrow Fig. 3b). However, many available nonoverlapping relays happen to be unconnected, so they are disenableable (reverse red arrows Fig. 3b).

The data in Wang et al. suggest that about half of the available relays that do not drive the CT cell (either because they are not connected or connected only silently) are silently connected. This is in line with estimates of the expected fraction of silent connections (see[16, 143]; Supplementary Fig. legend).

Identical logic applies to all the on-center relays (not shown). Indeed, it is possible that these use the same CT partner as the off-relays, since on and off relays corresponding to the same point in visual space never fire at the same time. This halves the number of required CT neurons, but requires that early in development a given CT cell partner with a pair of layer 4 cells.

In summary, the proofreading hypothesis explains the observations in the 4 classes of relays: overlapping and matching (connected, T-B); overlapping and nonmatching (nonconnected , B-T); non-overlapping and receiving feedback (some would be silently connected and would get T-B feedback, and some would not be connected but are available and would get B-T feedback, explaining the mixture of feedback effects seen); non-overlapping and non-available (no feedback necessary or observed).

The Wang et al. Supp. Fig. 3 shows the feedback circuitry that the authors deduce from their observations. It shows the full sets of on and off relays tiling a patch of visual space. They suggest that the RF properties of the shown layer 6 cell reflect the feedforward input it receives from relays, although current data do not yet support this inference (i.e. the RF properties could be entirely derived from layer 4 input). They propose that the pattern of feedback innervation is largely a mirror (phase-reversed) version of the presumed feedforward innervation: on relays that feedfoward (shown in red) receive T-B (i.e. phase reversed) feedback via TRN. The feedforward connections of the off-relays that (hypothetically) feedforward to generate the off-lobes of the layer 6 cell are not shown and the proposed feedback arrangements to these relays is somewhat ambiguous. The experimental results suggest that these neurons are also "matching" and therefore receive T-B input. However this is not shown in the diagram: indeed, it is not clear whether the putative offlobe lies to the right or left of the on-lobe. The simplest interpretation of the figure is that this layer 6 has an isolated on-lobe (like the cell shown in their Fig. 2d), and does not receive functional driving input from off relays.
The figure shows, in blue, the set of "nonmatching" overlapping off relays as getting direct feedback (i.e B-T). This is also phase-reversed in the sense that activation of these relays functionally hyperpolarizes (or at least, fails to depolarize) the layer 6 cell.

The functional connections discussed so far agree perfectly with those diagrammed in our Fig. 3b.

This Wang et al. Supp. Fig. 3 also shows feedback to a set of nonoverlapping cells (indeed, since it is implied that the layer 6 cell only has an isolated on-lobe, all overlapping relays lie along a single short line, corresponding to the red (matching) and blue (nonmatching) relays). The authors illustrate the situation they infer from their data: that almost all nonoverlapping relays that receive feedback lie orthogonal to the set of overlapping relays (see Fig. 6 in Wang et al.). However they show these orthogonal" relays as all receiving both B-T and T-B feedback. This is not correct: about half the nonoverlapping cells that get feedback get functional B-T feedback, and the other half get functional T-B feedback (as required for proofreading). It remains an interesting point whether many of the nonoverlapping relays that do not appear to get feedback may get balanced T-B and B-T feedback, thus appearing to get neither.
However, the result that mostly orthogonal nonoverlapping cells get feedback (of either sign) is not completely in accord with proofreading: in the simplest case we would expect that the determinant of whether nonoverlapping cells get feedback would be whether or not they make silent (or weak) connections or not, rather than their visual location. However, it is quite possible that these 2 factors are interrelated. It's also possible that the result in Fig. 6 reflects a sampling bias, since relays were selected based on whether their RFs either completely overlapped or did not overlap at all the recorded layer 6 RF. A more likely explanation of their Fig. 6 results, which would be completely consistent with proofreading, is that, as noted above, "parallel" nonoverlapping relays are likely to correspond to visual locations that are quite displaced relative to the CT RF center, and therefore to have intracortical axon arbors that do not quite reach the CT (or 4) cell dendrites – they would be "unavailable" and therefore would not get feedback. If the population of parallel nonoverlapping relays is more likely not to get feedback, this means that the population of nonparallel nonoverlapping cells will be relatively more likely to get feedback. (In Wang et al. Fig. 6, this expected reduction in feedback to nonoverlapping parallel relays is concealed by the fact that their "green" population also includes all the overlapping relays, all of which get feedback; this inclusion of nonoverlapping parallel relays may account for the observation that the "orthogonal" class is slightly more likely to get feedback than the parallel class). In this account, the lack of feedback to the nonparallel nonorthogonal relays would arise because their RF centers were mostly well-displaced form the CT RF center, and thus more likely to be "unavailable".

Nevertheless, the general point made in Wang Supp. Fig. 3, strongly supported by their data, is that many nonoverlapping cells receive feedback, and this can be of either sign. This is in agreement with the proofreading hypothesis.

Finally, we outline differences between the Hebb and antiHebb cases. In the Hebb case, the feedforward connections from relays to 6 would reflect the frequent occurrence of bright/dark edges of a particular orientation. The correlated firing of the corresponding relays would strengthen the appropriate feedforward connections; initial random biases would be amplified   to determine the final orientation to which the layer 6 (or its corresponding layer 4) cell were tuned (in the salt-and-pepper case; additional lateral interactions could also play a role in biassing tuning), as in the original von der Malsburg model. "nonmatching" relays would develop increasingly anticorrelated firing, and any connections they made would be eliminated. Nonoverlapping relays would tend to fire rather independently of layer 6 (or 4) cells, so any connection they happen to make would remain weak or silent. In the antiHebb case, the correlated firing would eliminate any tentative connections, and the anticorrelated firing would tend to strengthen them; the outcome would again be bilobed RFs but the outcome of the initial biases would be the contrary selection. Thus unless one was to follow the fate of individual layer 4 or 6 cells throughout the learning process, the general outcome is independent of the type of rule (see also Note N). However, in the antiHebb case there is a gradual overall decrease in the number of spike coincidences, CT cells will fire increasingly rarely, and relays will spend more and more time in tonic mode. These properties are highly desirable.

**FF**   In erroneous PCA learning the mutual information between input and output is (in the optimal, Gaussian, case) proportional to the cosine of the angle between the correct and erroneous learned vectors[20], which for weakly structured input, large n and very low error is proportional to reciprocal error (equation 4.3[20]).   The graceful failure seen with linear learning and catastrophic failure with nonlinear learning is similar to the graceful deterioration of linear associative memory nets as the number of loaded patterns increase, contrasted with the catastrophic failure seen with nonlinear nets when loading exceed a critical value.

**GG**    An alternative view of mind to connectionism is that evolution does the learning, and provides useful "modules", such as generative grammar[28,29]. One could view our demonstration of severe limits to the power of biological "connectoplasm"[28] as favoring this view. However, such genetic learning is subject to the same conceptual limitation as connectionist learning: too much error prevents both types of learning. In particular, even with proofreading, Darwinian evolution is restricted to small genomes, much of which is used to specify nonneural (or at least noncortical) machinery. Animals (especially humans) can do complex neural learning, and they could do it very effectively if they are endowed with proofreading. Since neocortex has circuits that could do proofreading, the onus is now placed on "innatists" to show that despite appearances this circuitry does not do proofreading. If the neocortex can proofread, objections to "connectoplasm" largely evaporate.

**HH**    Specifically, the next great transition, sex, boils down to a protocol for the exchange of information between genomes; language is a protocol for the exchange of information between brains. Penn et al.[144] has recently summarised evidence for a cognitive chasm between humans and other animals, especially for "higher-order" relational learning. However, the basic structure of human cortex is much the same as in other mammals, and especially close relatives. What could account for a general vast improvement in relational learning? The present hypothesis, that the neocortex is specialized to learn higher-order relations simply because it implements the necessary Hebbian proofreading, does not solve this puzzle but it places it in a new light.

## II Criticisms

Here we summarize weaknesses in our results and theory. As background, we first summarize the major weaknesses that we identified in an earlier version of the analysis and theory[22]; see also Elliott[145].

1.  In the early analysis, we studied the effect of crosstalk in a highly simplified model of feedforward learning, essentially 1-unit linear Hebbian learning from uncorrelated input signals; crosstalk only occurred between nearest neighbor inputs (i.e. connections were implicitly assumed to be made only of single synaptic clusters). We showed that graded crosstalk produces a graded blurring of the stable connection pattern, and derived an analytic formula in the simplest case. We pointed out that since there was no error catastrophe, there was no strong need for plasticity control. We speculated there might be an error catastrophe for nonlinear learning.

2.  The early version of proofreading was inefficient because it was only done presynaptically (via the burst/tonic transition). We also mistakenly argued that tonic = plastic, based on the idea that learning should occur in the default, tonic, mode, and particularly on the idea that if CT cells act as coincidence-detectors, their firing, which directly depolarises relays, should enable TC plasticity. Based on the analysis, we proposed that plasticity-enablement should be based on a *comparison* of coincidences at current and incipient connections. We pointed out that it would be difficult to ensure that the neighborhood relations at layer 4 and layer 6 connections might be identical.

The current version of our analysis and theory eliminates these difficulties, and provides better matches to available data.

1. We show here that in perhaps the simplest general case of nonlinear learning, introduction of modest crosstalk does produce a qualitative change at a sharp threshold, causing not a complete failure to learn, but a failure to learn anything interesting (default to socs). In the sense that learning defaults to erroneous linear learning, the neuron learns no more that what its inputs had already learned, and in this sense there is a complete learning collapse despite convergence to stable weights. This represents a key clarification of the "learning error catastrophe" concept.

2. We gradually realized that the linear neighborhood picture was likely to be wrong, at least in the case of thalamocortical (and corticocortical) connections). It's now clear that typically cortical connections are comprised of several, or even many, synapses, which are formed at points were dendrites and axons happen to approach within a spine length, and that spines appear and disappear throughout development and into maturity (though this varies between areas, over time and between labs and techniques. We shifted from a "completely local" to a "completely global" viewpoint: all connections would be neighbors of each other (though of course individual synapses still have closest neighbors, and, in the case that there are many synapses and much turnover, equally so. Of course this is unlikely to be the case at any one moment, so the pattern of errors would be described by a continuously shifting "instantaneous" matrix $E$. We surmise, but it remains to be proved, that one can roughly describe the overall effect of crosstalk, during prolonged learning, by a matrix $\mathbf{E} = <E>$. Since we expect that $E$ will reflect the happenstances of which spines appear/disappear at particular axodendritic appositions, whose locations are determined in an arbitrary, unknown and essentially random manner, this assumption should be approximately valid (it is closely related to the assumptions in molecular evolution models that mutations are position-independent, and that the relation between primary and tertiary protein structure is so complicated it can be approximated as random).  However, we also tested versions of $\mathbf{E}$ where the elements were randomly perturbed (by up to 30%) from their "equal-error-onto-all" values, and still saw an error catastrophe, at a similar threshold to the unperturbed case. Some authors have proposed that individual inputs could form synaptic "clusters", either because local dendritic segments behave as nonlinear units, or because "crosstalk" favors such clustering. However, we know of no evidence that such clustering occurs for thalamocortical connections, and while dendritic nonlinearities do occur, particularly in apical tufts (which thalamorecipient neurons typically lack), such 'clustered learning" seems unsuited to detecting very high order correlations (between very large numbers of inputs). Such "polyneurons" could be equivalent to entire networks, but only to networks that are sensitive to low-order correlations.

In the current version of proofreading (Figs 3b and Supp Fig 1) all incipient inputs are "neighbours" of currently connected inputs, and the correlations across current connections are used to disenable the plasticity of neighbors, a simpler type of comparison than in the earlier versions.

3. We realized that the accuracy of our proposed "correlation-based plasticity-control" mechanism for defeating hypothetical error catastrophes could be greatly improved by adding a postsynaptic limb to plasticity control, so that both pre- and postsynaptic control signals would have to be simultaneously present, picking out the relevant connection from the sea of irrelevant connections. Furthermore, this double control circuit closely resembled the actual connections of layer 6 CT output (branches to layer 4 and at thalamus). This could be termed "outer product" plasticity control, because it targets a particular entry in a connection matrix by choosing the conjunction of a particular row and a particular column. This combination of a relay and a layer 4 cell converging on a 6-cell, and that 6-cell feeding back to both relay and 4-cell nicely matches Callaway's summary of the connections of primate striate layer 6 cells as receiving copies of input and output, and providing output to both output and input [146].

4 Clearly it is impractical for layer 6 neurons to calculate the ratio of the correlation (= average causal coincidence rate) across all current and incipient connections, in order to control the plasticity of current connections when there are many available cells. Here we made a slight but crucial change in viewpoint: we realized that the mechanism we had postulated to control putative error catastrophes, plasticity control of connections across which correlation (= spike coincidence) is relatively weak, was formally identical to "proofreading", the main biological mechanism used to prevent DNA replication error catastrophes (the original analogy driving our work[48]). This was a remarkable simplification: instead of comparing rates of connection growth, using coincidence as the measure, one is just trying to improve coincidence detection! The comparison with coincidences across incipient connections becomes irrelevant, since these cannot grow until they form silent synapses. This shift of viewpoint is of course closely linked to the other shift outlined above[B]: instead of conflating spine addition to form silent synapses with Hebbian error as the basis for synaptic "mutation", we now separate the 2 processes, so that mutation=Hebbian error, and the creation of silent synapses increases error (because it increases crowding). Now it becomes clear that to minimize the consequences of error one has merely to make an independent assessment of connectional coincidences, and require that the 2 assessments (synaptic and neuronal) should concur. This preserves the essential logic and machinery of the older theory but radically simplifies the point of view (as well as highlighting strong connections to molecular biology: learning in both cases would be done by "pairing", of spikes or bases, and the key to successful learning would be accurate pairing).

Doubtless our current viewpoint also has many weaknesses. Candidates fall into 3 classes.

## A. Is Hebbian learning completely synapse- (and connection-) specific?

Perhaps Hebbian errors do not occur, or are so rare they do not interfere with learning. More specifically, perhaps our contention that chemical isolation and electrical coupling are incompatible is wrong, and the older (and widely current) view that spine necks have very low electrical resistance and very high "chemical resistance" is still correct, despite evidence that synaptic conductances and neck resistances may both be higher than originally estimated.

It's interesting to note that the common elementary textbook view is that specific polynucleotide pairing underlies accurate replication. In fact, the free-energy difference between correct and incorrect pairing is relatively low, and only contributes a factor of around 100 to the overall replication accuracy. Indeed, this must be the case, because if the binding were tighter by a factor of $10^7$, replication could never take place, because the 2 strands could not separate. Thus "read-out" (strand-separation) is just as important as "selectivity" – exactly analogous to the synaptic case: if the neck were so narrow that no chemical could escape, the synaptic conductance would have no effect on the neuron. Calculations suggest that the ratio of chemical to electrical isolation cannot exceed 1000 [138]. In both cases the lack of selectivity is fundamentally due to thermal agitation at body temperature.

If at least 0.1 % of the calcium that enters via unblocked NMDARs escapes to the shaft, the genie is out of the bottle: now error rates can only be kept low only by keeping synapses much further apart than the dendritic calcium length constant. Even if a strong connection can be implemented using a single synapse, this means that many times less than a thousand inputs can be learned from. Problem solving becomes much more difficult: if only 100 inputs can be learned from, then only 10X10 gray level patches can be attempted (and even less if color is added), and so forth.

It might be argued that layer 4 cells may have as few as 10 functional inputs (though the case of barrel cortex this is closer to $100^{120}$), but as we outline above, there are likely to be many more silent connections, which increase the error rate. The key is not the number of driving connections that generate the mature RF but the number of possible inputs that have to be searched through by the learning algorithm to find the mature RF (and that should continue to be available in case of cell death or changed input statistics).

Our view is consistent with the consensus that has emerged over the last decades: learning is *hard,* basically because of the curse of dimensionality. The antidote, massive neuronal parallelism, has its limits: analog Hebbian adjustments run into the other curse of all analog computation, thermal noise.

But if it could be shown that even at the strongest synapses the neck can be narrow enough that less than 1 part in a million of the entering calcium reaches the shaft, proofreading would not be necessary or even useful. At 1 part in 1000, it may not always be necessary, but could often be useful. At 1 part in 100, it becomes essential. At less than 1 part in 100, it's essential, but useful learning would occur so slowly (because almost all connections would have to be incipient) that often it would be of no use to an animal. This may be close to the true situation, simply because complex learning is so hard.

Quite apart from the question of what underlies crosstalk, studies in CA1 show it does occur[102,103]. In reference 103 the LTP threshold of around 10 synapses out of a total of around 10,000 Shaffer collateral synapses on that neuron was strongly affected, suggesting a high degree of specificity, with a $Q_{bio}$ value around 0.999, compared to the 0.88 $Q_c$ threshold value in Fig. 2. We argued above that a selfconsistent $Q_c$ value might be closer to 0.96. However, each incoming axon makes up to 10 synapses[147] so Q could drop to 0.99. However, particularly well-isolated synapses studied were selected[22], presumably overestimating Q. In addition, $Q_t$ gets much worse for low-kurtosis sources.

More generally, the question of whether the observed levels of accuracy are "negligible" or not can only be answered in the context of a specific model of the learning process (since direct experimental data on learning error catastrophes comparable to Crotty et al.[132] are not likely to be immediately forthcoming). We have provided prima facie evidence that for hoc learning the observed low level of inaccuracy may not be always adequate, and perhaps rarely.

**B Are our ICA numerical results convincing?**

Even if they are, perhaps our use of the ICA model is inappropriate, and more complex networks, such as density-estimation approaches (sparse coding nets, Helmholtz machines or Deep Belief Net etc.), are immune to crosstalk. This seems unlikely because error-free ICA learning is quite robust.

We only followed the collapse from hocs to socs over a range of error values in 4 specific cases. These particular examples of randomly-generated **M**s were chosen not because they were unusually error sensitive, but because there happened to be a large difference between the least principle component (which is set by the chance sampling errors that determine $C_s$) and the IC (set essentially by **M**), which was essential to test whether learning defaulted to an erroneous PC. In some of these cases all 3 of the weights showed clear, but modest, "jumps" at the threshold; in other cases (e.g. Fig. 2) the jump was large in 1 weight and smaller, though still apparently significant, in others. Note that we say the weights "jump" as a function of error: they take hundreds of thousands of epochs to shift from IC to PC. In all 4 cases, there was a sudden movement of the weight vector to the direction of the least eigenvector of **EC** at a threshold (as in Fig. 2). We believe that the direction of the eigenvectors of $C_L$, which is essentially set by chance fluctuations in the sample of vectors that determines $C_S$, is more or less random with respect to **M** itself, and therefore our choices of **M** are unbiased. Clearly however as we increased the sample size used to generate $C_S$ and therefore improve whitening, the threshold moved closer to the trivial value. As already noted, this makes even our average $b_c$ somewhat arbitrary. A very similar arbitrariness afflicts the Eigen molecular evolution model: although the critical error rate varies as $1/n$, the proportionality factor ln(s) is quite arbitrary. But given that some nontrivial threshold (which depends on the degree of subcortical decorrelation) exists, the key issue is that this threshold places a tight limit on the number of inputs a cortical cell can learn from, which can always be greatly increased by proofreading. Since the power of the cortex depends on how many inputs its individual neurons can learn from, proofreading would greatly increase the power of the cortex.

More specifically, one can think of our batch size as equivalent to the inverse of the biological learning rate used to do subcortical decorrelation. The learning rate essentially measures how big the strength changes produced by individual optimally timed spike pairs are. Typically in STDP experiments[87] this is around 20% for 70 to 100 optimally-timed spikes or k ~ 0.002, corresponding to a batch number of around 500. Thus the batch size we used seems biologically appropriate: it is unlikely that using real synapses the brain could do extremely good whitening. Furthermore, as we also noted above, subcortical decorrelation is further hampered by crosstalk, and selfconsistent values of $b_c$ are lower than brute values by a factor around 3.

The crosstalk threshold also depends on source statistics. Is our typical choice, Laplacian distributions, realistic? If one regards the key problem that a cortical layer 4 faces as learning from downstream cortical neurons, then the observation[148] that cortical spikes have roughly exponential distributions seems to justify this choice. The Laplacian is neither highly kurtotic nor very close to Gaussian.

We also saw sudden collapses away from the IC at a sharp threshold in a number of other cases, including n = 5, although we did not test these to see whether the collapse was to the erroneous PC. Collapses were also seen with various combinations of Laplace and Gauss sources, including 5 all Laplace sources. We also tested examples of $\mathbf{M_O}$ prepared by Gram-Schmidt orthogonalisation of $\mathbf{M}$. In these cases we found that even though $\mathbf{M_O}$ was almost perfectly orthogonal, provided the single Laplacian corresponded to the column of $\mathbf{M_O}$ that was left unaffected by the Gram-Schmidt deflation, collapses occurred at subtrivial crosstalk thresholds. These thresholds were widely scattered for different $\mathbf{M}$s. Finally, we have also done extensive tests using the multiunit Bell-Sejnowski rule, which does not require prewhitening[32]. Here the effect of crosstalk is slightly different: at a very sharp $\mathbf{M}$-dependent threshold learning becomes *unstable*, in the sense that a limit cycle emerges from a stable fixed point. We believe this is closely related to the collapse we report here: like all multiunit rules, the B-S rule includes an "antiredundancy" term which forces different output neurons to find different ICs. This means that neurons can *cooperate*: provided one of them can find an IC despite low levels of crosstalk (for example, because this row of $\mathbf{M^{-1}}$ is longer than other rows), it can provide additional clues to the others. Indeed, at the threshold, we found the network jumps between different assignments of neurons to ICs, but spends very long times between jumps at essentially correct ICs. We interpret this as reflecting the complete failure of 1 of the neurons to learn its natural IC using hocs present in its feedforward inputs, and relying instead on the strong antiredundancy signals from other neurons to save it from collapsing to a erroneous PC. Then, as error increases further, the jumps become more frequent, and the amount of time spent near correct assignments gradually decreases, until the weights move in smooth orbits.

It would of course be nice, though laborious, to obtain a more complete set of simulations using different n's, $\mathbf{M}$'s, $\mathbf{C}$'s, source statistics, learning rates, nonlinearities and $\mathbf{E}$'s, each over a range of crosstalk values. However, an analytic approach would be even more useful. Unfortunately such analysis is unlikely to be straightforward: a key ingredient in the current stability analyses is the fact that for the ICA generative model an unmixing matrix[149] or vector[79] is an exact equilibrium of the rule, so that in linearising around this point the outputs track the sources (source statistics set the stability condition). However, with crosstalk this will no longer be true. A more promising analytic route uses a statistical mechanical approach[80,150]. Here the key idea is to consider a large set of inputs; now, in the thermodynamic limit, the set of possible behaviours converges to a single overwhelmingly probable behaviour, in the manner that fluctuating forces exerted by varying molecular impacts on a surface can be replaced by an overall "pressure".

Linear square orthogonal noiseless mixing is particularly favorable for simple feedforward learning by single neurons: provided the sign of the nonlinearity is correct, the unmixing weight vector is a (often global) stable fixed point of the rule. This apparently remains true even in the presence of crosstalk. Fortunately, mixing does not have to be exactly orthogonal for the equilibrium to remain stable (such robustness in the vicinity of a hyperbolic attractor is implied by the Hartman-Grobman theorem), so it is not necessary (and would be biologically impossible) for the prewhitening to be exact. But if the input vectors are only approximately white (for example because the decorrelation matrix is learned using a slightly inaccurate Hebb rule[39,20]), the 1-unit rule fails. The fact that even the simplest most robust type of hoc learning fails with crosstalk suggests, but does not prove, that other more complex rules, which can handle more general generative models, would also fail with crosstalk, with one possible exception: if the more complex learning procedure could actually lower the effective crosstalk level (either by a type of proofreading strategy, or by improving whitening), it might allow the complex procedure to be used to solve the simple ICA problem (though whether it could still learn the more complex learning problems it was originally designed to handle is another matter). One obvious difficulty is that it has been generally difficult to prove the convergence of the more complex procedures even in the absence of crosstalk, and many results in this field remain empirical.

Even in the case of linear square mixing, 1-unit ICA has to be supplemented with additional procedures (for example, deflation) to ensure that different neurons learn different ICs ("antiredundancy"). But these procedures themselves must be learned using crosstalk-prone rules, and it is not clear that they would then remain effective, or that they could "rescue" failed feedforward learning. It is also not clear how important antiredundancy is in the brain, which has no shortage of neurons. However, the issues of whether crosstalk can prevent 1-unit learning or "antiredundancy" are rather different, and we regard the former as more fundamental. Although antiredundancy might not be important, this does not mean that redundancy somehow magically solves the hoc problem.

Density estimation techniques[75] can successfully learn from hocs for inputs that a generated in more complex ways than linear square mixing. A useful approach is to learn an approximate internal generative model, by adjusting generative weights until the joint distribution of the output of the internal and external generative models exactly match. This can be done iteratively, by using the current internal generative model to learn a recognition model, and the current recognition model to model the internal "sources" that drive the generative model. However, if the recognition model cannot be learned at all in the presence of crosstalk even when the internal generative model is perfect (suggested by our results), it is not clear that this iterative approach could be successful in the presence of crosstalk.

There is also another type of difficulty in extending the density estimation methods to the brain: even though there is extensive feedback, which could in principle implement an internal generative model,  the feedback often seems to be less precise than the feedforward paths. For example, cortical feedback to relays seems to be less precise than the thalamocortical feedforward path in several ways: the feedback is partly via TRN, which has high divergence and whose inhibitory synapses on relays are unlikely to be individually adjustable (inhibitory synapses are not Hebbian); the direct feedback is mainly modulatory rather than driving, and is located on distal dendrites where it is unlikely to control the timing of relay spikes. This suggests to us that learning in this pathway is more likely to be driven by socs than hocs. Similar difficulties attach to feedback from layer 6 to layer 4, and from higher cortical areas to lower cortical areas.

The root cause of the difficulty posed by crosstalk is that the brain cannot "learn" the instantaneous error matrix $E$ which is required to "unscramble" the effects of crosstalk, any more than a cell can use the sequences of its proteins to "correct" replication errors. So in essence we suggest that it does back-up coincidence detection; this is a rather inefficient way of preventing errors, comparable to using parity checks to defeat transmission noise. As mentioned above, there is an intriguing relation between density estimation methods and "proofreading": both require learning internal generative models, although the internal generative model used in proofreading is highly simplified: it corresponds to the transpose of the current feedforward connection matrix (which has entries 0 or 1), not to the inverse of the current feedforward weight matrix. This makes it easier to learn. Furthermore, the output of this model is NOT used "generatively"; instead it is used to control the plasticity of feedforward connections.From a neurobiology perspective, it would be more robust to learn feedforward weights online rather than feedback weights.

Another weakness of our numerical approach is that it is not clear what role machine precision, and Matlab floating point resolution, plays, and how this would affect the biological interpretation of our results. By analogy with the Eigen model we suspect that the higher the bit resolution of the learned information (in this case, the IC) the more accurate the learning rule must be, but we have not yet been able to pin this down. Biologically it seems plausible that one could partly overcome the effect of Hebbian imprecision by making the update rule, or the weight vectors themselves, binary or at least digital, and there is some evidence for this[97,158]. A rough estimate of the effective bit resolution of thalamocortical connections would be based on 1 AMPAR per coincidence, 1000 AMPARs per synapse, and 10 synapses per connection, yielding 13 bits per connection, which is much lower than our Matlab "double" precision (52 bits). However, if each synapse in a connection can be independently adjusted (for example, if only one synapse per connection is adjustable at a time[22]), then one could obtain ~100 bits per connection (and $10^5$ bits per neuron, which is about double the estimated capacity of Purkinje cells[143], which do not have proofreading).

Yet another weakness is that we did not look at the subGaussian source case, with a Hebb rule. However there are reasons for thinking this might be even more error-sensitive. The linear Hebb rule is stabilized by error. If the PC is stabilized, this would make the effect of socs even worse in the subGaussian case.

**C Proofreading**.

This brings us to the second main strand of our paper, the suggestion that the cortex defeats error by a proofreading strategy. Here the main weakness is that the proposed mechanism, while biologically plausible and consistent with much know anatomy and physiology, does not have strong direct experimental support. In particular, virtually nothing is known about mature thalamocortical plasticity (or even if it occurs), or how it might be regulated "metaplastically" by 6-4 and/or 6-T feedback. This weakness is however counterbalanced by a strength: there are no other clear and generally established theoretical proposals for the function of either the feedback loops from layer 6, or of the burst-tonic transition. Indeed, it is only recently that it has become more widely accepted that the bursting occurs at all in the waking state.

Some of the difficulties with early versions of proofreading have already been resolved (see above). Remaining difficulties can be grouped under 4 headings:

1. The feedforward anatomy onto layer 6 CT cells
2. The feedforward physiology onto these cells, including possible coincidence detection
3. Feedback anatomy onto layer 4 cells and thalamus
4. Feedback physiology, including burst/tonic modulation and plasticity regulation.

1. Feedforward Connections.

We postulate 2 sets of connections onto layer 6 CT cells: from a fixed "partner" layer 4 cell and from the set of relays that currently synapse on that partner. The second set of connections is well-known: relays that synapse in layer 4 always send branches to layer 6, and there is strong evidence for a monosynaptic connection. The issue of whether this monosynaptic connection is made primarily within layer 6 itself, or within layer 4, is less clear. A recent detailed study in cat striate cortex[151] showed that while there are many candidate "close approaches" in layer 4, none are actually occupied with EM-confirmed synapses, while there were considerable (but not numerous) confirmed synapses in layer 6. However, these synapses are much sparser than the relay synapses in layer 4 itself, suggesting that this monosynaptic pathway is relatively weak, and probably unable by itself to drive the CT cell. However, careful measurement of responses to whisker deflections in rat barrel cortex suggests that some layer 6 cells can respond with very short latencies (comparable to layer 4 latencies)[137] but see [118]. However, these fast layer 6 cells may not be CT cells.

Furthermore it is well known that in striate cortex many of the CT cells have simple receptive fields that closely resemble those of layer 4 cells vertically above them. Other CT cells are complex. It is established that the simple layer 4 RFs are at least partly (and possibly entirely) generated by the patterned convergence of appropriate relays [108, 109, 119], and it seems very plausible that this is also true for the very similar CT RFs. More generally, it has always been a puzzle why neurons distributed in a vertical column should have similar RFs, and why simple RFs should be duplicated in both 4 and 6; our hypothesis clarifies this by suggesting that the 4 RFs are singly simple and the 6 RFs are doubly simple: both simple but in crucially different ways.

The first connection is less well documented, though there are many reports that spiny stellate cells send descending axon branches into layer 6, and these branches develop very early as expected for the partner hypothesis[25]. There is also a single report of a monosynaptic spiny stellate-6 connection in cat striate cortex[162] though this was possibly on a nonCT cell. Once again however, there are also ample opportunities for CT dendrites to interact monosynaptically with thalamocortical afferents in layer 4. If these inputs are weak and distal, they might be entirely missed in dual recordings, especially since it is statistically extremely unlikely that one would record simultaneously from a connected pair (we postulate essentially no convergence or divergence in this pathway).

However, a very recent report[8] has revealed such a very early, specific connection, in a rather unusual way: this unidirectional 4-6 connection seems to be formed between clonal sister neurons that are generated by asymmetric division of a single progenitor (the reverse, 6-4 connection was not seen even though it is well-documented in the adult; in the proofreading hypothesis this reverse connection is modulatory and would not be seen using photoactivation). This finding is in remarkable accord with the proofreading hypothesis, which requires a specific and very early 4-6 partnering.

On balance then current data seems to fit this aspect of the proofreading scheme quite well.

2. Feedforward Physiology.

As already mentioned, the circuits generating layer 6 CT RFs are not fully understood: both the thalamocortical input and the layer 4 input are well suited to this task, but neither seems up to it alone. In proofreading, this is exactly what is expected: it would require the combination of both sets of inputs, with the correct timing, to fire the CT cell. Very recent data (Larkum, personal communication) suggest that layer 6 cells do possess nonlinear dendritic properties that could underlie such coincidence detection: the distal dendrites of layer 6 pyramidal neurons branch in layer 4 (as noted above) and possess voltage-dependent calcium channels that show "BAC-firing"[23]: a backpropagating AP in combination with distal layer 4 synaptic input triggers a burst of forward-propagating spikes. We suggest this burst of spikes carries the vital "proofreading confirmation" to the thalamic and layer 4 targets of the CT neuron. The initial backAP would be triggered by concerted thalamocortical input; this spike may not be emitted by the axon if there is strong but very brief chandelier inhibition, or it may not be enough to trigger confirmation.

More generally, there is considerable evidence that CT cells fire only rarely[117,118], especially during "natural" stimulation in awake animals, and that therefore TC input alone is usually not enough to fire them, in line with both anatomy (sparse inputs) and physiology (small epsps). Indeed, in order to study the RF properties of CT cells using extracellular recording it is often necessary to apply additional subliminal depolarizing current[118]. If layer 4 firing is caused by strong synchronous CT input, and layer 6 firing is caused by weaker CT synchronous input combined with weak layer layer 4 firing, why do CT cells usually fire much less than layer 4 cells (An exception is the work of de Kock et al. [137])? This brings us to another aspect of proofreading which can only be mentioned here: the role of layer 4 recurrent circuits. Most of the layer 4 spikes are presumably not triggered directly by synchronous TC spikes, but by recurrent amplification[153,154]. Since only the direct spikes coincide with TC input, one expects that layer 6 cells will fire much less often than layer 4 cells. Possibly in the conditions of the de Kock et al. experiments the recurrent spikes were not prominent (e.g. because strong whisker deflections were used, that do not require recurrent amplification). Because only spikes triggered by feedforward input are triggered by real world hocs, only these spikes (when they coincide with relay spikes) should be used for feedforward learning; conversely, if CT cells are firing lots of spikes, this activity could be used to down regulate recurrent amplification, so that "accidental" coincidences with recurrently-mediated layer 4 spikes can be minimized. In summary, we suggest that tight "causal" coincidences between relay and layer 4 spikes become increasingly rare as learning advances, so layer 6 cells fall increasingly silent (which of course increases proofreading accuracy). This may explain why relay bursting is also very rare in the alert awake state, especially in first order thalamic nuclei, which are more likely to have consummated their learning.

3. Feedback Anatomy.

The general scheme underlying Figs 3 and Supp. Fig. 1 are well accepted: CT cells feedback in thalamus to TRN and directly to distal dendrites of relay cells, in a topographic manner, and to layer 4 cells. Furthermore, all these connections are very prominent (feedback synapses typically outnumbering feedforward synapses by 10 to 1). Indeed, it is as much the abundance of the connections, as the paucity of the explanations, that constitute the layer 6 riddle which out work attempts to unravel.

However, much less information is available about the detailed anatomical pattern of the feedback. In particular, there seems to be no information as to whether the very strong layer 6 feedback to layer 4 spiny stellates, which is mediated by "drumstick" synapses ("*boutons terminaux*") like the direct feedback to relays, is convergent, divergent or 1 to 1 (roughly speaking excitatory neurons in layers 6 and 4 are equal in number, as required by our proofreading "partner" scheme, with the intriguing exception of primate striate cortex).

Several interesting papers address the pattern of feedback to thalamus[2,3,7,155]. In almost all cases these follow a "reciprocity rule": CT cells from a given cortical patch feedback to the home nucleus, and usually the general topographic location, of the relay cells that feedforward to that cortical patch. However, there are exceptions (notably, there are layer 6 inputs from motor cortex to the barreloid heads in VPM, which does not feedforward to motor cortex), which have led Deschenes et al.[155] to propose a closely related 'parity' rule: CT cells would feedback to relays that receive inputs from different branches of subcortical afferents. This emphasizes the role of the subcortical input to relays, rather than of relay input to cortex.

The rules of reciprocity and parity are in turn closely related to the rules postulated for proofreading, which requires that a CT cell feeds back to the set of relays that do *not* innervate, but *could* innervate, the layer 4 cell that partners with (and receives intracortical feedback from) the relevent layer 6 cell. It would also feedback, via TRN to all "available" relays. However, it is difficult to see why motor cortex layer 6 would supply barreloid relays (which do not supply motor cortex) under the proofreading scheme , or indeed under almost any conceivable scheme.

2  and 7 address the issue of whether CT input to TRN and to relays fits a "feedforward inhibition" or "lateral inhibition" pattern (i.e, whether a CT axon targets the same relays that receive TRN inhibition that is excited by that axon). The data seem to indicate a mixture of both types, which is in agreement with Supp. Fig. 1.

The most dramatic agreement between the postulated proofreading circuitry and data is the work of Wang et al. 1 summarized in Note CC.

4. Feedback Physiology.

A core feature of our approach has been the assumption that the burst-tonic transition plays a vital role in the waking state: it would convey an almost instant message from the relay to the TC connection it makes that "multiplexes" an additional signal onto the signals carried by the timing of individual spikes. We have compared this to sending the same message in green or red envelopes, with the color determining not the "content" of the message but the way the message is handled[22]. This is quite different from the traditional view that burst mode is simply an "idle" signal; conveying no specific message at all, and was first promoted by Murray Sherman. Indeed the unpopularity of this notion in traditional circles has made it difficult to publish our ideas. Sherman[125] has argued that the "burst" mode might serve as a "wake-up" call to cortex, since the bursts seem to be particularly effective in triggering cortical responses[156]. It's not clear however why an already awake cortex would need "waking up", or exactly on what basis this layer 6 "alarm" would be triggered. Since the burst is preceded by a long silent period, it is also not clear that the increased burst effectiveness would outweigh the necessary previous silence.

We take instead the view that any slight change in the effectiveness of information transmission or detectability[126] between tonic and burst mode is incidental to the true underlying purpose of the mode shift, which is to control the plasticity of the relevent connection, not its effectiveness. In this view neurons do 2 things: they process information and they learn (and learning improves information processing), so one must always ask whether the format of a signal varies in order to affect one or the other of these 2 (often incompatible) functions. Roughly speaking we postulate that tonic spikes carry detailed spatiotemporal information but cannot, by themselves, lead to learning, while bursts carry inevitably slightly degraded spatiotemporal information, but also enable learning.

However, there is rather little evidence for any form of TC plasticity in the adult, though remarkably little attention has been paid to the possibility that input bursting (perhaps together with postsynaptic mGluR activation) is required (there are many reports that output bursting necessary for plasticity[157]). The general view has been that TC plasticity is only important during a critical period refinement of TC connections, and that afterwards all cortical plasticity is intracortical (e.g. from 4 to 2/3, recurrent, or onto apical tufts). This is perhaps the single greatest weakness of our theory. On the other hand, the feedforward basis of the RFs of the first cortical neurons is perhaps the single strongest plank of our current understanding of cortical microcircuitry, and it seems odd that the foundation of the cortical edifice should be irrevocably laid during a rushed and fuzzy "critical" period. Under our theory intracortical learning would be particularly easy to study (it does not require proofreading) but that does not mean it is particularly important.

**General Weaknesses**

We have not attempted to prove that learning hocs is important, or that it is done using synapses rather than neurons. We take it for granted that mammals do learn to recognize objects, and that this involves learning high-order relations between sensory data. Hoc learning involves multiplication and traditionally this is thought to be done using Hebbian synapses. However, neurons can multiply and thus could learn a specific hoc. However, when we say a neuron learns from hocs, we mean that it learns an aspect of the overall pattern of many hocs: indeed, learning individual hocs seems doomed from the start because there are so many! The way an Oja neuron learns the first PC of the input distribution, as the leading eigenvector of the entire set of socs, is paradigmatic, though rather trivial: just as the PC is the line that best fits the data, the eigenvector of C summarises the essence of the socs (indeed, the 2 are the same). Another approach to ICA, based on higher-order cumulant tensors, applies this approach to hocs[79].

At a more general level, one might criticize our theory for being an overambitious and premature attempt to bring together disparate facts from unrelated fields, and wrapped up in oversimplified and over-abstract formalism. In particular, if none of the various components are completely solid, one might argue that any attempt to construct a grandiose edifice is doomed to failure. Even more specifically, this paper combines 2 startlingly different elements: a claim that nonlinear learning of hocs fails in an extremely oversimplified model neuron that is nevertheless paradigmatic of the type of complex learning cortical neurons do, and a claim that an elaborate "proofreading" circuit maps onto complex and even questionable data in ways that no other more conventional competing hypothesis can better.

Here of course tastes differ. We believe that unless there is a vigourous and detailed attempt to create new ways of thinking about cortex, neuroscience will continue to be make extremely slow process. Roughly speaking, despite the enormous output of papers since Hubel and Wiesel's magnificent start, there has been remarkably little improvement in deep understanding. We argue that perceiving the deep relations between masses of facts is an intrinsically difficult process, not only because of the combinatorial explosions involved, but also because of hardware limitations. The way to defeat the curse of dimensionality is not to use better algorithms, but better, and more massively parallel, wetware. We believe that the neocortex is above all a learning machine, and that to understand the machine one needs not only to analyse the different parts but to consider the interaction between lowest level (synapse biophysics) and highest level (complex learning) descriptions. This inevitably means taking nothing on trust: the experts at the lowest level cannot see what is required at the highest level, and vice versa. What is needed above all are fresh, and perhaps slightly naïve, perspectives.

What one grandiosely terms "mind" may be nothing more than the ability to learn accurately, and boils down to a cunning, and somewhat stale, synaptic trick.

## Section 5

### References to the Supplementary Information